\documentclass[11pt,journal,final,letterpaper,onecolumn,oneside]{IEEEtran}
\renewcommand{\baselinestretch}{1.3}
\IEEEoverridecommandlockouts
\usepackage{times,latexsym,amsfonts,amsmath,amssymb,mathrsfs,verbatim,cite,balance}
\usepackage{epsfig,epsf}
\usepackage{graphicx}

\usepackage{txfonts}
\usepackage{lscape}
\usepackage[usenames,dvips]{color}
\definecolor{myblue}{rgb}{0.,0,0}

\def\zZ{{\mathbb Z}}
\def\rR{{\mathbb R}}
\def\eE{{\mathbb E}}
\def\pP{{\mathbb P}}

\def\QED{\mbox{\rule[0pt]{1.5ex}{1.5ex}}}

\def\@begintheorem#1#2{\tmpitemindent\itemindent\topsep 0pt\rm\trivlist
    \item[\hskip \labelsep{\indent\it #1\ #2:}]\itemindent\tmpitemindent}
\def\@opargbegintheorem#1#2#3{\tmpitemindent\itemindent\topsep 0pt\rm \trivlist
    \item[\hskip\labelsep{\indent\it #1\ #2\
    \rm(#3):}]\itemindent\tmpitemindent}
\def\@endtheorem{\endtrivlist\unskip}

\newtheorem{theorem}{Theorem}
\newtheorem{definition}{Definition}
\newtheorem{fact}{Fact}
\newtheorem{proposition}{Proposition}
\newtheorem{lemma}{Lemma}
\newtheorem{corollary}{Corollary}
\newtheorem{remark}{Remark}

\newcommand{\supp}{\operatorname{supp}}
\title{The Infinite-message Limit of Two-terminal Interactive
  Source Coding}

\author{\authorblockN{Nan Ma and Prakash Ishwar\\}
\thanks{This material is based upon work
  supported by the US National Science Foundation (NSF) under awards
  (CAREER) CCF--0546598 and CCF-0915389. Any opinions, findings, and
  conclusions or recommendations expressed in this material are those
  of the authors and do not necessarily reflect the views of the
  NSF. Parts of this work were presented in Alleron'09, ITA'10, and
  ISIT'10.}
  \thanks{
N.~Ma is with Department of
Electrical Engineering and Computer Sciences at the University of
California, Berkeley, CA 94709.
e-mail: nanma@eecs.berkeley.edu.}
\thanks{
P.~Ishwar is with the Department of Electrical and Computer
Engineering at Boston University, Boston, MA 02215.
e-mail: pi@bu.edu.}

}
\begin{document}

\maketitle

\begin{abstract}
A two-terminal interactive function computation problem with
  alternating messages is studied within the framework of distributed
  block source coding theory.  {For any finite number of messages,
  a single-letter characterization of the sum-rate-distortion function
  was established in previous works} using  {standard}
  information-theoretic techniques.  This, however, does not
   {provide} a satisfactory characterization of the
  infinite-message limit, which is a new, unexplored dimension for
  asymptotic-analysis in distributed block source coding involving
  potentially  {an infinite number of} infinitesimal-rate messages.
\textcolor{myblue}{In this paper, the infinite-message sum-rate-distortion function,
viewed as a functional of the joint source pmf and the distortion levels,
is characterized as
the least element of a partially ordered family of functionals
having certain convex-geometric properties.}
 The new characterization
   {does not involve evaluating the infinite-message limit of a
  finite-message sum-rate-distortion expression.}
\textcolor{myblue}{This characterization
leads to a family of lower bounds for the infinite-message sum-rate-distortion expression
and a simple criterion to test the optimality of any
achievable infinite-message sum-rate-distortion expression.}
  For computing the  {samplewise}
  Boolean AND function of two  {physically separated} independent
  Bernoulli sources  {with zero Hamming distortion at one or both
  terminals}, the respective infinite-message minimum sum-rates are
  characterized in closed analytic form. These sum-rates are
   {shown to be} achievable using infinitely many
  infinitesimal-rate messages. The  {new convex-geometric
  characterization is used to develop} an iterative algorithm for
  evaluating any  {finite-message} sum-rate-distortion
  function.  {It is also used to} construct the  {first}
  examples which  {demonstrate} that  {for lossy source
  reproduction},  {two messages can} strictly improve the
  one-message Wyner-Ziv rate-distortion function  {settling an
  unresolved question from a 1985 paper}.   {It is shown that} a
   {single} backward message  {of} arbitrarily small rate can
  lead to an arbitrarily large gain  {in the sum-rate}.
\end{abstract}

%
%
%
%
%

\section{\textcolor{myblue}{Introduction}}\label{section:intro}

\begin{figure}[!htb]
  \begin{center}
    \scalebox{0.5}{\input{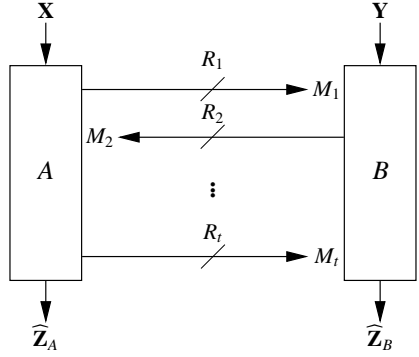}}
  \end{center}
  \caption{\sl Interactive distributed  {block} source coding with $t$
    alternating messages.
    \label{fig:structure}}
\vspace{-1ex}
\end{figure}

In this paper we study a two-terminal interactive function computation
problem with alternating messages (Figure~\ref{fig:structure}) within a
distributed block source coding framework.
 {Here, $(X(1),Y(1)), \ldots, (X(n), Y(n))$ are $n$ iid samples of
a two-component discrete memoryless stationary source with joint pmf
$p_{XY}(x,y)$, $(x, y) \in \mathcal X \times \mathcal Y$, $|\mathcal
X \times \mathcal Y|< \infty$. The $n$ samples of the first
component ${\mathbf X} := (X(1),\ldots,X(n))$ are available at
terminal $A$ whereas the $n$ samples of the second component
${\mathbf Y} := (Y(1),\ldots,Y(n))$ are available at a different
terminal $B$. The two component sources are, in general,
statistically dependent.} Terminal $A$ is required to produce a
sequence $\widehat{{\mathbf Z}}_A \in {\mathcal Z}_A^n$,
{$|{\mathcal Z}_A| < \infty$,} such that $d_{A}^{(n)}({\mathbf
X},{\mathbf Y},\widehat{{\mathbf Z}}_A) \leq D_A$ where
$d_{A}^{(n)}$ is a distortion function of $3n$ variables. Similarly,
terminal $B$ is required to produce a sequence $\widehat{{\mathbf
Z}}_B \in {\mathcal Z}_B^n$,  {$|{\mathcal Z}_B| < \infty$,} such
that $d_{B}^{(n)}({\mathbf X},{\mathbf Y},\widehat{{\mathbf Z}}_B)
\leq D_B$.
%
%
To achieve the desired objective, $t$ coded messages,
$M_1,\ldots,M_t$, of respective bit rates (bits per source sample),
$R_1,\ldots,R_t$, are sent alternately from the two terminals
starting with some terminal.  {Each} message sent from a terminal
can depend on the source samples at that terminal and on all the
previous messages (which are available to both terminals). There is
enough memory at both terminals to store all the source samples and
messages. After $t$ messages, terminal $A$ produces a sequence
$\widehat{{\mathbf Z}}_A \in {\mathcal Z}_A^n$ and terminal $B$
produces a sequence $\widehat{{\mathbf Z}}_B \in {\mathcal Z}_B^n$.
The sum-rate-distortion function ${R_{sum,t}}(D_A, D_B)$ is the
infimum of the sum of all rates $\sum_{i=1}^t R_i$ for which the
following criteria hold: $\pP(d_A^{(n)}({\mathbf X},{\mathbf
Y},\widehat{{\mathbf Z}}_A) > D_A)$ and $\pP(d_B^{(n)}({\mathbf
X},{\mathbf Y},\widehat{{\mathbf Z}}_B) > D_B)$ $\rightarrow 0$ as
$n \rightarrow \infty$.

When the {distortion} criterion is of the form of vanishing
probability of block-error
(\textcolor{myblue}{Section~\ref{section:comp}}), or of the form of
expected per-sample distortion (Section~\ref{subsection:rdproblem}),
for any {\it {finite} number} $t$, a single-letter characterization of
the set of all feasible coding rate-distortion tuples (the
rate-distortion region) and the {$t$-message} sum-rate-distortion
function ${R_{sum,t}}(D_A,D_B)$, was {established} in previous {works}
\cite{ISIT08,Journal_interactive}, {in terms of minimizing certain
  conditional mutual information quantities involving auxiliary random
  variables satisfying certain conditional independence and
  cardinality constraints.}  {This} does not, \textcolor{myblue}{in
  general,} {provide} a satisfactory characterization of the {\it
  infinite-message limit} $R_{sum,\infty}(D_A,D_B) :=
\lim_{t\rightarrow \infty} R_{sum,t}(D_A,D_B)$. The two main
objectives of this paper are to provide: {(i)} a characterization of
$R_{sum,\infty}(D_A,D_B)$ which is not in terms of evaluating the
infinite-message limit of a finite-message sum-rate-distortion
expression, and (ii) an iterative algorithm to evaluate
it. Understanding the sum-rate-distortion function in the limit where
potentially an infinite number of alternating messages are allowed to
be exchanged will shed light on the fundamental benefit of cooperative
interaction in two-terminal problems. While asymptotics involving
blocklength, rate, quantizer step-size, and network size have been
explored in the distributed block source coding literature,
asymptotics involving an infinite number of messages, each with
potentially infinitesimal rate, has not been studied. The number of
messages is a relatively unexplored resource and a new dimension for
asymptotic analysis.

\subsection{Contributions}

By viewing the sum-rate-distortion function as a functional of the
joint source distribution and distortion levels, a new
convex-geometric blocklength-free single-letter characterization of
the infinite-message sum-rate-distortion function is developed. The
new characterization is ``limit-free'' in that it does not involve
evaluating an infinite-message limit. Instead, it is in terms of the
least element of a family of partially-ordered
functionals defined by the coupled per-sample distortion criteria. The
new characterization is free of auxiliary random variables and
therefore does not involve any Markov chain constraints or cardinality
bounds. The new characterization leads to a simple criterion to test
the optimality of any achievable infinite-message sum-rate-distortion
expression.

For computing, {with zero Hamming distortion}, the {samplewise}
Boolean AND function of two independent Bernoulli {component} sources
at one or both terminals, the respective infinite-message minimum
sum-rates are fully characterized in closed analytic form. These
optimal sum-rates are shown to be achievable using a novel
construction for an infinite sequence of auxiliary random variables
that, in the limit, correspond to using infinitely many
infinitesimal-rate messages.

The functional viewpoint {is used to develop} an iterative algorithm
for evaluating any finite-message sum-rate-distortion function {which
  includes, as special cases, the rate-distortion, conditional
  rate-distortion, and Wyner-Ziv rate-distortion functions.}
\textcolor{myblue}{In the algorithm, the complexity of computation in
  each iteration does not grow with iteration number.}

{The new convex-geometric
characterization is also used to affirmatively answer the following
question that was left unresolved in\cite{Kaspi1985}: For lossy
source reproduction, can the two-message sum-rate-distortion
function be {\em strictly} smaller than the one-message Wyner-Ziv
rate-distortion function? Explicit examples are constructed to
demonstrate that the ratio and the difference between the
one-message and two-message rate-distortion functions can be
arbitrarily large and simultaneously the ratio of the backward rate
to the forward rate in the two-message sum-rate can be arbitrarily
small. These are the first known examples that explicitly
demonstrate the benefit of interaction for distributed lossy source
reproduction.}

\textcolor{myblue}{Results for the sum-rate-distortion function are
  also extended to the \emph{weighted} sum-rate-distortion function
  where the rates of messages sent from $A$ to $B$ and from $B$ to $A$
  are weighted to account for communication costs that are different
  in different directions. The weighted sum-rate-distortion function
  can be further used to characterize the directed sum-rate-distortion
  \emph{region}, which represents the tradeoff between the two
  directed sum-rates from $A$ to $B$ and from $B$ to $A$, and the
  target distortion levels.}


\subsection{Related work}

Related interactive computation problems have been studied
extensively in the area of communication complexity
~\cite{Yao1979,CommComplexity} where the main focus is on {\it
exact}  {(error-free)} computation, without regard for the
statistical dependencies in samples across terminals, and where
computing efficiency is gauged in terms of the {\it
order-of-magnitude} of the {\it total number of bits} exchanged; not
bit-rate (notable exceptions to this main focus are
\cite{Orlitskyworst1,Orlitskyworst2}). Two-way distributed block
source coding where the goal is to {\it reproduce} the sources with
a non-zero per-sample distortion, as opposed to computing functions,
was studied by Kaspi~\cite{Kaspi1985} who characterized the
$t$-message sum-rate-distortion function in each direction. Orlitsky
and Roche~\cite{OrlitskyRoche} studied two-terminal samplewise
function computation with a vanishing block-error probability and
characterized the feasible rates and the minimum sum-rate for two
alternating messages ($t=2$). A more detailed account of related
work appears in \cite{Journal_interactive}.

\textcolor{myblue}{ In \cite{Journal_col}, the infinite message limit
  of the minimum sum-rate for function computation was studied for a
  collocated network. The problem formulation there differs from that
  in this paper in the following ways. (i) A collocated network
  contains multiple source nodes and a sink node that has no
  observations, whereas the two-terminal problem network there is no
  sink node. (ii) In a collocated network the topology of
  communication is noiseless broadcast, (iii) The sources are mutually
  independent in \cite{Journal_col}, but are allowed to be arbitrarily
  related here. (iv) The function computation is lossless in
  \cite{Journal_col} but is allowed to be lossy here. Due to these
  differences, the results in this paper are not corollaries of the
  results in \cite{Journal_col} and vice versa.  The methodology in
  Sections~\ref{section:general} and \ref{section:itera} is similar to
  the counterparts in \cite{Journal_col} in spirit but is for a
  totally different problem.}

\subsection{Paper-organization and notation}

\textcolor{myblue}{For clarity of exposition,
results are first developed for the lossless function computation problem with the vanishing probability
of block-error criterion\footnote{If the per-sample distortion function  {is chosen} to be the Hamming
distortion with respect to a function of $X$ and $Y$ and the
distortion level  {is set} to zero, the characterization of the
sum-rate-distortion function essentially reduces to the
characterization of the minimum sum-rate  {for computing a function
of $X$ and $Y$} with a vanishing probability of block-error
\cite[Proposition~3]{Journal_interactive}. In this sense, the
per-sample distortion criterion is more general than the vanishing
probability of block error criterion.} in
Sections~\ref{section:comp}--\ref{section:weighted}.} In
Section~\ref{section:comp},
after formulating the problem, {the key} results from
\cite{ISIT08,Journal_interactive} that are needed for the subsequent
development are summarized. The
new convex-geometric characterization of the infinite-message
minimum sum-rate is developed in Section~\ref{section:general}.
An iterative algorithm for evaluating any finite-message minimum
sum-rate is then developed in Section~\ref{section:itera}. The
infinite-message minimum sum-rates for two examples are evaluated
in closed analytic form in Section~\ref{section:examples}.
\textcolor{myblue}{
The characterizations of the weighted minimum sum-rate and the directed sum-rate region are presented in Section~\ref{section:weighted}.}
\textcolor{myblue}{In Section~\ref{section:rd}, the results for lossless computation are generalized to the lossy rate-distortion problem with the
per-sample distortion criterion.
Then, in Section~\ref{section:Kaspi}} we use the
tools developed in Section~\ref{section:rd} to demonstrate that
 {in distributed lossy source reproduction problems, two-message}
interaction can  {significantly} improve the  {non-interactive}
Wyner-Ziv rate-distortion function.

Vectors are denoted by boldface letters; the dimension
will be clear from the context. The acronym `iid' stands for
independent and identically distributed and `pmf' stands for
probability mass function. With the exception of the symbols $R, D, N,
A$, and $B$, random quantities are denoted in upper case and their
specific instantiations in lower case. For integers $i, j$, with $i
\leq j$, $V_i^j$ denotes the sequence of random variables $V_i,
\ldots, V_j$. For $i \geq 1$, $V_1^i$ is abbreviated to $V^i$.  If $j
< i$ then ``$V_i^j$'' is the void expression ``''. More generally, if
{a quantity $Q_i$ is defined only for indices $i$ that belong to} a
subset {$\mathcal{S}$} of integers then for all integers $i$ not in
{$\mathcal{S}$}, ``$Q_i$'' $=$ ``''. For a set ${\mathcal S}$,
${\mathcal S}^n$ denotes the $n$-fold Cartesian product ${\mathcal S}
\times \ldots \times {\mathcal S}$. The support-set of a pmf $p$ is
the set over which it is strictly positive and is denoted by
$\supp(p)$. If $\supp(q) \subseteq \supp(p)$ then we write $q \ll
p$. The set of all pmfs on alphabet $\mathcal A$, i.e., the
probability simplex in $\rR^{|\mathcal {A}|}$, is denoted by
$\Delta(\mathcal A)$. $X\sim \mbox{Ber}(p)$ means
$p_{X}(1)=1-p_X(0)=p$, and $h_2(p)$ denotes its entropy in bits.
$\mbox{Unif}_{\mathcal A}$ denotes the uniform distribution in the set
$\mathcal A$. $X \Perp Y$ means $X$ and $Y$ are independent. The
indicator function of set $\mathcal{S}$ which is equal to one if $x\in
\mathcal{S}$ and is zero otherwise, is denoted by
$1_\mathcal{S}(x)$. Symbols $\wedge$ and superscript $c$ represent
Boolean AND and complement respectively. \textcolor{myblue}{A function
  $f: \mathcal{A}\rightarrow \rR$ is said to majorize another function
  $g: \mathcal{A}\rightarrow \rR$, $f$ if $\forall x\in \mathcal{A},
  f(x)\geq g(x)$.}  The hypograph of a function $f(x)$ on a set
$\mathcal A$ is given by $\mbox{hypo}_{\mathcal{A}}f:=\{(x,\phi): x\in
\mathcal{A}, \phi \leq f(x)\}$. The convex hull of a set $\mathcal A$
is denoted by $\mbox{ch}(\mathcal{A})$. For any $a\in [0,1]$, $\bar
a:=1-a$. For the erasure symbol $e$, $\bar e := e$.

\section{\textcolor{myblue}{Interactive function computation problem}}\label{section:comp}



 {Let $f_A: {\mathcal X} \times {\mathcal Y} \rightarrow {\mathcal
Z}_A$ and $f_B: {\mathcal X} \times {\mathcal Y} \rightarrow {\mathcal
Z}_B$ be functions of interest at terminals $A$ and $B$ respectively.
The desired outputs at terminals $A$ and $B$ are ${\mathbf Z}_A$ and
${\mathbf Z}_B$ respectively, where for $i=1,\ldots,n$, $Z_A(i) :=
f_A(X(i), Y(i))$ and $Z_B(i) := f_B(X(i), Y(i))$.}

\vspace{1ex}

  A two-terminal interactive distributed source code (for function
  computation) with initial terminal $A$ and parameters
  $(t,n,|{\mathcal M}_1|,\ldots,|{\mathcal M}_t|)$ is the tuple
  $(e_1,\ldots,e_t,g_A,g_B)$ of $t$ block encoding functions
  $e_1,\ldots,e_t$ and two block decoding functions $g_A, g_B$, of
  blocklength $n$, where for $j=1,\ldots,t$,
  \begin{eqnarray*}
    (\mbox{Enc.}j) &e_j:& \left\{
    \begin{array}{r@{\ , \quad}l}
      {\mathcal X}^{n} \times \bigotimes_{i=1}^{j-1} {\mathcal M}_i
      \rightarrow {\mathcal M}_j &\mbox{if }j\mbox{ is odd}\\
      {\mathcal Y}^{n} \times \bigotimes_{i=1}^{j-1} {\mathcal M}_i
      \rightarrow {\mathcal M}_j &\mbox{if }j\mbox{ is even}
    \end{array}
    \right.,\\
    (\mbox{Dec.}A) &g_A:&{\mathcal X}^{n} \times \bigotimes_{j=1}^{t}
    {\mathcal M}_j \rightarrow {\mathcal Z}_A^n,\\
    (\mbox{Dec.}B) &g_B:&{\mathcal Y}^{n} \times \bigotimes_{j=1}^{t}
    {\mathcal M}_j \rightarrow {\mathcal Z}_B^n.
  \end{eqnarray*}
  The output of $e_j$, denoted by $M_j$, is called the $j$-th message,
  and $t$ is the number of messages. The outputs of $g_A$ and $g_B$
  are denoted by $\widehat{\mathbf Z}_A$ and $\widehat{\mathbf Z}_B$
  respectively. For each $j$, $(1/n) \log_2 |{\mathcal M}_j|$ is
  called the $j$-th block-coding rate (in bits per sample). The sum of
  all the individual rates $(1/n) \sum_{j=1}^t\log_2 |{\mathcal M}_j|$
  is called the sum-rate.


  A rate tuple ${\mathbf R} = (R_1, \ldots, R_t)$ is admissible for
  $t$-message interactive function computation with initial terminal
  $A$ if, $\forall \epsilon > 0$, $\exists~ N(\epsilon,t)$ such that
  $\forall n> N(\epsilon,t)$, there exists an interactive distributed
  source code with initial terminal $A$ and parameters
  $(t,n,|{\mathcal M}_1|,\ldots,|{\mathcal M}_t|)$ satisfying
  \begin{eqnarray*}
    &&\frac{1}{n}\log_2 |{\mathcal M}_j| \leq R_j + \epsilon,\ j =
    1,\ldots, t,\\
    && \pP({\mathbf Z}_A \neq \widehat{\mathbf Z}_A) \leq \epsilon,\
    \pP({\mathbf Z}_B \neq \widehat{\mathbf Z}_B) \leq \epsilon.
  \end{eqnarray*}
%

We note that of interest here are the probabilities of block error
$\pP(\mathbf Z_A \neq \mathbf {\widehat Z}_A)$ and $\pP(\mathbf Z_B
\neq \mathbf {\widehat Z}_B)$ which are multi-letter distortion
functions. The set of all admissible rate tuples, denoted by
${\mathcal R}^A_t$, is called the operational rate region for
$t$-message interactive function computation with initial terminal
$A$.  The rate region is closed and convex due to the way it has been
defined. The minimum sum-rate is given by
\begin{equation*}
R^A_{sum,t} = \min_{\mathbf R \in \mathcal R_t^A} \sum_{j=1}^t R_j.
\end{equation*}
For initial terminal $B$, the rate region and the minimum sum-rate
are denoted by ${\mathcal R}^B_t$ and $R^B_{sum,t}$ respectively.


We allow the number of messages $t$ to be equal to $0$.  When $t=0$,
there is no message transfer and the initial terminal is irrelevant.
Thus for $t=0$, in the notation for the minimum sum-rate, we omit
the superscript and denote the minimum sum-rate  {by} $R_{sum,0}$.

For a given initial terminal, for $t=0$ and $t=1$, function
computation may not be feasible for general $p_{XY}$, $f_A$, $f_B$. If
the computation is infeasible, $\mathcal R^A_t$ is empty and we set
$R^A_{sum,t}=+\infty$. If for some specific $p_{XY}, f_A$, $f_B$, the
computation is feasible, then $R^A_{sum,t}$ will be finite. We note
that for $t\geq 2$, the computation is always feasible and
$R^A_{sum,t}$ is finite.

For all $j \leq t$, null messages, i.e., messages for which
$|{\mathcal M}_j|=1$, are permitted.
Hence, a $(t-1)$-message interactive code is a special case of a
$t$-message interactive code. Thus, $R_{sum,(t-1)}^A\geq R_{sum,t}^A$
and $R_{sum,(t-1)}^A \geq R_{sum,t}^B$ (see
\cite[Proposition~1]{ISIT08} for a detailed discussion). Therefore,
$\lim_{t \rightarrow \infty} R^A_{sum,t} = \lim_{t \rightarrow \infty}
R^B_{sum,t} =: R_{sum,\infty}$. The limit $R_{sum,\infty}$ is the
infinite-message minimum sum-rate.

Depending on the specific joint source pmf $p_{XY}$ and functions
$f_A$ and $f_B$, it may be possible to reach the infinite-message
limit $R_{sum,\infty}$ with finite $t$ (see end of
Section~\ref{subsection:oneside} for {an example}).

For all finite $t$, a single-letter characterization of the
operational rate region $\mathcal{R}_t^A$ and the minimum sum-rate
$R_{sum,t}^A$ were respectively provided in Theorem~1 and
Corollary~1 of \cite{ISIT08}.


\vspace{1ex}
\begin{fact}{\label{fact:finite}\it (Characterization of $R^{A}_{sum,t}$ \cite[Corollary~1]{ISIT08})}
\begin{equation}
  R^{A}_{sum,t} = \min_{p_{U^t|XY} \in\; \mathcal{P}_t^A(p_{XY})}
  [I(X;U^t|Y)+I(Y;U^t|X)].\label{eqn:minsumrate}
\end{equation}
\textcolor{myblue}{where $\mathcal{P}_t^A(p_{XY})$ is the set of all $p_{U^t|XY}$ such that (i) $H(f_A(X,Y)|X,U^t) = \:H(f_B(X,Y)|Y,U^t) = 0$, (ii) for
$i=1,\ldots,t$, if $i$ is odd, $U_i-(X,U^{i-1})-Y$ forms a Markov
chain, otherwise $U_i-(Y,U^{i-1})-X$ forms a Markov chain, and (iii) $\mathcal U_1,\ldots,\mathcal U_t$ are finite alphabets whose
cardinalities are bounded as follows}
\begin{equation}
  |\mathcal{U}_j| \leq \left\{
  \begin{array}{cc}
    |\mathcal{X}| \left(\prod_{i=1}^{j-1} |\mathcal{U}_i|\right) + t -
    j + 3 , & j\mbox{ odd}, \\
    |\mathcal{Y}| \left(\prod_{i=1}^{j-1} |\mathcal{U}_i|\right) + t -
    j + 3, & j\mbox{ even}.
  \end{array}
  \right.\label{eqn:cardinality}
\end{equation}
\end{fact}
\vspace{1ex}

\textcolor{myblue}{Since the constraint set of (\ref{eqn:minsumrate})
  is a compact set in a finite dimensional Euclidean space and the
  objective function is continuous, a minimizer exists in
  $\mathcal{P}_t^A(p_{XY})$ and (\ref{eqn:minsumrate}) is a finite
  dimensional optimization problem. Although the characterization of
  the minimum sum-rate in (\ref{eqn:minsumrate}) does not explicitly
  provide an actual code, it can be achieved by a sequence of
  ``Wyner-Ziv like'' codes that are based on random coding and
  binning.
The optimality of this coding strategy, i.e., the converse, is proved
using standard information inequalities, in particular Fano's
inequality, suitably defining auxiliary random variables in terms of
source components and message variables, and using Carath\'{e}odory's
theorem to establish bounds on the cardinalities of the auxiliary
random variables.}

\textcolor{myblue}{The conditional entropy constraints
  $H(f_A(X,Y)|X,U^t) = H(f_B(X,Y)|Y,U^t) = 0$ allow us to almost
  completely abstract out the function computation aspects of the
  problem and focus on the rates.  This bears resemblance to the
  formulation of rate regions in the network coding literature
  (cf. \cite{YeungISIT} and \cite[Chapter 21]{networkcoding}).}

The characterization of $R_{sum,t}^A$ in (\ref{eqn:minsumrate}) does
not directly inform us how quickly $R_{sum,t}^A$ converges to
$R_{sum,\infty}$, i.e., bounds on the rate of convergence are
unavailable for general $p_{XY}$, $f_A$, and $f_B$. In the absence of
such bounds, one pragmatic approach to estimate $R_{sum,\infty}$ is to
compute $R_{sum,t}^A$ {for increasing values of $t$ -- by using a
  computer to numerically solve (with some computer precision) the
  finite-dimensional optimization problem in (\ref{eqn:minsumrate}) --
  until the difference between $R_{sum,t-1}^A$ and $R_{sum,t}^A$ is
  smaller than some small number.} Although (\ref{eqn:minsumrate})
provides a single-letter characterization for $R_{sum,t}^A$ for each
finite $t$, as $t$ increases, an increasing number of auxiliary random
variables $U^t$ are involved in the optimization problem. In fact, due
to (\ref{eqn:cardinality}), the upper bounds for $|\mathcal U_t|$
increase exponentially with respect to $t$. Therefore, the dimension
of the optimization problem in (\ref{eqn:minsumrate}) explodes as $t$
increases. Each iteration is computationally much more demanding than
the previous one. To make matters worse, there appears to be no
obvious way of re-using the computations done for evaluating
$R_{sum,t-1}^A$ when evaluating $R_{sum,t}^A$, i.e., every time $t$ is
increased, a new optimization problem needs to be solved all over
again.  Finally, if we need to estimate $R_{sum,\infty}$ for a
different joint source pmf $p_{XY}$ (but for the same functions $f_A$
and $f_B)$, we would need to repeat this entire process for the new
$p_{XY}$.

\section{\textcolor{myblue}{Characterization of $R_{sum,\infty}(p_{XY})$}}\label{section:general}

In this section, we take a new fundamentally different approach. We
develop a general blocklength-free characterization of
$R_{sum,\infty}$ which does not involve taking a limit as
$t\rightarrow \infty$. Instead of developing the characterization of
$R_{sum,\infty}$ for a fixed joint source pmf $p_{XY}$ -- which is a
single nonnegative real number -- we characterize the entire
infinite-message minimum sum-rate surface $R_{sum,\infty}(p_{XY})$ --
which is a functional of the joint source pmf $p_{XY}$ -- in a single
concise description. This leads to a simple test for checking if a
given achievable sum-rate functional of $p_{XY}$ coincides with
$R_{sum,\infty}(p_{XY})$. It also provides a whole new family of lower
bounds for $R_{sum,\infty}$.

The key new object needed to develop the new characterization is the rate-reduction functional.

\subsection{The rate reduction functional $\rho^A_{t}(p_{XY})$}
\label{subsection:ratered}

If the goal is to {\it losslessly reproduce} the sources,
i.e., $f_A(x,y)=y, f_B(x,y)=x$, the minimum sum-rate is equal to
$H(X|Y)+H(Y|X)$ and this can be achieved by Slepian-Wolf coding. The
sum-rate needed for computing functions can only be smaller than
that needed for reproducing sources losslessly.  The reduction in
the minimum sum-rate for function computation in comparison to
source reproduction is given by
\begin{equation}
  \rho^A_{t}  :=  H(X|Y) + H(Y|X) - R^{A}_{sum,t}
   =  \max_{p_{U^t|XY} \in\;
    \mathcal{P}_t^A(p_{XY})}[H(X|Y,U^t)+H(Y|X,U^t)].\label{eqn:rho}
\end{equation}
%
For interactive distributed source codes with initial terminal $B$,
the minimum sum-rate and rate reduction are denoted by
$R^{B}_{sum,t}$ and $\rho^{B}_{t}$ respectively. A quantity which
plays a key role in the characterization of $R_{sum,\infty}$ is
$\rho^A_{0}$ corresponding to the ``rate reduction'' for zero
messages (there are no auxiliary random variables in this case).
Since the initial terminal has no significance when $t = 0$,
$\rho^A_{0} = \rho^B_{0} =: \rho_0$. Let
\begin{equation*}
  \mathcal P_{f_A f_B}:=\{p_{XY}\in \Delta(\mathcal X \times \mathcal
  Y): H(f_A(X,Y)|X)=H(f_B(X,Y)|Y)=0\}.
\end{equation*}
Error-free computations can be performed without any message transfers
if, and only if, $p_{XY}\in\; \mathcal P_{f_A f_B}$. Thus,
\begin{equation*}
  R_{sum,0}= \left\{
  \begin{array}{cc}
    0, & \mbox{if } p_{XY}\in\; \mathcal P_{f_A f_B}, \\
    + \infty, & \mbox{otherwise,}
  \end{array}
  \right.
\end{equation*}
\begin{equation}\label{eqn:rho0}
  \rho_{0}= \left\{
  \begin{array}{cc}
    H(X|Y)+H(Y|X), & \mbox{if } p_{XY}\in\; \mathcal P_{f_A f_B},\\
    - \infty, & \mbox{otherwise.}
  \end{array}
  \right.
\end{equation}

\vspace{1ex}
\begin{remark}\label{rem:simplexboundary}
If $f_A(x,y)$ is not a function of $x$ alone and $f_B(x,y)$ is not a
function of $y$ alone, then for all $p_{XY}\in \mathcal P_{f_A f_B}$,
we have $\supp(p_{XY}) \neq \mathcal X \times \mathcal Y$.  Such
$p_{XY}$ can only lie on the boundary of the probability simplex
$\Delta(\mathcal X \times \mathcal Y)$.
\end{remark}
\vspace{1ex}

Evaluating $R^{A}_{sum,t}$ is equivalent to evaluating the rate
reduction $\rho^A_{t}$.  Notice, however, that in (\ref{eqn:rho}), all
the auxiliary random variables appear only as conditioned random
variables whereas this is not the case in (\ref{eqn:minsumrate}). As
discussed in \textcolor{myblue}{Remark~\ref{rem:proofcomments2} in}
Section~\ref{subsection:mainthm}, this difference is critical as it
enables us to characterize $\rho_{\infty}:= \lim_{t \rightarrow
  \infty} \rho_t^A = \lim_{t \rightarrow \infty} \rho_t^B$ which then
gives us a characterization of $R_{sum,\infty}$ as $R_{sum,\infty} =
H(X|Y)+H(Y|X)-\rho_{\infty}$. The rate reduction functional is the key
to the characterization.

\textcolor{myblue}{\subsection{Characterization of $\rho_\infty$ for
    special families of joint source pmfs
\label{subsection:wholesimplex}}}
\textcolor{myblue}{ Generally speaking, $R^A_{sum,t}$, $\rho^A_t$,
  $R_{sum,0}$ and $\rho_{\infty}$ are functionals of $p_{XY}$, $f_A$,
  and $f_B$. We will view $R^A_{sum,t}(p_{XY})$, $\rho^A_t(p_{XY})$,
  $R_{sum,\infty}(p_{XY})$ and $\rho_{\infty}(p_{XY})$ as functionals
  of $p_{XY}$ with $f_A$ and $f_B$ fixed {in order} to emphasize the
  dependence {on} $p_{XY}$.  Ideally, we would like to characterize
  and evaluate $\rho_\infty(p_{XY})$ for a {\em single} specified
  joint source pmf $p_{XY}$ as in the point-\-to-\-point and Wyner-Ziv
  rate-distortion functions. As will become clear in the sequel, it is
  easier to characterize $\rho_{\infty}$ for a whole {\em family} of
  joint source pmfs $\mathcal{P}_{XY}$ that is {\em closed} (in a
  sense that will be made precise in
  Section~\ref{subsection:pmffamily}) than for a single joint source
  pmf $p_{XY}$. In this section, we will state and discuss the
  characterization of $\rho_\infty$ for two special but important
  families of joint source pmfs namely, the family of all product pmfs
  $\mathcal{P}_{XY} = \{p_Xp_Y| p_X \in \Delta(\mathcal{X}), p_Y \in
  \Delta(\mathcal{Y})\}$ (cf.~Proposition~\ref{prop:product}) and the
  family of all joint source pmfs $\mathcal{P}_{XY} =
  \Delta(\mathcal{X} \times \mathcal{Y})$
  (cf.~Proposition~\ref{prop:wholesimplex}). These propositions are
  special cases of a general result contained in
  Theorem~\ref{thm:functioncomp} that will be stated and proved in
  Section~\ref{subsection:mainthm}.  These propositions are intended
  to help understand the general result to follow without the
  additional complexity.}

\textcolor{myblue}{We begin by stating the characterization of the
  infinite-message limit for the family of all {\em independent}
  sources. This simplifies the characterization because independent
  sources are completely specified by the marginal distributions $p_X$
  and $p_Y$.}

\vspace{1ex}
\textcolor{myblue}{\begin{proposition}\label{prop:product}
Let $\mathcal{F}_1$ be the set of functionals
$\rho :  \{p_X p_Y|p_X\in \Delta(\mathcal X),p_Y\in
\Delta(\mathcal Y)\} \rightarrow \rR$ satisfying the
  following three conditions:
    \begin{enumerate}
    \item $\rho_0$-majorization: $\forall p_{XY}\in\; \mathcal P_{XY}$,
      $\rho(p_{XY}) \geq \rho_0(p_{XY})$.
    \item Concavity with respect to $p_X$ given $p_{Y}$:
      $\forall p_{Y}$, $\rho(p_X p_{Y} )$ is concave with respect to $p_X$.
    \item Concavity with respect to $p_Y$ given $p_{X}$:
      $\forall p_{X}$, $\rho(p_{X} p_Y)$ is concave with respect to $p_Y$.
    \end{enumerate}
Then the functional $\rho_\infty$ is the least element of the set
$\mathcal {F}_1$ with majorization as the partial ordering relation,
i.e., (i) $\rho_{\infty} \in\; \mathcal {F}_1$, (ii) for all $\rho
\in\; \mathcal {F}_1$ and all $p_X, p_Y$, we have
$\rho_{\infty}(p_Xp_Y) \leq \rho(p_Xp_Y)$.
\end{proposition}}
\vspace{1ex}

\textcolor{myblue}{The set of functionals $\mathcal{F}_1$ is partially
  ordered with respect to majorization. In general, a partially
  ordered set may not have a least
  element. Proposition~\ref{prop:product} asserts that $\mathcal
  {F}_1$ has a least element and that it is precisely
  $\rho_{\infty}$. When a least element exists, it is necessarily
  unique.}

\textcolor{myblue}{Proposition~\ref{prop:product} provides a
  limit-free characterization of $\rho_{\infty}$ in that there is no
  parameter $t$ which needs to be sent to infinity. Unlike in
  Fact~\ref{fact:finite}, this characterization is free of auxiliary
  random variables and their associated conditional entropy
  constraints, Markov chains, and cardinality bounds. In
  Fact~\ref{fact:finite}, the dependency of the minimum sum-rate on
  the desired functions is captured only through the conditional
  entropy constraints $H(f_A(X,Y)|X,U^t) = H(f_B(X,Y)|Y,U^t) =
  0$. These constraints depend on not only the joint source pmf
  $p_{XY}$ but also on the auxiliary random variables $U^t$. In
  contrast, in Proposition~\ref{prop:product}, this dependency is
  captured by the simpler constraints $H(f_A(X,Y)|X) = H(f_B(X,Y)|Y) =
  0$ which appear as part of the definition of $\mathcal P_{f_A f_B}$
  or equivalently $\rho_0$. These simpler constraints are completely
  free of the auxiliary random variables $U^t$ and can be directly
  checked for any given $p_{XY}$. If $f_A$ and $f_B$ are changed, then
  $\mathcal P_{f_A f_B}$ is changed. This changes $\rho_0$ which, in
  turn, changes $\mathcal {F}_1$ and therefore $\rho_{\infty}$. The
  characterization in Proposition~\ref{prop:product} is implicit and
  nonconstructive because it does not directly provide an algorithm
  for finding the least element $\rho_\infty$ of $\mathcal
  {F}_1$. However, we shall see that the corollaries to
  Theorem~\ref{thm:functioncomp} in Section~\ref{subsection:mainthm}
  lead to a simple optimality test for any achievable sum-rate
  functional and an iterative algorithm for evaluating $\rho_\infty$
  (cf. Section~\ref{section:itera}).}

%

\textcolor{myblue}{The infinite-message rate-reduction functional for
  independent sources turned out to be concave with respect to $p_X$
  for each fixed $p_Y$ and also concave with respect to $p_Y$ for each
  fixed $p_X$. In extending this characterization to
  $\Delta(\mathcal{X} \times \mathcal{Y})$, the family of {\em all}
  joint source pmfs, some additional qualifications are
  needed. Specifically, the concavity with respect to the $X$-marginal
  distribution $p_X$ holds for each fixed {\em conditional}
  distribution $p_{Y|X}$ as opposed to each fixed $Y$-marginal
  distribution $p_Y$. Likewise, concavity with respect to the
  $Y$-marginal $p_Y$ holds for each fixed conditional $p_{X|Y}$ as
  opposed to each fixed $X$-marginal. The counterpart of
  Proposition~\ref{prop:product} for $\Delta(\mathcal{X} \times
  \mathcal{Y})$ is given by the following proposition.}

\vspace{1ex}
\textcolor{myblue}{\begin{proposition}\label{prop:wholesimplex} Let
    $\mathcal{F}_2$ be the set of functionals $\rho:
    \Delta(\mathcal{X} \times \mathcal{Y}) \rightarrow \rR$ satisfying
    the following three conditions:
    \begin{enumerate}
    \item $\rho_0$-majorization: $\forall p_{XY}\in\; \Delta(\mathcal{X} \times \mathcal{Y})$,
      $\rho(p_{XY}) \geq \rho_0(p_{XY})$.
    \item Concavity with respect to $p_X$ given $p_{Y|X}$:
      $\forall p_{Y|X}$, $\rho(p_{Y|X} p_X)$ is concave with respect to $p_X$.
    \item Concavity with respect to $p_Y$ given $p_{X|Y}$:
      $\forall p_{X|Y}$, $\rho(p_{X|Y} p_Y)$ is concave with respect to $p_Y$.
    \end{enumerate}
Then the functional $\rho_\infty$ is the least element of the set
$\mathcal {F}_2$ with majorization as the partial ordering relation,
i.e., (i) $\rho_{\infty} \in\; \mathcal {F}_2$, (ii) for all $\rho
\in\; \mathcal {F}_2$ and all $p_{XY}\in\; \Delta(\mathcal{X} \times
\mathcal{Y})$, we have $\rho_{\infty}(p_{XY}) \leq \rho(p_{XY})$.
\end{proposition}}
\vspace{1ex}



\textcolor{myblue}{Observing the similarity of the characterizations
  in Propositions~\ref{prop:product} and \ref{prop:wholesimplex} it is
  natural to wonder whether there are other families of joint source
  pmfs for which similar characterizations hold. In other words, can
  the results of these propositions be unified and generalized in some
  way? What are the key properties required of a family of joint
  source pmfs for such a characterization to hold? This is an
  important question because one would like to have a characterization
  for as small a family as possible, ideally, for a single specified
  joint source pmf. To answer this question, we must first examine the
  key properties that are shared by the family of all product pmfs and
  the family of all joint pmfs. We will do this in the next subsection
  and then state and prove the general result in
  Section~\ref{subsection:mainthm}.}

\subsection{Marginal-\-perturbations-\-closed family of
 joint pmfs $\mathcal{P}_{XY}$}\label{subsection:pmffamily}

\textcolor{myblue}{The family of all product pmfs and the family of
  all pmfs share the following common property: if a joint pmf belongs
  to the family, then so does any other joint pmf that shares the same
  conditional pmf. Put another way (and made precise below), these
  families of joint pmfs are {\em closed} with respect to marginal
  perturbations. This type of closedness property is needed for the
  concavity conditions 2) and 3) of Propositions~\ref{prop:product}
  and \ref{prop:wholesimplex} to even make sense. The notion of
  marginal perturbations is made precise in the following definition.}

\vspace{1ex}
\begin{definition}{\it ($X$-marginal and $Y$-marginal perturbation sets
$\mathcal{P}_{Y|X}(p_{XY})$ and $\mathcal{P}_{X|Y}(p_{XY})$)}
\label{def:xperturb} The set of $X$-marginal perturbations of a pmf
$p_{XY}\in \Delta(\mathcal{X}\times\mathcal{Y})$ is defined as
\begin{equation*}
  \mathcal{P}_{Y|X}(p_{XY}) := \{p'_{XY} \in \Delta(\mathcal{X}\times
  \mathcal{Y}): p'_{XY} \ll p_{XY}, p'_{XY}p_X = p_{XY}p'_X\}
  \label{eqn:xperturb}
\end{equation*}
where $p_X$ and $p'_X$ denote the $X$-marginals of $p_{XY}$ and
$p'_{XY}$ respectively. Similarly, let
\begin{equation*}
  \mathcal{P}_{X|Y}(p_{XY}) := \{p'_{XY} \in \Delta(\mathcal{X}\times
  \mathcal{Y}): p'_{XY} \ll p_{XY}, p'_{XY}p_Y = p_{XY}p'_Y\}
  \label{eqn:yperturb}
\end{equation*}
denote the set of $Y$-marginal perturbations of $p_{XY}$ where $p_Y$
and $p'_Y$ denote the $Y$-marginals of $p_{XY}$ and $p'_{XY}$
respectively.
\end{definition}
\vspace{1ex}


Essentially, $\mathcal{P}_{Y|X}(p_{XY})$ is the collection of all
joint pmfs $p'_{XY}$ which have the same conditional pmf $p_{Y|X}$,
{that is,} $p'_{XY} = p_{Y|X}\cdot p'_X$ on $\supp(p'_{XY})$.  The
subtlety is that the conditional pmf $p'_{Y|X}$ of the joint pmf
$p'_{XY}$ is well-defined only on $\supp(p'_X)\times \mathcal{Y}$.
Corresponding statements can be made for $\mathcal{P}_{X|Y}(p_{XY})$.
The sets $\mathcal{P}_{Y|X}(p_{XY})$ and $\mathcal{P}_{X|Y}(p_{XY})$
are nonempty as they contain $p_{XY}$.

\vspace{1ex}
\begin{remark}\label{rem:margpertbsets}
For all $p_{XY}$: (i) $\mathcal{P}_{Y|X}(p_{XY})$ and
$\mathcal{P}_{X|Y}(p_{XY})$ are convex sets of joint pmfs; (ii) if
$p'_{XY} \in\; \mathcal{P}_{Y|X}(p_{XY})$ then
$\mathcal{P}_{Y|X}(p'_{XY}) \subseteq \mathcal{P}_{Y|X}(p_{XY})$; and
(iii) if $p'_{XY} \in\; \mathcal{P}_{X|Y}(p_{XY})$ then
$\mathcal{P}_{X|Y}(p'_{XY}) \subseteq \mathcal{P}_{X|Y}(p_{XY})$.
\end{remark}
\vspace{1ex}

\textcolor{myblue}{The family of joint pmfs $\mathcal{P}_{XY}$ which
  is closed with respect to $X$-marginal and $Y$-marginal
  perturbations can be formally described as follows.}

\vspace{1ex}
\begin{definition}{\it (Marginal-\-perturbations-\-closed family of
 joint pmfs $\mathcal{P}_{XY}$)}\label{def:distributionset} A family
  of joint pmfs $\mathcal P_{XY}\subseteq \Delta(\mathcal X \times
  \mathcal Y)$ will be called marginal-perturbations-closed if for all
  $p_{XY} \in\; \mathcal{P}_{XY}$, $\mathcal{P}_{Y|X}(p_{XY})\cup
  \mathcal{P}_{X|Y}(p_{XY}) \subseteq \mathcal{P}_{XY}$.
\end{definition}
\vspace{1ex}

The family of all product pmfs and the family of all pmfs are examples
of marginal-\-perturbations-\-closed families of joint pmfs. Another
example is the set of all joint pmfs with supports contained in a
specified subset of $\mathcal X \times \mathcal Y$, i.e., $\mathcal
P_{XY}=\Delta(S)$ where $S \subseteq \mathcal X \times \mathcal Y$.
If $q_X q_Y$ belongs to any marginal-perturbations-closed family with
$\supp(q_X)=\mathcal X$ and $\supp(q_Y)=\mathcal Y$, then the family
will also contain $\Delta(\mathcal{X})\times \Delta(\mathcal{Y})$,
that is, all product pmfs on $\mathcal X \times \mathcal Y$.


\textcolor{myblue}{The characterizations in
  Propositions~\ref{prop:product} and \ref{prop:wholesimplex} can be
  generalized to all marginal-perturbations-closed families of joint
  pmfs. The statement and proof of this general result and related
  corollaries is the subject matter of the next subsection.}

\subsection{Main result}\label{subsection:mainthm}

\textcolor{myblue}{In Propositions~\ref{prop:product} and
  \ref{prop:wholesimplex} we introduced two families of functionals
  $\mathcal{F}_1$ and $\mathcal{F}_2$. They can be unified as
  follows.}

\vspace{1ex}
\begin{definition}{\it (Marginal-\-perturbations-\-concave,
    $\rho_0$-\-majorizing family of functionals
    $\mathcal{F}(\mathcal{P}_{XY})$)}\label{def:F}
  Let $\mathcal P_{XY}$ be any marginal-perturbations-closed family of
  joint source pmfs on $\Delta(\mathcal X \times \mathcal Y)$. The set
  of marginal-perturbations-concave, $\rho_0$-majorizing family of
  functionals $\mathcal{F}(\mathcal{P}_{XY})$ is the set of all the
  functionals $\rho: \mathcal P_{XY} \rightarrow \rR$ satisfying the
  following three conditions:
    \begin{enumerate}
    \item $\rho_0$-majorization: $\forall p_{XY}\in\; \mathcal P_{XY}$,
      $\rho(p_{XY})\geq \rho_0(p_{XY})$.
    \item Concavity with respect to $X$-marginal perturbations:
      $\forall p_{XY}\in\; \mathcal{P}_{XY}$, $\rho$ is concave on
      $\mathcal{P}_{Y|X}(p_{XY})$.
    \item Concavity with respect to $Y$-marginal perturbations:
      $\forall p_{XY}\in\; \mathcal{P}_{XY}$, $\rho$ is concave on
      $\mathcal{P}_{X|Y}(p_{XY})$.
    \end{enumerate}
\end{definition}
\vspace{1ex}

\begin{remark}\label{rem:jointfamily}
 Since $\rho_0(p_{XY})= -\infty$ for all $p_{XY}\notin \mathcal P_{f_A
 f_B}$, condition 1) of Definition~\ref{def:F} is trivially satisfied
 for all $p_{XY}\in\; \mathcal{P}_{XY} \setminus \textcolor{myblue}{\mathcal{P}}_{f_A f_B}$ (we use
 the convention that $\forall a\in \rR$, $a > -\infty$). Thus the
 statement that $\rho$ majorizes $\rho_0$ on the set $\mathcal
 {P}_{XY}$ is equivalent to the statement that $\rho$ majorizes
 $H(X|Y)+H(Y|X)$ on the set $\mathcal P_{f_A f_B}\bigcap \mathcal
 {P}_{XY}$.
\end{remark}
\vspace{1ex}

\begin{remark}\label{rem:jointnonconvex}
 Conditions 2) and 3) do not imply that $\rho$ is concave on
 $\mathcal{P}_{XY}$. In fact, $\mathcal{P}_{XY}$ itself may not be
 convex. For example, the set $\mathcal P_{XY}=\{p_X p_Y|p_X\in
 \Delta(\mathcal X),p_Y\in \Delta(\mathcal Y)\}$ is not convex.
\end{remark}
\vspace{1ex}

We now state and prove the main result of this paper.

\vspace{1ex}
\begin{theorem}\label{thm:functioncomp}
The functional $\rho_\infty$ is the least element of the set
$\mathcal F(\mathcal P_{XY})$ with majorization as the partial
ordering relation, i.e.,
  (i) $\rho_{\infty} \in\; \mathcal F(\mathcal P_{XY})$, (ii) for all
  $\rho \in\; \mathcal F(\mathcal P_{XY})$ and all $p_{XY}\in\; \mathcal
  P_{XY}$, we have $\rho_{\infty}(p_{XY}) \leq \rho(p_{XY})$.
\end{theorem}
\vspace{1ex}

\textcolor{myblue}{This theorem provides a limit-free characterization
  of $R_{sum,\infty}$ for any marginal-\-perturbations-\-closed family
  of pmfs. In terms of the optimal coding strategy, $R_{sum,\infty}$
  can be approached by a sequence of ``Wyner-Ziv like'' codes, just as
  in the achievability proof of Fact~\ref{fact:finite}. This is
  illustrated in Appendix~\ref{app:achievabililty_example2} for the
  example studied in Section~\ref{subsection:oneside}.}

\begin{proof}
To prove Theorem~\ref{thm:functioncomp} we will use a Lemma that
establishes a connection between a $t$-message interactive coding
problem and several related $(t-1)$-message interactive coding
subproblems. Intuitively, to construct a $t$-message interactive code
with initial terminal $A$, we need to begin by choosing the first
message. This corresponds to choosing the auxiliary random variable
$U_1$.  Then for each realization $U_1 = u_1$, constructing the
remaining part of the code becomes a $(t-1)$-message subproblem with
initial terminal $B$ with the same desired functions, but with a
different source pmf $p_{XY|U_1}(\cdot,\cdot|u_1)\in\;
\mathcal{P}_{Y|X}(p_{XY})$.  Lemma~\ref{lemma:connection} connects the
rate reduction of the original problem to the rate reduction of the
subproblems.  {We can repeat this procedure recursively to construct a
  $(t-1)$-message interactive code with initial terminal $B$.  After
  $t$ steps of recursion, we will be left with the trivial $0$-message
  problem.}

{
\begin{lemma}\label{lemma:connection}
\begin{itemize}
\item[(i)] For all $t\in \zZ^+$ and all $p_{XY}\in \mathcal{P}_{XY}$,
\begin{equation}
  \rho^A_t(p_{XY})= \max_{p_{U_1|X}}\left\{ \sum_{u_1\in\;
    \supp(p_{U_1})} p_{U_1}(u_1)\:
  \rho^B_{t-1}(p_{XY|U_1}(\cdot,\cdot|u_1))\right\}.\label{eqn:convexify}
\end{equation}
\item[(ii)] For all $t\in \zZ^+$ and all $q_{XY} \in
  \mathcal{P}_{XY}$, $\rho^A_t$ is concave on
  $\mathcal{P}_{Y|X}(q_{XY})$.
\item[(iii)] For all $t\in \zZ^+$ and all $q_{XY}\in
 \mathcal{P}_{XY}$, if $\rho: \mathcal{P}_{XY} \rightarrow \rR$ is
 concave on $\mathcal{P}_{Y|X}(q_{XY})$ and if for all $p_{XY}\in
 \mathcal{P}_{Y|X}(q_{XY})$, $\rho^B_{t-1}(p_{XY}) \leq \rho(p_{XY})$,
 then for all $p_{XY} \in \mathcal{P}_{Y|X}(q_{XY})$,
 $\rho^A_t(p_{XY}) \leq \rho(p_{XY})$.
\textcolor{myblue}{\item[(iv)] The results of parts (i) -- (iii) above
  also hold if $A$ is swapped with $B$ and simultaneously,
  $\mathcal{P}_{Y|X}$ and $p_{U_1|X}$ are replaced by
  $\mathcal{P}_{X|Y}$ and $p_{U_1|Y}$ respectively.}
\end{itemize}
\end{lemma}

\textcolor{myblue}{The proof of Lemma~\ref{lemma:connection} is
  presented in Appendix~\ref{app:lemma1proof}. Here we will focus on
  explaining the intuition underlying the proof of the Lemma. Due to
  (\ref{eqn:rho}), the functional $\rho_t^A$ can be expressed as the
  maximum of $H(X|Y,U^t)+H(Y|X,U^t)$ where the maximum is over all
  choices of auxiliary random variables $U^t$. Then the following two
  properties can be established: (a) by conditioning on the random
  variable $U_1$, $H(X|Y,U^t)+H(Y|X,U^t)$ can be written as a convex
  combination of
  $\{H(X|Y,U_2^t,U_1=u_1)+H(Y|X,U_2^t,U_1=u_1)\}_{u_1\in
    \mathcal{U}_1}$; (b) for any fixed choice of $U_1$ and each
  realization $U_1 = u_1$, the maximum of
  $H(X|Y,U_2^t,U_1=u_1)+H(Y|X,U_2^t,U_1=u_1)$ over all choices of
  $U_2^t$ is $\rho_{t-1}^{B}$, that is, the rate reduction of the
  $(t-1)$-message subproblem with initial terminal $B$.  From (a) and
  (b) it would follow that for each fixed choice of $U_1$, $\rho_t^A$
  is bounded from below by a convex combination of
  $\rho_{t-1}^{B}$. Moreover, for the optimal choice of $U_1$ this
  lower bound can be shown to be tight. This line of reasoning would
  establish part (i) of the lemma. Part (i) would, in turn, imply that
  $\rho_t^A$ is a concave functional (part (ii) of the lemma) and is
  the least concave functional that majorizes $\rho_{t-1}^{B}$ (part
  (iii) of the lemma).}
\vspace{1ex}
}

%

We are now ready to prove Theorem~\ref{thm:functioncomp}.

\noindent \emph{Proof of part (i) of
Theorem~\ref{thm:functioncomp}:}
  We need to verify that $\rho_{\infty}$ satisfies all three
  conditions in Definition~\ref{def:F}:

  1) Since $\forall p_{XY}\in\; \mathcal{P}_{XY}$,
  $R_{sum,\infty}(p_{XY})\leq R_{sum,0}(p_{XY})$, we have
  $\rho_{\infty}(p_{XY})\geq \rho_0(p_{XY})$. Thus $\rho_{\infty}$ is
  $\rho_0$-majorizing.

  2) Due to part (ii) of Lemma~\ref{lemma:connection}, {for all $t\in
    \zZ^+$ and all $q_{XY} \in \mathcal{P}_{XY}$,} $\rho^A_t$ is
  concave on $\mathcal{P}_{Y|X}(q_{XY})$.  {Since for all $q_{XY} \in
    \mathcal{P}_{XY}$, $\lim_{t\rightarrow\infty}\rho^A_t(q_{XY}) =
    \rho_{\infty}(q_{XY})$, it follows that $\rho_{\infty}$ is concave
    on $\mathcal{P}_{Y|X}(q_{XY})$.}

   { 3) Due to parts (ii) and (iv) of Lemma~\ref{lemma:connection},
     for all $t\in \zZ^+$ and all $q_{XY} \in \mathcal{P}_{XY}$,
     $\rho^B_t$ is concave on $\mathcal{P}_{X|Y}(q_{XY})$. Since for
     all $q_{XY} \in \mathcal{P}_{XY}$,
     $\lim_{t\rightarrow\infty}\rho^B_t(q_{XY}) =
     \rho_{\infty}(q_{XY})$, it follows that $\rho_{\infty}$ is
     concave on $\mathcal{P}_{X|Y}(q_{XY})$.}

Thus, $\rho_{\infty} \in\; \mathcal F(\mathcal P_{XY})$.

\noindent \emph{Proof of part (ii) of Theorem~\ref{thm:functioncomp}:}
It is sufficient to show that: $\forall \rho \in\; \mathcal F(\mathcal
P_{XY})$, $\forall p_{XY}\in\; \mathcal P_{XY}$, and $\forall t\in
\zZ^+\bigcup\{0\}$, $\rho^A_t(p_{XY})\leq \rho(p_{XY})$ and
$\rho^B_t(p_{XY})\leq \rho(p_{XY})$. We prove this by induction on
$t$.  For $t = 0$, the result is true by condition 1) in
Definition~\ref{def:F}:
$\rho_0^A(p_{XY})=\rho_0^B(p_{XY})=\rho_0(p_{XY})\leq \rho(p_{XY})$.
Now assume that for an arbitrary $t\in\zZ^+$,
$\rho^A_{t-1}(p_{XY})\leq \rho(p_{XY})$ and $\rho^B_{t-1}(p_{XY})\leq
\rho(p_{XY})$ hold.  {From} parts (iii) and (iv) of
Lemma~\ref{lemma:connection}, {it follows that}
$\rho^A_{t}(p_{XY})\leq \rho(p_{XY})$ and $\rho^B_{t}(p_{XY})\leq
\rho(p_{XY})$.  {This} completes the proof of
Theorem~\ref{thm:functioncomp}.
\end{proof}
\vspace{1ex}
\vspace{1ex}
\begin{remark}\label{rem:proofcomments1}
In the proof of Theorem~\ref{thm:functioncomp}, the
marginal-perturbations-closed property of $\mathcal{P}_{XY}$ is used
in Remark~\ref{rem:lemma1i}  {in Appendix~\ref{app:lemma1proof}},
which is in turn used in the proof of part (iii) of
Lemma~\ref{lemma:connection}.
\end{remark}
\vspace{1ex}

\begin{remark}
It can be verified that the functional $(H(X|Y) + H(Y|X))$ belongs to
$\mathcal{F}(\Delta(\mathcal{X}\times\mathcal{Y}))$.  Although both
$(H(X|Y) + H(Y|X))$ and $\rho_{\infty}(p_{XY})$ are concave on
$X$-marginal and $Y$-marginal perturbation sets of $p_{XY}$, this
alone does not guarantee \textcolor{myblue}{the convexity or concavity
  of} $R_{sum,\infty}(p_{XY}) = (H(X|Y) + H(Y|X)) -
\rho_{\infty}(p_{XY})$ on the marginal perturbation sets of $p_{XY}$.
\textcolor{myblue}{An independent argument, however, can be used to
  prove that $R_{sum,\infty}$ is concave:}
\end{remark}

\vspace{1ex}
\begin{proposition}\label{prop:concavity} \textcolor{myblue}{(i) For
all $t\in \zZ^+$, $R_{sum,t}^A(p_{XY})$ is a concave functional of
$p_{XY}$ for $p_{XY}\in \Delta(\mathcal{X} \times \mathcal{Y})$. (ii)
$R_{sum,\infty}(p_{XY})$ is a concave functional of $p_{XY}$ for
$p_{XY}\in \Delta(\mathcal{X} \times \mathcal{Y})$.}
\end{proposition}
\vspace{1ex}

\textcolor{myblue}{The proof of this result is presented in
  Appendix~\ref{app:proofconcavity}. We note that the concavity in
  Proposition~\ref{prop:concavity} is not with respect to marginal
  perturbations.}

\begin{remark}\label{rem:proofcomments2}
For each $t$, $\rho^A_t$ is the maximum of $(H(X|Y,U^t)+H(Y|X,U^t))$,
where $U^t$ appear only as conditioned random variables.  This enables
us to use the ``law of total conditional entropy'' (which corresponds
to convexification) and arrive at (\ref{eqn:convexify}) {(see
  Appendix~\ref{app:lemma1proof})}. Notice, however, that
$R_{sum,\infty}$ is the minimum value of $(I(X;U^t|Y) + I(Y;U^t|X))$
over all $U^t$ \textcolor{myblue}{where $U^t$ do not appear as
  conditioned random variables. Therefore, we cannot use the same
  technique to express $R_{sum,t}^A$ as a convex combination of
  $R_{sum,t-1}^{B}$.} Due to these reasons, although evaluating
$\rho_{\infty}$ is equivalent to evaluating $R_{sum,\infty}$, the rate
reduction functional is the key to the characterization as remarked in
Section~\ref{subsection:ratered}.
\end{remark}
\vspace{1ex}

Lemma~\ref{lemma:connection} implies that $\rho^A_t$ is the least
functional that is concave on $X$-marginal perturbation sets and
majorizes $\rho^B_{t-1}$, and $\rho^B_t$ is the
least functional that is concave on $Y$-marginal perturbation sets
and majorizes $\rho^A_{t-1}$. These  {implications lead} to the
following corollary.

\vspace{1ex}
\begin{corollary}\label{cor:hypograph}
\emph{(Constructing $\rho^A_t$ and $\rho^B_t$ from $\rho^B_{t-1}$ and
  $\rho^A_{t-1}$ {respectively})} For all $t\in \zZ^+$ and {all}
$p_{XY}\in \mathcal{P}_{XY}$, we have (i)
$\mbox{ch}\left(\mbox{hypo}_{\mathcal P_{Y|X}(p_{XY})}
\rho_{t-1}^B\right) = \mbox{hypo}_{\mathcal P_{Y|X}(p_{XY})} \rho_t^A$
and (ii) $\mbox{ch}\left(\mbox{hypo}_{\mathcal P_{X|Y}(p_{XY})}
\rho_{t-1}^A\right) = \mbox{hypo}_{\mathcal P_{X|Y}(p_{XY})}
\rho_t^B$.
\end{corollary}
\vspace{1ex}

In the convex optimization literature, $(-\rho_1^A)$ is also called
the double Legendre-Fenchel transform or convex biconjugate of
$(-\rho_{t-1}^B)$ \cite{ConvexAnalysis}.
Corollary~\ref{cor:hypograph} enables us to evaluate $\rho^A_t$ and
$\rho^B_t$ for arbitrary $t$ using an iterative algorithm described in
Section~\ref{section:itera}.

Corollary~\ref{cor:concavityoptimality} establishes a connection
between the concavity property of $\rho^A_t$ {(respectively
  $\rho^B_t$)} and the optimality of $\rho^A_t$ {(respectively
  $\rho^B_t$)}.
\vspace{1ex}
\begin{corollary}\label{cor:concavityoptimality}
\emph{(Concavity and optimality of $\rho^A_t$  {and $\rho^B_t$})}
For all $t\in \zZ^+$, the following three conditions are equivalent:
\begin{itemize}
\item[(i)] $\forall p_{XY}\in \mathcal{P}_{XY},
\rho^A_t(p_{XY})=\rho_{\infty}(p_{XY})$,
\item[(ii)] $\forall p_{XY}\in \mathcal{P}_{XY},
\rho^A_t(p_{XY})=\rho^B_{t+1}(p_{XY})$,
\item[(iii)] $\forall p_{XY}\in \mathcal{P}_{XY}$, $\rho^A_t$ is
concave on $\mathcal{P}_{X|Y}(p_{XY})$.
\end{itemize}
 {For all $t\in \zZ^+$, the following three conditions are also
equivalent:
\begin{itemize}
\item[(iv)] $\forall p_{XY}\in \mathcal{P}_{XY},
\rho^B_t(p_{XY})=\rho_{\infty}(p_{XY})$,
\item[(v)] $\forall p_{XY}\in \mathcal{P}_{XY},
\rho^B_t(p_{XY})=\rho^A_{t+1}(p_{XY})$,
\item[(vi)] $\forall p_{XY}\in \mathcal{P}_{XY}$, $\rho^B_t$ is
concave on $\mathcal{P}_{Y|X}(p_{XY})$.
\end{itemize}
}
\end{corollary}
\begin{proof}
 {We will prove the equivalence of conditions (i), (ii) and
  (iii). The proof of the equivalence of conditions (iv), (v), and
  (vi) is analogous.}  Condition (i) implies (ii) because
  $\rho^A_t\leq \rho^B_{t+1}\leq \rho_{\infty}$. Condition (ii)
  implies (iii)  {due to} parts (ii) and (iv) of
  Lemma~\ref{lemma:connection}. Condition (iii) implies (i) due
  to the following  {reasons:}  {From
  Lemma~\ref{lemma:connection}(ii), Condition (iii), and the fact that
  $\rho^A_t\geq \rho_0$, we have $\rho^A_t \in
  \mathcal{F}(\mathcal{P}_{XY})$.} By Theorem~\ref{thm:functioncomp},
  $\rho^A_t\geq \rho_{\infty}$.   {But the inequality}
  $\rho^A_t\leq \rho_{\infty}$ always holds.  {Therefore we have
  $\rho^A_t=\rho_{\infty}$.}
\end{proof}
\vspace{1ex}

 {Due to Lemma~\ref{lemma:connection},} for all $t\in \zZ^+$, $\rho_t^A$
always satisfies conditions 1) and 2) in Definition~\ref{def:F}
($\rho_0$-majorization and concavity with respect to $X$-marginal
perturbations), but not necessarily condition 3); $\rho_t^B$ always
satisfies conditions 1) and 3) in Definition~\ref{def:F}
($\rho_0$-majorization and concavity with respect to $Y$-marginal
perturbations), but not necessarily condition 2). By
Theorem~\ref{thm:functioncomp}, $\rho_{\infty}$ satisfies all three
conditions of Definition~\ref{def:F}  {and is not larger than any
$\rho$ which satisfies all three conditions}. By
Corollary~\ref{cor:concavityoptimality}, once $\rho_t^A$ satisfies
condition 3) in Definition~\ref{def:F}, which is also condition
(iii) in Corollary~\ref{cor:concavityoptimality}, then
$\rho_t^A=\rho^B_{t+1}=\rho_{\infty}$.   {Similarly, once $\rho_t^B$
satisfies condition 2) in Definition~\ref{def:F}, then
$\rho_t^B=\rho^A_{t+1}=\rho_{\infty}$.  Thus, $\rho_t^A$ and
$\rho_t^B$ equal $\rho_{\infty}$ iff they satisfy all three
conditions. If all three conditions are not satisfied (two are
always satisfied), it is beneficial to increase the number of
messages. For example, if $\rho_t^A$ is not concave on a
$Y$-marginal perturbation set, then for some $p_{XY}$,
$\rho_t^A(p_{XY}) < \rho_{t+1}^B(p_{XY})$ (and $\rho_{t+1}^B(p_{XY})
\leq \rho_{t+2}^A(p_{XY})$) because otherwise $\rho_t^A =
\rho_{t+1}^B$ and by Corollary~\ref{cor:concavityoptimality},
$\rho_t^A = \rho_{\infty}$.}
 {In summary,} the concavity of $\rho_t^A$ on $Y$-marginal
perturbation sets is equivalent to the optimality of $\rho_t^A$
 {and the concavity of $\rho_t^B$ on $X$-marginal perturbation sets
is equivalent to the optimality of $\rho_t^B$. Moreover, if a single
additional message is not beneficial then no number of additional
messages will be beneficial.}

Since every $\rho\in\; \mathcal F(\mathcal P_{XY})$ gives an upper
bound for $\rho_{\infty}$, $(H(X|Y)+H(Y|X)-\rho)$ gives a lower bound
for $R_{sum,\infty}$. This fact provides a simple method to test if an
achievable sum-rate functional is optimal.
\vspace{1ex}
\begin{corollary}\label{cor:optimalitytest}
\emph{(Optimality test for an achievable sum-rate)} Let $R^*$ be a
sum-rate functional which is achievable using an arbitrary number of
messages. If $\rho^* := (H(X|Y) + H(Y|X) - R^*)\in \mathcal
F(\mathcal P_{XY})$, then $R^*=\textcolor{myblue}{R_{sum,\infty}}$.
\end{corollary}
\begin{proof}
If $R^*$ is a sum-rate functional which is achievable then $\forall
p_{XY} \in \mathcal{P}_{XY}$, $R^*(p_{XY}) \geq
R_{sum,\infty}(p_{XY})$. If $\rho^*\in \mathcal F(\mathcal P_{XY})$
then by Theorem~\ref{thm:functioncomp}, $\rho^*\geq \rho_{\infty}$
and thus $R^*\leq \textcolor{myblue}{R_{sum,\infty}}$. Therefore $R^*=\textcolor{myblue}{R_{sum,\infty}}$.
\end{proof}
\vspace{1ex}

By Corollary~\ref{cor:optimalitytest}, in order to certify that an
achievable sum-rate is optimal, we only need to verify majorization
and concavity properties of the corresponding rate reduction
functional.   {The nontrivial part of the test is to verify the
concavity properties.} We will demonstrate this test on two examples
in Section~\ref{section:examples}.


\section{Iterative algorithm for computing $R_{sum,t}^A(\cdot)$
  and $R_{sum,\infty}(\cdot)$}\label{section:itera}

Although Theorem~\ref{thm:functioncomp} provides a characterization
of $\rho_{\infty}$ and $R_{sum,\infty}$ that is not obtained by
taking a limit, it does not directly provide an algorithm to
evaluate $R_{sum,\infty}$.   {If an expression for an achievable
sum-rate as a function of $p_{XY}$ is available, we can use
Corollary~\ref{cor:optimalitytest} to test whether it coincides with
the infinite-message limit. In some cases, as in
Section~\ref{section:examples}, we may get lucky and the achievable
sum-rate expression will pass the test. In such cases, we will
obtain a closed-form expression for the infinite-message limit. In
other cases, a computer-based numerical evaluation may be the only
recourse.} To efficiently represent and search for the least element
of $\mathcal F(\mathcal P_{XY})$ is nontrivial because each element
is a functional; not a scalar.  Corollary~\ref{cor:hypograph},
however, inspires an iterative algorithm for evaluating
$R_{sum,t}^A$ and $R_{sum,\infty}$.  Corollary~\ref{cor:hypograph}
states that $\rho^A_t$ can be constructed  {on any given
$X$-marginal perturbation set} by taking  {the} convex hull of the
hypograph of $\rho^B_{t-1}$ on  {the} $X$-marginal perturbation set.
To determine $\rho_{t}^A(p_{XY})$ for all $p_{XY}\in\; \mathcal
P_{XY}$, we can, in principle, first choose a cover for $\mathcal
P_{XY}$ made up of $X$-marginal perturbation sets, say $\{\mathcal
P_{Y|X}(p_{XY})\}_{p_{XY}\in\; \mathcal A}$, where $\mathcal A
\subseteq \mathcal P_{XY}$, and then take convex hulls in every
 {$X$-marginal} perturbation set in the cover.  {This idea
leads to the following ``alternating marginal concavification''
algorithm.}

\vspace{1ex}
\textbf{Algorithm to evaluate $R_{sum,t}^A$ and $R_{sum,t}^B$}
\begin{itemize}

\item \textbf{Initialization:} Choose a marginal-perturbations-closed
  family $\mathcal P_{XY}$ containing all joint source pmfs of
  interest.  Define $\rho_0^A(p_{XY}) = \rho_0^B(p_{XY}) =
  \rho_0(p_{XY})$ by equation (\ref{eqn:rho0}) in the domain $\mathcal
  P_{XY}$. Choose a cover for $\mathcal P_{XY}$ made up of
  $X$-marginal perturbation sets, denoted by $\{\mathcal
  P_{Y|X}(p_{XY})\}_{p_{XY}\in\; \mathcal A}$, where $\mathcal
  A\subseteq \mathcal P_{XY}$. Also choose a cover for $\mathcal
  P_{XY}$ made up of $Y$-marginal perturbation sets, denoted by
  $\{\mathcal P_{X|Y}(p_{XY})\}_{p_{XY}\in\; \mathcal B}$, where
  $\mathcal B\subseteq \mathcal P_{XY}$.

\item \textbf{Loop:} For $\tau=1$ through $t$ do the following:\\
  For every $p_{XY}\in\; \mathcal A$, do the following in the set
  $\mathcal P_{Y|X}(p_{XY})$:

  \begin{itemize}

  \item Construct $\mbox{hypo}_{\mathcal P_{Y|X}(p_{XY})}\rho_{\tau-1}^B$.

  \item Let $\rho_\tau^A$ be the upper boundary of
    $\mbox{ch}\left(\mbox{hypo}_{\mathcal P_{Y|X}(p_{XY})} \rho_{\tau-1}^B\right)$.

  \end{itemize}

  For every $p_{XY}\in\; \mathcal B$, do the following in the set
  $\mathcal P_{X|Y}(p_{XY})$:

  \begin{itemize}

  \item Construct $\mbox{hypo}_{\mathcal P_{X|Y}(p_{XY})}
    \rho_{\tau-1}^A$.

  \item Let $\rho_\tau^B$ be the upper boundary of
    $\mbox{ch}\left(\mbox{hypo}_{\mathcal P_{X|Y}(p_{XY})} \rho_{\tau-1}^A\right)$.

  \end{itemize}

   Optimality test:\\
  if $\forall p_{XY}\in \mathcal{P}_{XY}$,
  $\rho^A_{\tau}(p_{XY})=\rho^B_{\tau-1}(p_{XY})$, then set $\rho_t^A
  = \rho_t^B = \rho^A_{\tau}$ and exit the loop.\\
  if $\forall p_{XY} \in \mathcal{P}_{XY}$, $\rho^B_{\tau}(p_{XY}) =
  \rho^A_{\tau-1}(p_{XY})$, then set $\rho_t^A = \rho_t^B =
  \rho^B_{\tau}$ and exit the loop.

  \item \textbf{Output:}
    $R_{sum,t}^A(p_{XY})=H(X|Y)+H(Y|X)-\rho_t^A(p_{XY})$ and
    $R_{sum,t}^B(p_{XY})=H(X|Y)+H(Y|X)-\rho_t^B(p_{XY})$.

\end{itemize}
\vspace{1ex}

To make  {computer-based numerical evaluation} feasible, $\mathcal
P_{XY}$ has to be discretized. Once discretized, however, in each
iteration, the amount of computation is the same and is fixed by the
discretization step-size. Also note that results from each iteration
are re-used in the following one. Therefore, for large $t$, the
complexity to compute $R_{sum,t}^A$ grows linearly with respect to
$t$.

$R_{sum,\infty}$ can also be evaluated to any precision, in principle,
by running this iterative algorithm for $t = 1, 2, \ldots$, until some
stopping criterion is met, e.g., the maximum difference between
$\rho_{t-1}^A$ and $\rho_t^A$ on $\mathcal P_{XY}$ falls below some
threshold. Developing stopping criteria with precision guarantees
requires some knowledge of the rate of convergence which is not
established in this paper; the rate may, however, be empirically
estimated. For the example presented in
Section~\ref{subsection:oneside}, the empirical rate of convergence
and the impact of the discretization step-size {on} the iterative
evaluation is discussed.  When the objective is to evaluate
$R_{sum,\infty}(p_{XY})$ for all pmfs in $\mathcal P_{XY}$, this
iterative algorithm is much more efficient than using
(\ref{eqn:minsumrate}) to solve for $R_{sum,t}^A$ for each $p_{XY}$
for $t = 1, 2, \ldots$, an approach which {literally} follows the
definition of $R_{sum,\infty}$ as the limit of $R_{sum,t}^A$ as
$t\rightarrow \infty$. Our iterative algorithm is based on
Theorem~\ref{thm:functioncomp} which is a {\em limit-free}
characterization of $R_{sum,\infty}$.

\section{Examples}\label{section:examples}

%
%

\textcolor{myblue}{For samplewise computation of the Boolean AND
  function of independent Bernoulli sources at both terminals, an
  achievable sum-rate functional was derived in closed analytic form
  in a previous work. Using Corollary~\ref{cor:optimalitytest} in
  Section~\ref{subsection:twoside}, we will show that this sum-rate
  functional is, in fact, optimal. In Section~\ref{subsection:oneside}
  we will present and derive a closed-form analytic expression for an
  achievable sum-rate functional for another closely related problem
  wherein the Boolean AND function of independent Bernoulli sources is
  required to be computed at only one terminal. This coding strategy
  is based on a sequence of Wyner-Ziv codes (with random coding and
  binning) and uses, in the limit, an infinite number of
  infinitesimal-rate messages. We will also establish its optimality
  providing another illustration of the use of the optimality test in
  Corollary~\ref{cor:optimalitytest}. For this example, we will also
  illustrate a numerical implementation of the iterative algorithm
  described in Section~\ref{section:itera}.}

\subsection{{\bf $R_{sum,\infty}$} for independent binary sources and
  the Boolean AND function computed at both
  terminals}\label{subsection:twoside}

In \cite[Sec.~IV.F]{ISIT08}, we studied the samplewise computation
of the Boolean AND function at both terminals for independent
Bernoulli sources, i.e., $\mathcal X = \mathcal Y = \{0,1\}$, $X
\Perp Y$, $X \sim \mbox{Ber}(p)$, $Y \sim \mbox{Ber}(q)$, and
$f_A(x,y) = f_B(x,y) = x \wedge y$. An interesting interactive
coding scheme was described in \cite{ISIT08} where the individual
rate for each message vanished as the number of messages went to
infinity. The (achievable) infinite-message sum-rate of this scheme,
denoted by $R^*$, was evaluated in closed form as
%
\begin{equation}\label{eqn:example1rate}
  R^*(p,q) = \left\{
  \begin{array}{ll}
    h_2(p) + p h_2(q) + p \log_2 q + p(1-q) \log_2 e, & \mbox{if } 0 \leq p \leq q \leq 1, \\
    R^*(q,p), & \mbox{if } 0 \leq q \leq p \leq 1.
  \end{array}
  \right.
\end{equation}
 {Actually, the expression for $R^*(p,q)$ in
(\ref{eqn:example1rate}) was only derived for the case $0 < p \leq q <
1$ in \cite[Sec.~IV.F]{ISIT08}. From this, the expression for
$R^*(p,q)$ for the case $0 < q \leq p < 1$ follows immediately by
interchanging $p$ and $q$. If $pq = 0$ then the samplewise Boolean AND
function is always $0$ and no message needs to be sent from any
terminal. Thus for all $p, q \in [0,1]$, $R^*(p,0) = R^*(0,q) = 0$. If
$(1-p)(1-q) = 0$ then either $p = 1$, in which case all samples of the
$X$-source are equal to one and the samples of the Boolean AND
function coincide with the samples of the $Y$-source, or $q = 1$, in
which case the samples of the Boolean AND function coincide with those
of the $X$-source. If $p =1$ then a single message from $B$ to $A$ of
rate $h_2(q)$ would be sufficient to compute the samplewise Boolean
AND function at both terminals and if $q = 1$ then a single message
from $A$ to $B$ of rate $h_2(p)$ would suffice.}

Since $R^*(p,q)$ is an achievable sum-rate, $R^* \geq R_{sum,\infty}$.
Using Theorem~\ref{thm:functioncomp}, we shall now prove that $R^*$
is, in fact, equal to $R_{sum,\infty}$. We will verify that $\rho^* :=
H(X|Y) + H(Y|X) - R^* $ belongs to $\mathcal F(\mathcal P_{XY})$ for
the product pmf family $\mathcal P_{XY}$, which will imply, by
Theorem~\ref{thm:functioncomp}(ii), that $\rho^* \geq \rho_{\infty}$,
i.e., $R^* \leq R_{sum,\infty}$. We note that $R_{sum,\infty}$ is not
evaluated using Theorem~\ref{thm:functioncomp}. Only part (ii) of
Theorem~\ref{thm:functioncomp} is used as a converse proof to show
that the achievable sum-rate $R^*$ is $R_{sum,\infty}$.

Since the sources are independent, we take the
marginal-perturbations-closed family to be $\mathcal P_{XY}=\{p_X
p_Y|p_X\in \Delta(\mathcal X),p_Y\in \Delta(\mathcal Y)\}$. For each
product pmf $p_X p_Y$, the $X$-marginal and $Y$-marginal perturbation
sets are $\mathcal P_{Y|X}(p_X p_Y)=\{p_X' p_Y:p_X'\ll p_X\}$ and
$\mathcal P_{X|Y}(p_X p_Y)=\{p_X p_Y':p_Y'\ll p_Y\}$ respectively.

{{\bf Parametric representation of joint source pmfs:} Since both
  sources are binary, we can use the scalars $p = p_X(1)$ and $q =
  p_Y(1)$ to represent the marginal pmfs $p_X$ and $p_Y$ respectively.
  The product pmf $p_X p_Y$ can be represented by a point $(p,q)\in
  [0,1]^2$. In the remainder of this section and in
  Section~\ref{subsection:oneside}, for convenience we shall abuse
  notation and write $R_{sum,t}^A(p,q)$ $\rho_{sum,t}^A(p,q)$,
  $R_{sum,t}^B(p,q)$, $\rho_{sum,t}^B(p,q)$, etc., instead of
  $R_{sum,t}^A(p_Xp_Y)$ $\rho_{sum,t}^A(p_Xp_Y)$,
  $R_{sum,t}^B(p_Xp_Y)$, $\rho_{sum,t}^B(p_Xp_Y)$, etc.}

 {In this parametric representation,} for all pmfs $(p,q) \in
(0,1)^2$, the $X$-marginal and $Y$-marginal perturbation sets are
the line segments $[0,1]\times \{q\}$ and $\{p\}\times [0,1]$
respectively. For all pmfs $(0,q)$, where $q\in(0,1)$, the
$X$-marginal and $Y$-marginal perturbation sets are  {represented
by} $(0,q)$ and $\{0\}\times [0,1]$ respectively. For the pmf
$(0,0)$, both the $X$-marginal and $Y$-marginal perturbation sets
are $(0,0)$. The marginal perturbation sets of remaining pmfs
 {$(p,q)$} on the boundary of $[0,1]^2$ can be derived  {by
exploiting the symmetry in the problem} (swap $p$ and $q$; then swap
symbols $0$ and $1$).

It is  {straightforward} to  {confirm} that
\begin{equation*}
  R_{sum,0}(p,q)= \left\{
  \begin{array}{cc}
    0, & \mbox{if } (p,q)\in\; \mathcal P_{f_A f_B}, \\
    + \infty, & \mbox{otherwise,}
  \end{array}
  \right.
\end{equation*}
where $\mathcal P_{f_A f_B}=\{(p,q): p=0 \mbox{ or } q=0 \mbox{ or }
p=q=1\}$. It is also  {straightforward} to verify that for all
$(p,q)$, $R^*(p,q) \leq R_{sum,0}(p,q)=0$, or equivalently,
$\rho^*(p,q)\geq \rho_{0}(p,q)$. By taking the first and
second-order partial derivatives of $\rho^*(p,q) = h_2(p) + h_2(q) -
R^*(p,q)$ with respect to $p$ and $q$, we can verify that for any
fixed $q$, $\rho^*(p,q)$ is concave with respect to $p$  {(and
therefore also with respect to $p_X$)}, and for any fixed $p$,
$\rho^*(p,q)$ is concave with respect to $q$  {(and therefore also
with respect to $p_Y$)}.  Therefore, $\rho^*(p,q)$ is concave in
every $X$-marginal and $Y$-marginal perturbation set. Therefore,
$\rho^*(p,q)\in\; \mathcal F(\mathcal P_{XY})$.  {From
\textcolor{myblue}{
Corollary~\ref{cor:optimalitytest} it follows that}
$R_{sum,\infty}(p,q) = R^*(p,q)$.

%
%

\subsection{{\bf $R_{sum,\infty}$} for independent binary sources and
  Boolean AND function computed at only terminal $B$}
  \label{subsection:oneside}

 {Consider} the problem in Section~\ref{subsection:twoside}
 {with one modification: the Boolean AND function needs to be
computed only at terminal $B$, i.e., $f_A(x,y) = 0$ and $f_B(x,y) =
x \wedge y$. The joint source distribution is unchanged:} $X \Perp
Y$, $X \sim \mbox{Ber}(p)$, $Y \sim \mbox{Ber}(q)$.  {The following
sum-rate $R^*(p,q)$ is shown to be achievable in
Appendix~\ref{app:achievabililty_example2} using the technique that
was developed in \cite[Sec.~IV.F]{ISIT08}.}
\begin{equation}
  R^*(p,q) = \left\{
  \begin{array}{ll}
    h_2(p) + p h_2(q) + p \log_2 q + p(1-2 q) \log_2 e, &    \mbox{if } 0 \leq p \leq q \leq 1/2, \\
%
%
    R^*(q,p), & \mbox{if } 0 \leq q \leq p \leq 1/2, \\
    R^*(1-p,q),& \mbox{if } 0 \leq q \leq 1/2 \leq p \leq 1, \\
    h_2(p),& \mbox{if } 1/2 \leq q \leq 1.
  \end{array}
  \right.\label{eqn:Rstaroneside}
\end{equation}
%
%
%
 {Using the parametric representation, (abuse of) notation, and
method in Section~\ref{subsection:twoside}}, it can verified that
$\mathcal P_{f_Af_B} = \{(p,q): p = 0 \mbox{ or } q = 0 \mbox{ or } p
= 1\}$ and $\rho^*(p,q) = (h_2(p) + h_2(q) - R^*(p,q))$ belongs to
$\mathcal F(\mathcal P_{XY})$, where $\mathcal P_{XY} = [0,1]^2$
is the same marginal-perturbations-closed family used in
Section~\ref{subsection:twoside}. Therefore, $R^* = R_{sum,\infty}$.

{\em Iterative algorithm:}  {We will use this example to demonstrate
how the iterative algorithm discussed in Section~\ref{section:itera}
can be implemented on a computer}.

\begin{itemize}
  \item \textbf{Initialization:} Choose $\mathcal
P_{XY}=[0,1]^2$. Choose $\mathcal A=\{(1/2,q)\}_{q\in[0,1]}$, which
leads to a cover for $\mathcal P_{XY}$ made up of $X$-marginal
perturbation sets $\{ [0,1]\times \{q\}\}_{q\in[0,1]}$. Similarly,
choose $\mathcal B=\{(p,1/2)\}_{p\in[0,1]}$, which leads to a cover
made up of $Y$-marginal perturbation sets $\{\{p\}\times
[0,1]\}_{p\in[0,1]}$.

 {For computer-based numerical evaluation}, discretize
$\mathcal{P}_{XY}$ into an $N \times N$ grid  {$\mathcal{P}_{XY}^*
:= \{(i/(N-1), j/(N-1)): i, j = 0, \ldots, N-1\}$. Correspondingly
discretize} the two covers are into the collection of the columns
and the collection of the rows of $\mathcal{P}_{XY}^*$.
Compute $\rho_0^A(p,q)=\rho_0^B(p,q)=\rho_0(p,q)$  {using equation}
(\ref{eqn:rho0}) for all $(p,q)\in \mathcal{P}_{XY}^*$ as follows:
\begin{equation}
  \rho_0(p,q)= \left\{
  \begin{array}{cc}
    h_2(p)+h_2(q), & \mbox{if } p=0 \mbox{ or }q=0\mbox{ or }p=1, \\
    - \infty, & \mbox{otherwise.}
  \end{array}
  \right.
\label{eqn:rho0example}
\end{equation}
 \item \textbf{Loop:} For $\tau=1$ through $t$ do the following.
    \begin{itemize}
      \item For every  {$q\in\{j/(N-1): j = 0, \ldots, N-1\}$}, do
    the following. Let  {$\mathcal{H}_q(\rho_{\tau-1}^B) := \{
    (p, \rho_{\tau-1}^B(p,q) ): p = i/(N-1), i = 0, \ldots, N-1,
    \rho_{\tau-1}^B(p,q) \neq -\infty \}$}. Take the convex hull
    of $\mathcal{H}_q(\rho_{\tau-1}^B)$ and denote it by
    $\mbox{ch}(\mathcal{H}_q(\rho_{\tau-1}^B))$.  {For every
    $p\in \{i/(N-1): i =0, \ldots, N-1 \}$,}
    \begin{equation*}
      \rho_{\tau}^A(p,q):= \left\{
      \begin{array}{cc}
        - \infty, & \mbox{if } \left\{\ \rho: (p,\rho)\in
        \mbox{ch}\left(\mathcal{H}_q(\rho_{\tau-1}^B)\right)\right\}=
        \varnothing , \\
        \max \left\{\ \rho: (p,\rho)\in
        \mbox{ch}\left(\mathcal{H}_q(\rho_{\tau-1}^B)\right)\right\},
        &\mbox{otherwise.}
      \end{array}
      \right.
    \end{equation*}
    By the above definition, $(p,\rho_{\tau}^A)$ is on the upper
    boundary of $\mbox{ch}(\mathcal{H}_q(\rho_{\tau-1}^B))$,
    taking the symbol $-\infty$ into consideration.
      \item For every  {$p \in \{ i/(N-1): i = 0, \ldots, N-1 \}$},
    do the following. Let  {$\mathcal{H}_p(\rho_{\tau-1}^A) :=
    \{(q, \rho_{\tau-1}^A(p,q)): q = j/(N-1), j = 0, \ldots, N-1,
    \rho_{\tau-1}^A(p,q) \neq -\infty\}$.} Take the convex hull of
    $\mathcal{H}_p(\rho_{\tau-1}^A)$ and denote it by
    $\mbox{ch}(\mathcal{H}_p(\rho_{\tau-1}^A))$. For every  {$q
    \in \{j/(N-1): j = 0, \ldots, N-1\}$,}
    \begin{equation*}
      \rho_{\tau}^B(p,q):= \left\{
      \begin{array}{cc}
        - \infty, & \mbox{if } \left\{\ \rho: (q,\rho)\in
        \mbox{ch}\left(\mathcal{H}_p(\rho_{\tau-1}^A)\right)\right\}=
        \varnothing ,\\
        \max \left\{\ \rho: (q,\rho)\in
        \mbox{ch}\left(\mathcal{H}_p(\rho_{\tau-1}^A)\right)\right\},
        &\mbox{otherwise.}
      \end{array}
      \right.
    \end{equation*}
      \item  {Optimality test:}
     {if $\forall (p,q) \in \mathcal{P}^*_{XY}$,
      $\rho^A_{\tau}(p,q) = \rho^B_{\tau-1}(p,q)$ then set
      $\rho_t^A = \rho_t^B = \rho^A_{\tau}$ and exit the loop. \\
      if $\forall (p,q) \in \mathcal{P}^*_{XY}$,
      $\rho^B_{\tau}(p,q) = \rho^A_{\tau-1}(p,q)$ then set
      $\rho_t^A = \rho_t^B = \rho^B_{\tau}$ and exit the loop.}
    \end{itemize}
  \item \textbf{Output:}
     {$R_{sum,t}^{A}(p,q)=h_2(p)+h_2(q)-\rho_t^{A}(p,q)$ and
    $R_{sum,t}^{B}(p,q)=h_2(p)+h_2(q)-\rho_t^{B}(p,q)$} for all
    $(p,q)\in\mathcal{P}^*_{XY}$.
\end{itemize}

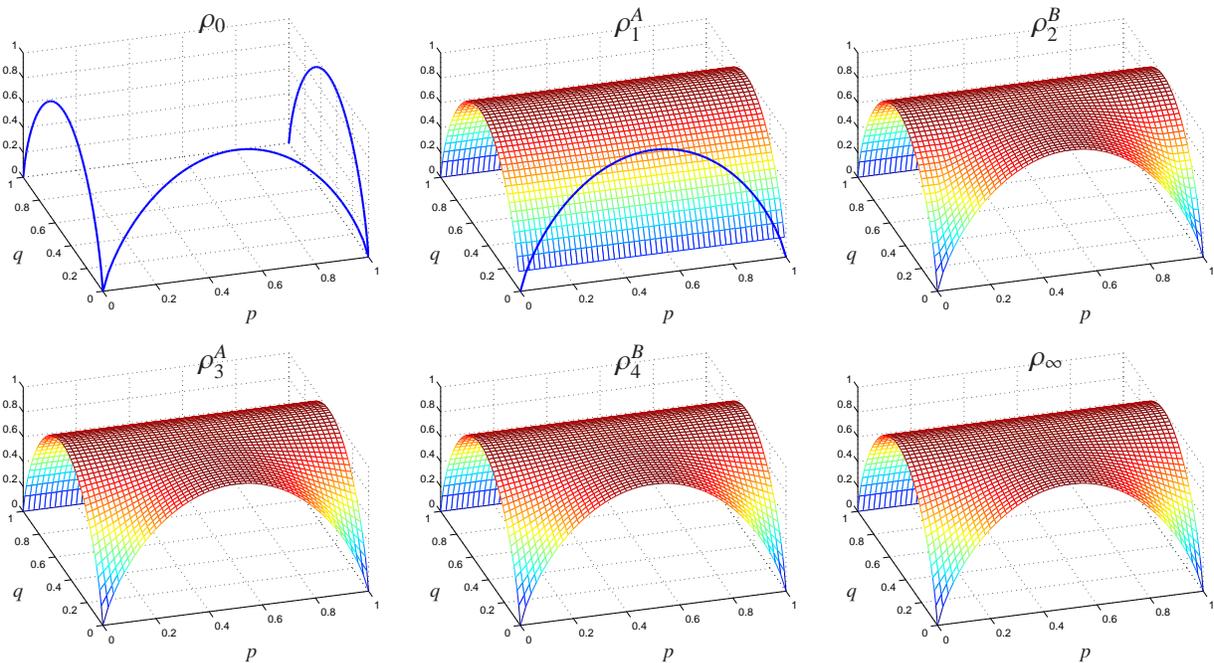
\begin{figure*}[!htb]
\centering
\includegraphics[scale = 0.4]{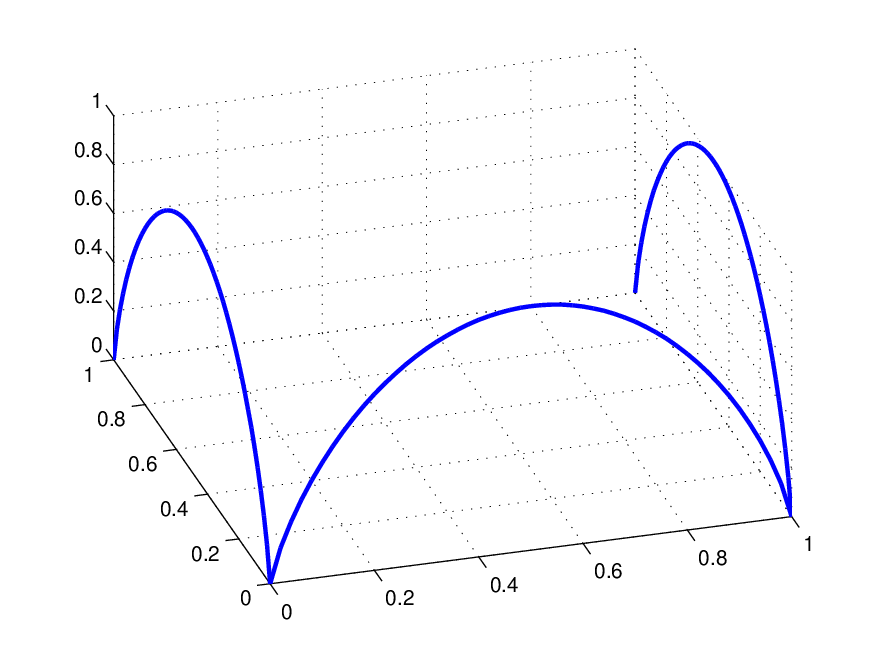}\hspace{-0.5cm}
\includegraphics[scale = 0.4]{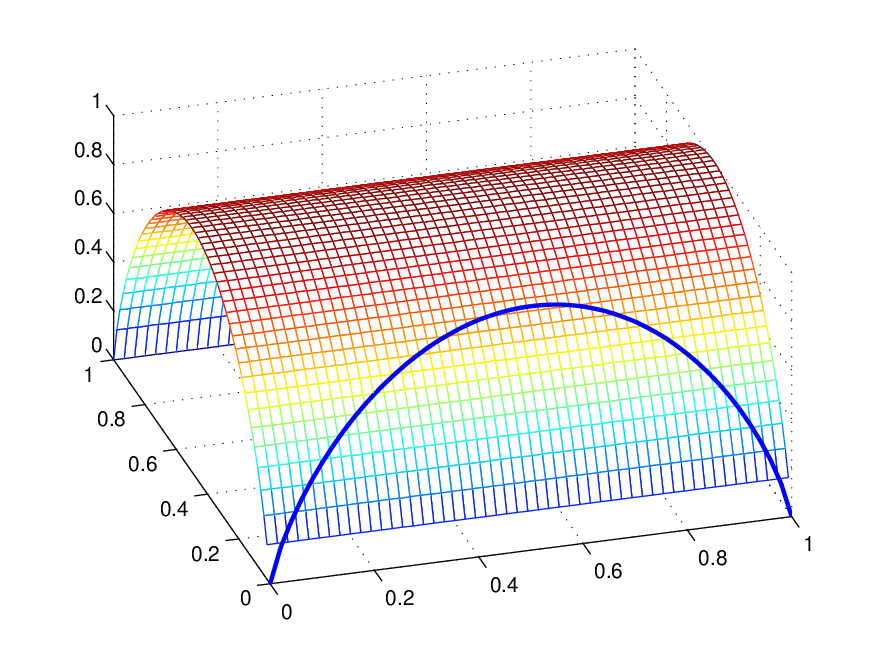}\hspace{-0.5cm}
\includegraphics[scale = 0.4]{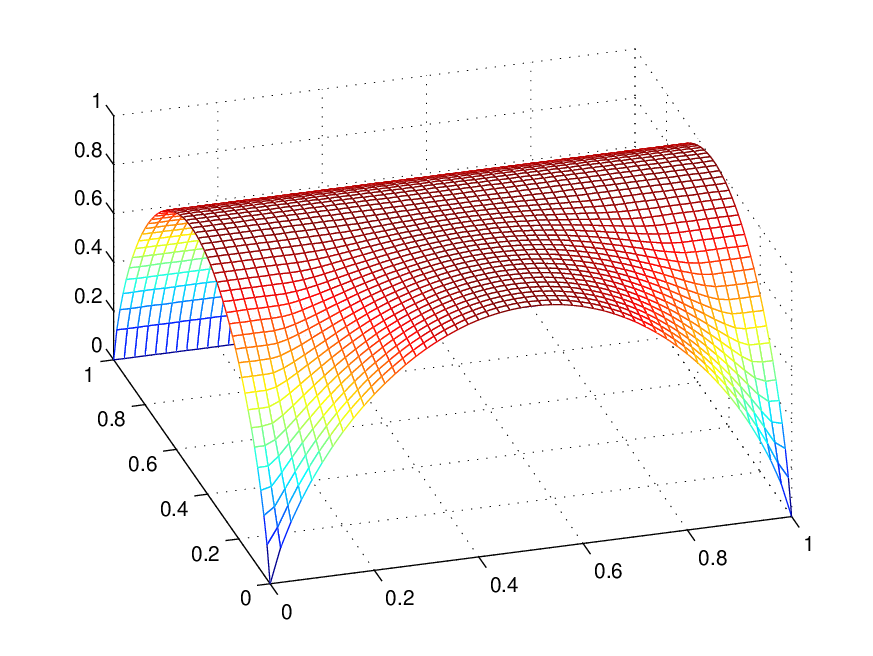}
\includegraphics[scale = 0.4]{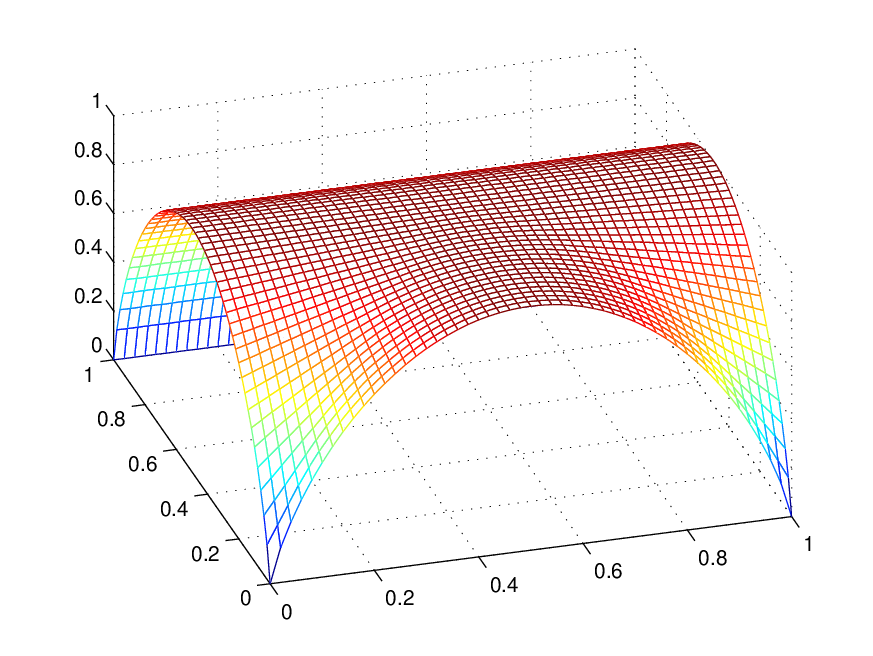}\hspace{-0.5cm}
\includegraphics[scale = 0.4]{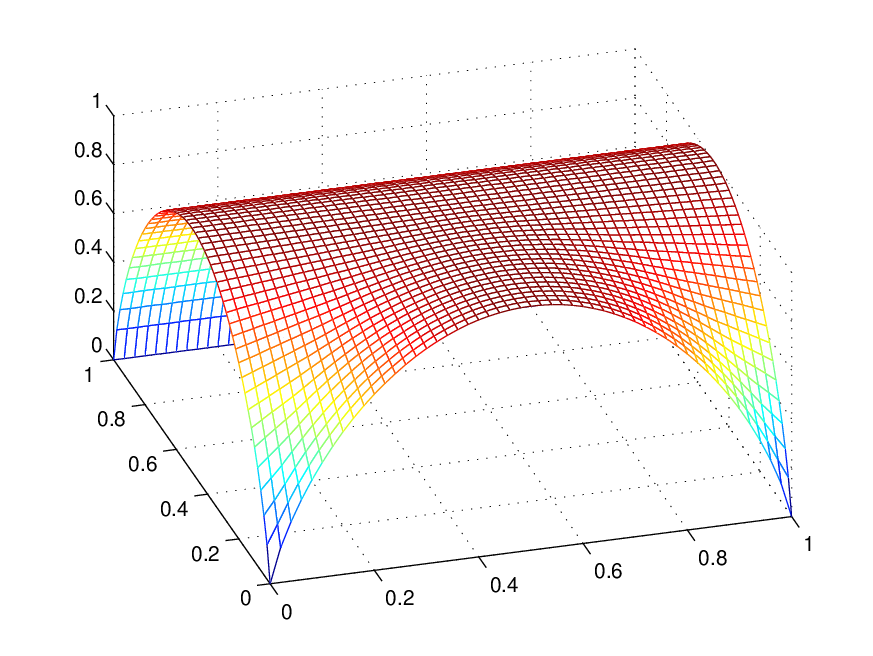}\hspace{-0.5cm}
\includegraphics[scale = 0.4]{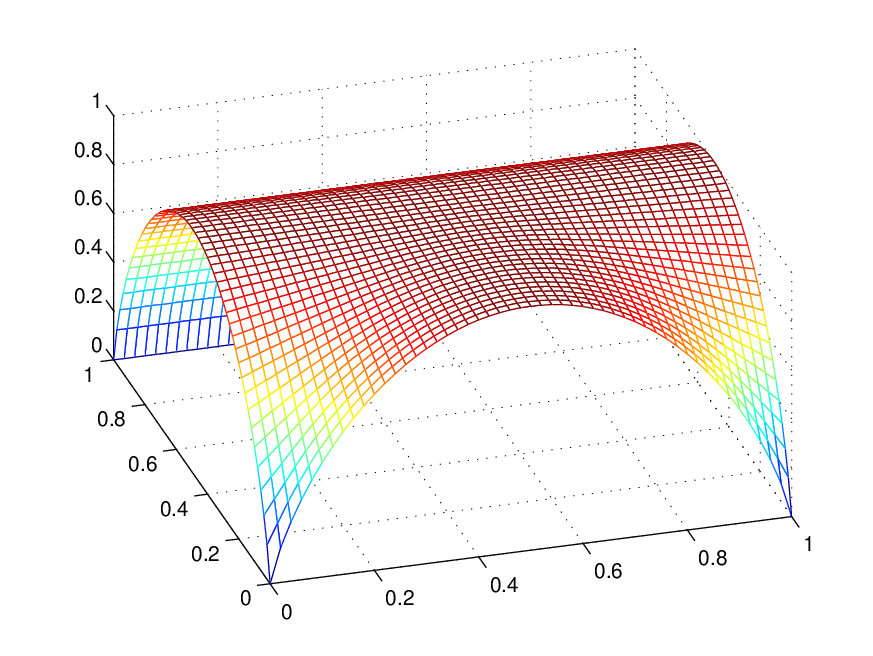}
\begin{picture}(14,0)
\put(0,0) {\put(-0.8,6){\makebox(0,0){$\scriptstyle q$}}
\put(2.3,5.2){\makebox(0,0){$\scriptstyle p$}}
\put(1.8,9.1){\makebox(0,0){$\rho_0$}}}

\put(5.54,0) {\put(-0.8,6){\makebox(0,0){$\scriptstyle q$}}
\put(2.3,5.2){\makebox(0,0){$\scriptstyle p$}}
\put(1.8,9.1){\makebox(0,0){$\rho_1^A$}}}

\put(11.08,0) {\put(-0.8,6){\makebox(0,0){$\scriptstyle q$}}
\put(2.3,5.2){\makebox(0,0){$\scriptstyle p$}}
\put(1.8,9.1){\makebox(0,0){$\rho_2^B$}}}

\put(0,-4.49) {\put(-0.8,6){\makebox(0,0){$\scriptstyle q$}}
\put(2.3,5.2){\makebox(0,0){$\scriptstyle p$}}
\put(1.8,9.1){\makebox(0,0){$\rho_3^A$}}}

\put(5.54,-4.49) {\put(-0.8,6){\makebox(0,0){$\scriptstyle q$}}
\put(2.3,5.2){\makebox(0,0){$\scriptstyle p$}}
\put(1.8,9.1){\makebox(0,0){$\rho_4^B$}}}

\put(11.08,-4.49) {\put(-0.8,6){\makebox(0,0){$\scriptstyle q$}}
\put(2.3,5.2){\makebox(0,0){$\scriptstyle p$}}
\put(1.8,9.1){\makebox(0,0){$\rho_{\infty}$}}}
\end{picture}
\caption{\sl  {Mesh plots of the rate reduction function for
  different number of messages $t$ for the example in
  Section~\ref{subsection:oneside}. Plots for $\rho_0$ and
  $\rho_\infty$ were generated using the closed-form expressions in
  (\ref{eqn:rho0example}) and (\ref{eqn:Rstaroneside}). The remaining
  plots were generated numerically on a computer using the
  ``alternating marginal concavification'' algorithm described in
  Section~\ref{subsection:oneside}.\label{fig:gallery}}
}
\end{figure*}

 {Figure~\ref{fig:gallery} shows mesh plots of the rate reduction
function for different values of $t$. While the plots for $\rho_0$ and
$\rho_\infty$ were generated using the closed-form expressions in
(\ref{eqn:rho0example}) and (\ref{eqn:Rstaroneside}), the remaining
plots were generated numerically on a computer by the iterative
algorithm.  Figure~\ref{fig:gallery} enables one to explicitly
visualize the ``alternating marginal concavification'' process:
$\rho^A_1$ is obtained from $\rho_0$ by finding the ``lowest''
surface, not lower than $\rho_0$, which is concave along the $p$-axis
for each value of $q$; $\rho^B_2$ is obtained from $\rho^A_1$ by
finding the ``lowest'' surface, not lower than $\rho^A_1$, which is
concave along the $q$-axis for each value of $p$; and so on.}

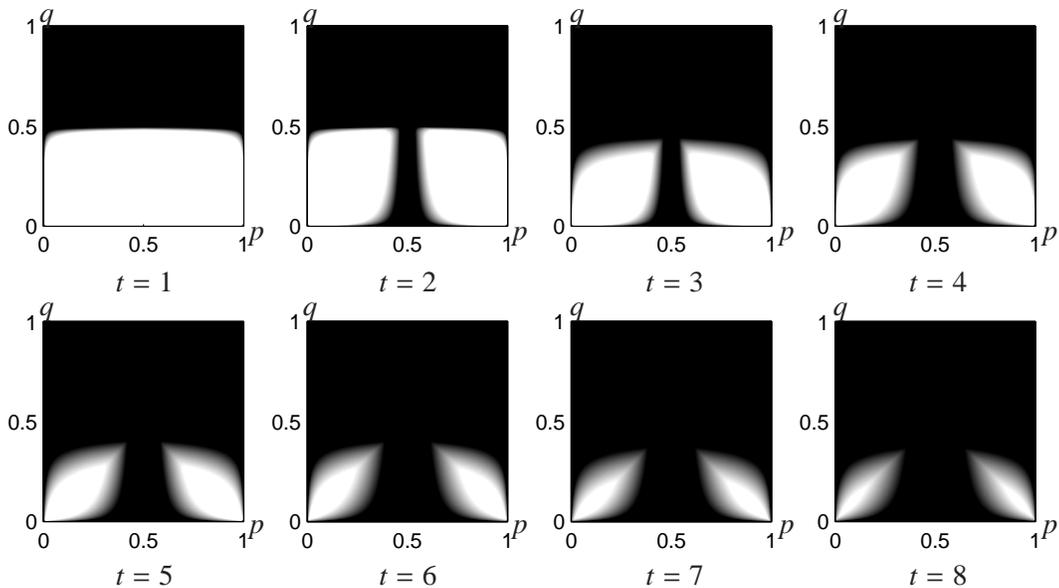
\begin{figure*}[!htb]
\centering
\includegraphics[scale = 0.8]{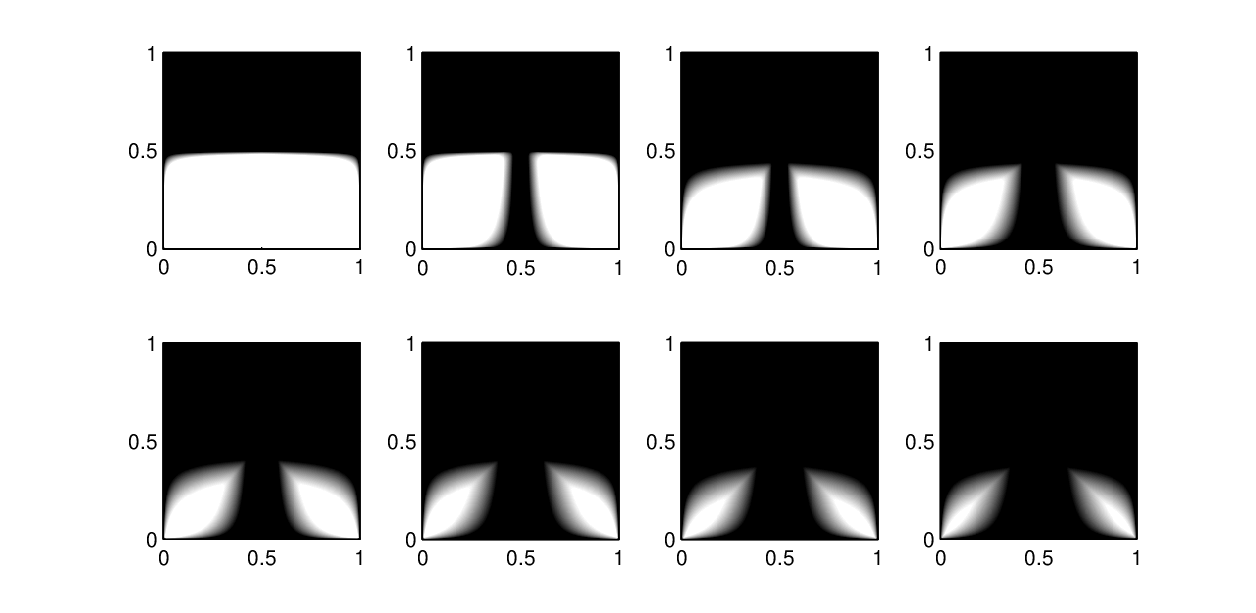}
\begin{picture}(14,0)
\put(0,0.1){ \put(0,0){ \put(2.05,4.7){\makebox(0,0){$t=1$}}
\put(3.53,5.3){\makebox(0,0){$p$}}
\put(0.75,8.2){\makebox(0,0){$q$}}}
\put(3.5,0){\put(2.05,4.7){\makebox(0,0){$t=2$}}
\put(3.53,5.3){\makebox(0,0){$p$}}
\put(0.75,8.2){\makebox(0,0){$q$}}}
\put(7.05,0){\put(2.05,4.7){\makebox(0,0){$t=3$}}
\put(3.53,5.3){\makebox(0,0){$p$}}
\put(0.75,8.2){\makebox(0,0){$q$}}}
\put(10.55,0){\put(2.05,4.7){\makebox(0,0){$t=4$}}
\put(3.53,5.3){\makebox(0,0){$p$}}
\put(0.75,8.2){\makebox(0,0){$q$}}}
\put(0,-3.9){\put(2.05,4.7){\makebox(0,0){$t=5$}}
\put(3.53,5.3){\makebox(0,0){$p$}}
\put(0.75,8.2){\makebox(0,0){$q$}}}
\put(3.5,-3.9){\put(2.05,4.7){\makebox(0,0){$t=6$}}
\put(3.53,5.3){\makebox(0,0){$p$}}
\put(0.75,8.2){\makebox(0,0){$q$}}}
\put(7.05,-3.9){\put(2.05,4.7){\makebox(0,0){$t=7$}}
\put(3.53,5.3){\makebox(0,0){$p$}}
\put(0.75,8.2){\makebox(0,0){$q$}}}
\put(10.55,-3.9){\put(2.05,4.7){\makebox(0,0){$t=8$}}
\put(3.53,5.3){\makebox(0,0){$p$}}
\put(0.75,8.2){\makebox(0,0){$q$}}}}
\end{picture}
\caption{\sl
 {For the example in Section~\ref{subsection:oneside}, the
differences $(\rho_\infty - \rho_t)$, for $t = 1, 2, \ldots$, are
plotted as grayscale images where the brightness at any coordinate
$(p,q)$ is proportional to $\log(\rho_\infty(p,q) -
\rho_t(p,q))$. Brighter (whiter) shades correspond to larger
differences between $\rho_\infty$ and $\rho_t$ while a pure-black
shade corresponds to a difference that is smaller than $10^{-4}$.}
\label{fig:contourplot}}
\end{figure*}

 {For $t\geq 2$, the points at which successive rate reduction
functions differ from $\rho_\infty$ are only barely distinguishable
(visually) in Figure~\ref{fig:gallery}. To better visualize the rate
of convergence, in Figure~\ref{fig:contourplot} we plot the
differences $(\rho_\infty - \rho_t) = (R_{sum,t} - R_{sum,\infty})$,
$t = 1, 2, \ldots$, as grayscale images where the brightness at any
coordinate $(p,q)$ is proportional to $\log(\rho_\infty(p,q) -
\rho_t(p,q))$. Brighter (whiter) shades correspond to larger
differences between $\rho_\infty$ and $\rho_t$ while a pure-black
shade corresponds to a difference that is smaller than $10^{-4}$.}

{\textbf{Reaching $R_{sum,\infty}$ with finite $t$:} For some joint
  source pmfs, the limit $R_{sum,\infty}$ can possibly be reached by
  $R_{sum,t}$ with a finite $t$. Specifically,}
 {
\begin{itemize}
 \item[(i)] For all $(p,q) \in \mathcal P_{f_Af_B}$, $R_{sum,0} = 0 =
   R_{sum,\infty}$ and no message needs to be sent.
 \item[(ii)] For all $(p,q)\in (0,1)\times[1/2,1]$, $R_{sum,\infty} =
   h_2(p)$ and this sum-rate can be achieved with $t = 1$ message from
   $A$ to $B$. Thus $R^A_{sum,1} = R_{sum,\infty}$. However, note that
   $R^B_{sum,1} = \infty$ and because $(p,q) \notin \mathcal
   P_{f_Af_B}$, $R_{sum,0} = \infty$.
 \item[(iii)] For $(p,q)\in \{1/2\} \times (0,1/2)$, $R_{sum,\infty} =
   h_2(q)$. In \cite[Sec.~V.C]{OrlitskyRoche} it was shown that this
   sum-rate can be achieved with $t = 2$ messages, the first from $B$
   to $A$ and the second from $A$ to $B$. Thus $R^B_{sum,2} =
   R_{sum,\infty}$. We note that $R^B_{sum,1} = \infty$ and in
   \cite[Sec.~IV.C]{ISIT08} it was shown that $R^A_{sum,1} = \log_2 2
   = 1$.
\end{itemize}
}
 {Thus for the distributions $(p,q)$ discussed above,
$R_{sum,\infty}$ can be reached with $t = 0$, $1$ or $2$
messages. However, Figure~\ref{fig:contourplot} also shows that for
values of $(p,q)$ that are close to the line segments $p = q < 1/2$
and $(1-p) = q < 1/2$, even when $t = 8$, $R_{sum,t}$ is not very
close to $R_{sum,\infty}$. We conjecture that for the example
considered in this section, there exist values of $(p,q)$ close to the
line segments $p = q < 1/2$ and $(1-p) = q < 1/2$ for which
$R_{sum,t}(p,q)$ is {\em strictly} smaller than $R_{sum,\infty}(p,q)$
for all {\em finite} values of $t$, i.e., an infinite number of
messages are {\em necessary} to reach the infinite-message limit.}

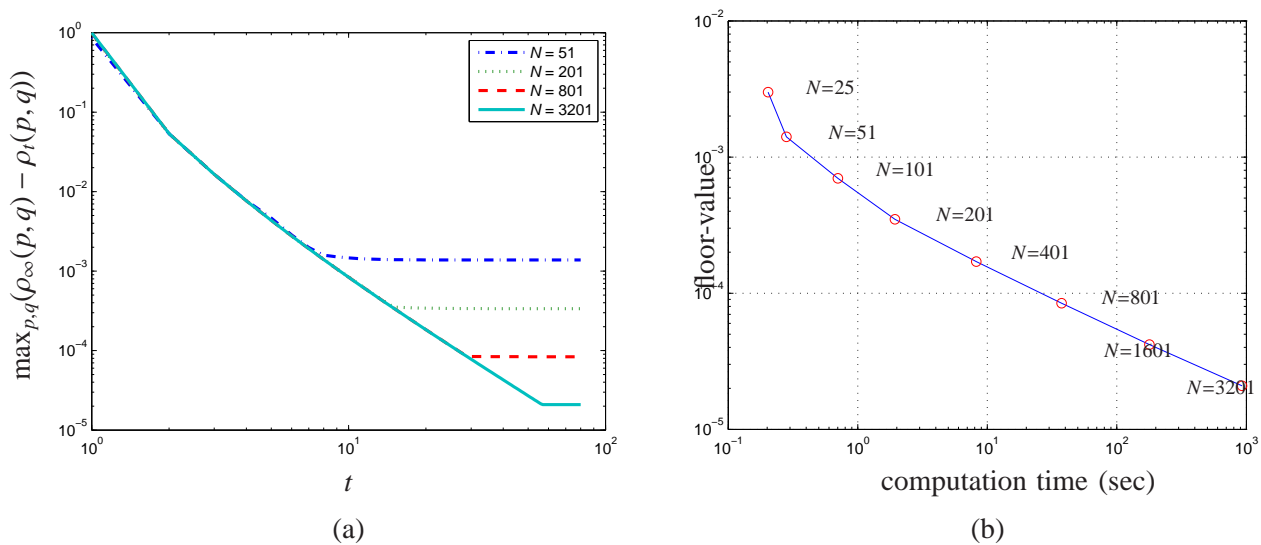
\begin{figure*}[!htb]
\centering
\includegraphics[scale = 0.5868]{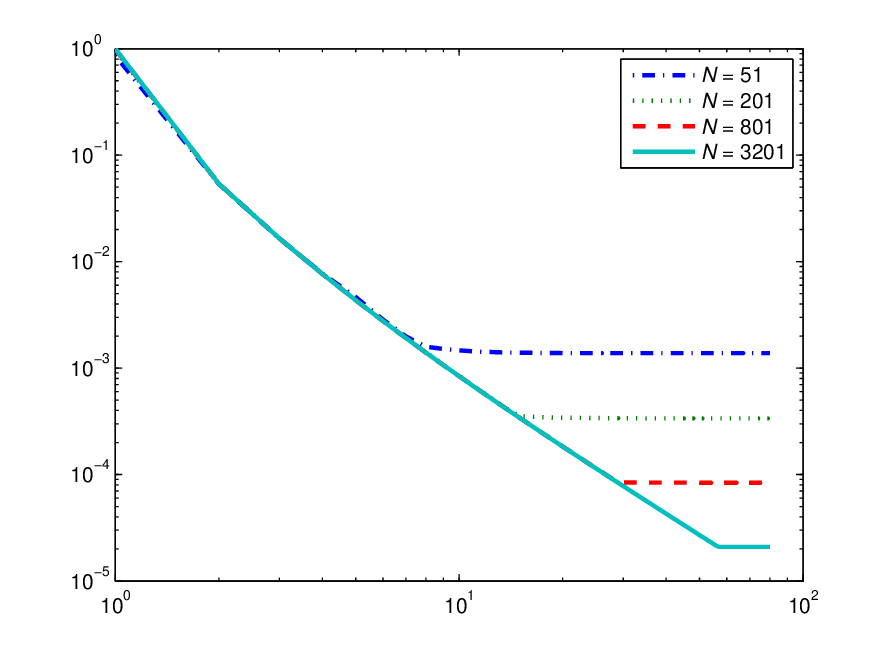}\hspace{-0.5cm}
\includegraphics[scale = 0.60]{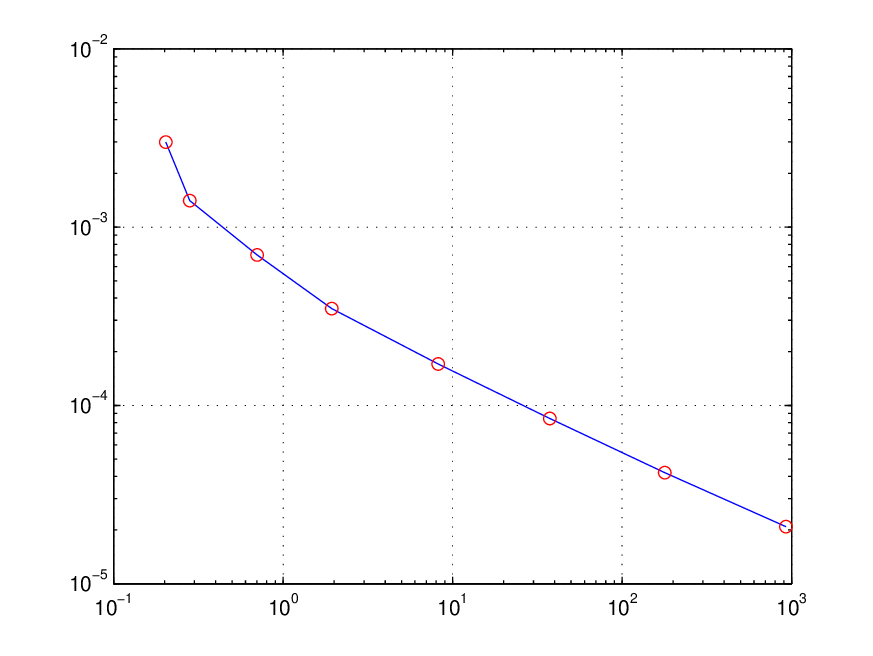}
\begin{picture}(14,0)
\put(2.9,0){\makebox(0,0){(a)}} \put(2.9,0.63){\makebox(0,0){$t$}}
\put(11.4,0){\makebox(0,0){(b)}}
\put(11.8,0.63){\makebox(0,0){computation time (sec)}}
\put(9.3,5.9){{\makebox(0,0){$\scriptstyle N=25$}}}
\put(9.6,5.3){{\makebox(0,0){$\scriptstyle N=51$}}}
\put(10.3,4.8){{\makebox(0,0){$\scriptstyle N=101$}}}
\put(11.1,4.2){{\makebox(0,0){$\scriptstyle N=201$}}}
\put(12.1,3.7){{\makebox(0,0){$\scriptstyle N=401$}}}
\put(13.3,3.1){{\makebox(0,0){$\scriptstyle N=801$}}}
\put(13.4,2.4){{\makebox(0,0){$\scriptstyle N=1601$}}}
\put(14.5,1.9){{\makebox(0,0){$\scriptstyle N=3201$}}}
\put(-1.6,2){\rotatebox{90}{$\max_{p,q} (\rho_\infty(p,q) - \rho_t(p,q))$}}
\put(7.5,3.2){\rotatebox{90}{floor-value}}
\end{picture}
\caption{\sl
 {(a) Plot showing how $\max_{p,q}(\rho_\infty(p,q)-\rho_t(p,q))$
monotonically decreases as the number of messages $t$ increases and
reaches a floor whose value depends on the number of grid points
$N^2$ in the discretization of $\mathcal{P}_{XY}$ (see
Section~\ref{subsection:oneside}). (b) Plot showing the relationship
between the floor-value (due to discretization) in plot (a) and the
computation time needed to reach the floor-value. When $N$ is
doubled, the floor-value is roughly halved and the computation time
needed to reach the floor-value is roughly quadrupled.}
\label{fig:discretization}
}
\end{figure*}

{\textbf{Discretization step-size and floor-value:}
  \textcolor{myblue}{The time complexity of the algorithm depends on
    the way $\mathcal{P}_{XY}$ is discretized into
    $\mathcal{P}^*_{XY}$. In this particular example, the time
    complexity as a function of the discretization parameter $N$ and
    the number of iterations $t$ can be analyzed as follows.  Since
    finding the convex hull of $N$ points in a plane takes $O(N \log
    N)$ operations \cite{convexhull}, and in each iteration $N$ convex
    hulls need to be computed, the time complexity to complete $t$
    iterations is $O(t N^2 \log N )$. }

Ideally, for all $(p,q) \in [0,1]^2$, as $t$ increases, the difference
between $\rho_\infty(p,q)$ and $\rho_t(p,q)$ should keep decreasing
until it becomes zero. However, due to the discretization of
$\mathcal{P}_{XY}$, the difference does not monotonically decrease to
zero but saturates around a discretization-induced floor-value.
Figure~\ref{fig:discretization}(a) shows how
$\max_{p,q}(\rho_\infty(p,q)-\rho_t(p,q))$ decreases as $t$ increases
and reaches a floor whose value depends on $N$. \textcolor{myblue}{In
  this numerical experiment, as $t$ increases,
  $\max_{p,q}(\rho_\infty(p,q)-\rho_t(p,q))$ decreases roughly as
  $1/t^2$ before reaching a floor value.} The finer is the
discretization (larger the $N$), the lower is the floor-value.
Figure~\ref{fig:discretization}(b) shows how the floor-value behaves
in relation to the computation time\footnote{The computation time is
  obtained using Matlab R2009a, a Pentium 4 CPU with clock rate
  2.80GHz and 2GB RAM.} needed to reach it and the corresponding value
of $N$.  Roughly speaking, when $N$ is doubled, the floor-value is
halved and the computation time needed to reach the floor-value is
quadrupled.  These are purely empirical observations of the
convergence-behavior restricted to the example of this section. A
theoretical analysis of the rate of convergence of the proposed
iterative algorithm for general problems is not established here and
is left for future work.}

\section{Weighted minimum sum-rate and directed sum-rate
region}\label{section:weighted}

 {The discussion thus far has focused on} the minimum sum-rate
 {as the overall measure of efficiency}, where the rates of
messages from terminal $A$ to $B$ and from terminal $B$ to $A$ are
added up with the same unit weight. This  {is appropriate for the
scenario} where the costs of communication in  {both} directions are
the same. In some applications, however, the  {communication costs}
in different directions  {can be different}. For example, in the
communication between a cellphone and a base station, the cellphone
has a stringent battery constraint whereas the base station does
not. Therefore each bit sent by the cellphone is more expensive.
 {In this section, we show how results from previous sections can
be extended to such an asymmetric communication scenario.}

\subsection{Weighted minimum sum-rate}\label{subsection:weightedsumrate}
 {Unequal communication costs from $A$ to $B$ and $B$ to $A$ can be
addressed by introducing nonnegative weights $k_{AB}$ and $k_{BA}$
respectively. For initial terminal $A$ and nonnegative weights
$(k_{AB},k_{BA})$, the weighted minimum sum-rate and the weighted rate
reduction can be defined, respectively, as follows}
\begin{align}
R_{sum,t}^{A,k_{AB},k_{BA}} &:=\min_{\mathbf R \in \mathcal
R_t^A}\left( \sum_{i\; odd}k_{AB} R_i + \sum_{i\; even}k_{BA}
R_i\right),\label{eqn:weightedsumrate}\\
\rho_{t}^{A,k_{AB},k_{BA}} &:=
k_{AB}H(X|Y)+k_{BA}H(Y|X)-R_{sum,t}^{A,k_{AB},k_{BA}},\nonumber
\end{align}
 {where, $\mathcal R_t^A$ is the set of all admissible rate
tuples ${\mathbf R}$. Comparing
these definitions with those of $R_{sum,t}^{A}$ and $\rho_{t}^{A}$,
it follows that the single-letter characterizations of
$R_{sum,t}^{A,k_{AB},k_{BA}}$ and $\rho_{t}^{A,k_{AB},k_{BA}}$ can
be obtained from (\ref{eqn:minsumrate}) and (\ref{eqn:rho}) by
scaling the terms corresponding to the rates of messages from $A$ to
$B$ and from $B$ to $A$ by the weights $k_{AB}$ and $k_{BA}$
respectively:}
\begin{align*}
R_{sum,t}^{A,k_{AB},k_{BA}}&=\min_{p_{U^t|XY} \in\;
\mathcal{P}_t^A(p_{XY})} [k_{AB} I(X;U^t|Y)+ k_{BA} I(Y;U^t|X)],\\
\rho_{t}^{A,k_{AB},k_{BA}}&= \max_{p_{U^t|XY} \in\;
\mathcal{P}_t^A(p_{XY})}[k_{AB} H(X|Y,U^t)+ k_{BA} H(Y|X,U^t)].
\end{align*}
 {For initial terminal $B$, the operational definitions and
single-letter characterizations of $R_{sum,t}^{B,k_{AB},k_{BA}}$ and
$\rho_{t}^{B,k_{AB},k_{BA}}$ are analogous.}

 {As in the symmetric communication scenario, the infinite-message
limits $R_{sum,\infty}^{k_{AB},k_{BA}}$ and
$\rho_\infty^{k_{AB},k_{BA}}$ are respectively defined as the limits
of $R_{sum,t}^{A,k_{AB},k_{BA}}$ and $\rho_{t}^{A,k_{AB},k_{BA}}$ as
$t\rightarrow \infty$. The functional
$\rho_\infty^{k_{AB},k_{BA}}(p_{XY})$ can be characterized by the
following extension of Theorem~\ref{thm:functioncomp}.
\vspace{1ex}
\begin{theorem}\label{thm:functioncompweighted}
If $\mathcal{F}^{k_{AB},k_{BA}}(\mathcal{P}_{XY})$ denotes a family
of functionals that satisfies Definition~\ref{def:F} with $\rho_0$
replaced by $\rho_0^{k_{AB},k_{BA}}$ then
$\rho_{\infty}^{k_{AB},k_{BA}}$ is the least element of the set
$\mathcal F^{k_{AB},k_{BA}}(\mathcal P_{XY})$ with majorization as
the partial ordering relation.
%
\end{theorem}
} \vspace{1ex}

 The proof of Theorem~\ref{thm:functioncompweighted}
 {is parallel to that of} Theorem~\ref{thm:functioncomp}
 {with} the weights $k_{AB}$ and $k_{BA}$ applied,
 {respectively,} to all the terms corresponding to rates from $A$
to $B$ and from $B$ to $A$.

Similarly,  {with $\rho_0$ replaced by $\rho_0^{k_{AB},k_{BA}}$ in
the initialization step,} the iterative algorithm described in
Section~\ref{section:itera} can be  {suitably modified} to compute
the weighted minimum sum-rate.

\subsection{Directed sum-rate
region}\label{subsection:directedregion} A finer characterization of
the tradeoff between the rates in two directions is the two
dimensional directed sum-rate region defined by
$\mathcal{R}_{sum,t}^A := \{(\sum_{i\; odd}R_i, \sum_{i\; even}
R_i): \mathbf R \in \mathcal{R}_t^A\}$. The first component of
$\mathcal{R}_{sum,t}^A$ is the directed sum-rate of all messages
from $A$ to $B$. The second component is the directed sum-rate of
all messages from $B$ to $A$. This characterization is coarser than
the complete $t$-dimensional rate region $\mathcal{R}_t^A$ where
each component corresponds to the rate of a single message, but
finer than the minimum sum-rate $R_{sum,t}^A$.

The weighted sum-rate can be used to characterize the directed
sum-rate region as follows.
\vspace{1ex}

\begin{theorem}\label{thm:finiteweightedregion}
For all $t\in\{0\} \bigcup \zZ^+$,
$\mathcal{R}^A_{sum,t}=\left\{(R_{AB},R_{BA}):\forall \theta \in
[0,\pi/2], R_{AB} \cos \theta + R_{BA} \sin \theta \geq
R_{sum,t}^{A,\cos \theta, \sin \theta}\right\}$.
\end{theorem}

\vspace{1ex}
\begin{proof}
When the function computation problem is infeasible, (i)
$\mathcal{R}^A_{sum,t}$ is empty and (ii) $\forall \theta\in
[0,\pi/2]$, $R_{sum,t}^{A,\cos \theta, \sin \theta}=\infty$, so that
the set appearing in the right side of the equation in
Theorem~\ref{thm:finiteweightedregion} is empty. Therefore
Theorem~\ref{thm:finiteweightedregion} holds when the computation is
infeasible.

When the function computation problem is feasible,
$\mathcal{R}^A_{sum,t}$ is nonempty. Since $\mathcal{R}^A_{t}$ is
convex and closed, a linear projection of this set to a lower
dimensional space is also convex and closed. Since
$\mathcal{R}^A_{sum,t}$ is a linear projection of
$\mathcal{R}^A_{t}$ to $\rR^2$, it is convex and closed. Since every
closed convex subset of $\rR^n$ is the intersection of the
halfspaces that contain it \cite[Proposition~B.15]{Bertsekas},
$\mathcal{R}^A_{sum,t}$ is the intersection of all the halfspaces of
$\rR^2$ that contain $\mathcal{R}^A_{sum,t}$.

Every halfspace in $\rR^2$ can be written in the following form:
$\{(R_{AB},R_{BA}): R_{AB} \cos \theta + R_{BA} \sin \theta \geq
c\}$, where $\theta\in [0,2 \pi)$ describes the orientation of the
boundary of the halfspace and $c\in \rR$ describes the displacement
of the boundary from the origin. Among all the halfspaces with a
common orientation parameter $\theta$ in the range $[0,\pi/2]$, the
one with the displacement parameter $c = R_{sum,t}^{A,\cos \theta,
\sin \theta}$ is the tightest halfspace that contains
$\mathcal{R}^A_{sum,t}$. This follows by first noting that $\cos
\theta$ and $\sin \theta$ are both nonnegative for $\theta \in
[0,\pi/2]$ and then setting $k_{AB} =\cos \theta$ and $k_{BA} =\sin
\theta$ in (\ref{eqn:weightedsumrate}). There is no halfspace that
contains $\mathcal{R}^A_{sum,t}$ with an orientation parameter
$\theta$ in the range $(\pi/2,2 \pi)$, because in this range, either
$\cos \theta < 0$ or $\sin \theta < 0$ and for sufficiently large
values of $R_{AB}$ or $R_{BA}$\footnote{Since
$\mathcal{R}^A_{sum,t}$ is nonempty, there exists
$(R_{AB}^*,R_{BA}^*)\in \mathcal{R}^A_{sum,t}$. According to the
definition of a rate region, $\forall R_{AB}\geq R_{AB}^*$ and
$\forall R_{BA}\geq R_{BA}^*$, we have $(R_{AB},R_{BA})\in
\mathcal{R}^A_{sum,t}$.} the inequality $R_{AB} \cos \theta + R_{BA}
\sin \theta \geq c$ will be violated for any value of $c$. Therefore
the intersection of all the halfspaces that contain
$\mathcal{R}^A_{sum,t}$ is the intersection of the halfspaces with
orientation $\theta \in [0,\pi/2]$ and displacement
$c=R_{sum,t}^{A,\cos \theta, \sin \theta}$.
\end{proof}

Since null messages are permitted, we have
$\mathcal{R}_{sum,t}^A\subseteq \mathcal{R}_{sum,t+1}^A$ and
$\mathcal{R}_{sum,t}^A\subseteq \mathcal{R}_{sum,t+1}^B$ (see
\cite[Proposition~1]{ISIT08} for a detailed discussion). We can
consider the directed sum-rate region in the limit $t$ going to
infinity. Define $\mathcal{R}_{sum,\infty}$ as the closure of
$\bigcup_{t=1}^\infty \mathcal{R}_{sum,t}^A$. We can characterize
$\mathcal{R}_{sum,\infty}$ using $R_{sum,\infty}^{\cos \theta, \sin
\theta}$ as follows.

\begin{theorem}\label{thm:infiniteweightedregion}
$\mathcal{R}_{sum,\infty}=\left\{(R_{AB},R_{BA}):\forall \theta \in
[0,\pi/2], R_{AB} \cos \theta + R_{BA} \sin \theta \geq
R_{sum,\infty}^{\cos \theta, \sin \theta}\right\}$.
\end{theorem}
\begin{proof}
Since $\left\{\mathcal{R}_{sum,t}^A\right\}_{t=1}^{\infty}$ is a
sequence of expanding convex sets, it can be verified that the union
of them is also convex. Since $\mathcal{R}_{sum,\infty}$ is defined
as the closure of
$\left\{\mathcal{R}_{sum,t}^A\right\}_{t=1}^{\infty}$, it is both
closed and convex.  {The rest of the proof can be completed using
the arguments from Theorem~\ref{thm:finiteweightedregion}.}
\end{proof}

Combining Theorems~\ref{thm:functioncompweighted} and
\ref{thm:infiniteweightedregion}, we have a characterization of
$\mathcal{R}_{sum,\infty}$ as a functional of $p_{XY}$. We can also
numerically evaluate $\rho_{\infty}^{\cos \theta, \sin\theta}$ using
 {an iterative algorithm that would be analogous to that in
Section~\ref{section:itera}. This would, in turn, provide a means to
numerically evaluate $\mathcal{R}_{sum,\infty}$.}


\section{Extension to interactive rate-distortion
problem}\label{section:rd}

\subsection{Problem formulation} \label{subsection:rdproblem}

In \cite{Journal_interactive} we studied the interactive coding
problem with per-sample distortion criteria. Let $d_A:
\mathcal{X}\times\mathcal {Y} \times \mathcal {Z}_A\rightarrow
\mathcal D$ and $d_B: \mathcal{X}\times\mathcal {Y} \times \mathcal
{Z}_B\rightarrow \mathcal D$ be bounded single-letter distortion
functions, where $\mathcal D := [0,d_{\max}]$. The fidelity of
function computation can be measured by the per-sample average
distortion
\[d_A^{(n)}(\mathbf x, \mathbf y, \hat{\mathbf z}_A):=\frac{1}{n}\sum_{i=1}^{n}
d_A(x(i), y(i), \hat{z}_A(i)), \ \ d_B^{(n)}(\mathbf x, \mathbf y,
\hat{\mathbf z}_B):=\frac{1}{n}\sum_{i=1}^{n} d_B(x(i), y(i),
\hat{z}_B(i)).\] Of interest here are the expected per-sample
distortions $E[d_A^{(n)}(\mathbf X, \mathbf Y, \widehat{\mathbf
    Z}_A)]$ and $E[d_B^{(n)}(\mathbf X, \mathbf Y, \widehat{\mathbf
    Z}_B)]$. We note that although the desired functions $f_A$ and
$f_B$ do not explicitly appear in these fidelity criteria, they are
subsumed by $d_A$ and $d_B$ because they accommodate general
relationships between the sources and the outputs of the decoding
functions. The performance of $t$-message interactive coding for
function computation is measured as follows.

\vspace{1ex}
\begin{definition}\label{def:ratedistortion}
A rate-distortion tuple ${(\mathbf R,\mathbf D)} = (R_1, \ldots, R_t,
D_A, D_B)$ is admissible for $t$-message interactive function
computation with initial terminal $A$ if, $\forall \epsilon > 0$,
$\exists~ N(\epsilon,t)$ such that $\forall n> N(\epsilon,t)$, there
exists an interactive distributed source code with initial terminal
$A$ and parameters $(t,n,|{\mathcal M}_1|,\ldots,|{\mathcal M}_t|)$
satisfying
\begin{eqnarray*}
&&\frac{1}{n}\log_2 |{\mathcal M}_j| \leq R_j + \epsilon,\ j =
1,\ldots, t,\\
&& E[d_A^{(n)}(\mathbf X, \mathbf Y, \widehat{\mathbf Z}_A)]\leq D_A
+ \epsilon,\ E[d_B^{(n)}(\mathbf X, \mathbf Y, \widehat{\mathbf
Z}_B)] \leq D_B+\epsilon.
\end{eqnarray*}
\end{definition}
\vspace{1ex}

The set of all admissible rate-distortion tuples, denoted by
${\mathcal {RD}}^A_t$, is called the operational rate-distortion
region for $t$-message interactive function computation with initial
terminal $A$. The rate-distortion region is closed and convex due to
the way it has been defined. The sum-rate-distortion function
$R^A_{sum,t}(\mathbf D)$ is given by $\min \left(\sum_{j=1}^t
R_j\right)$ where the minimization is over all $\mathbf R$ such that
$(\mathbf R, \mathbf D) \in\; \mathcal {RD}^A_t$. For initial terminal
$B$, the rate-distortion region and the minimum sum-rate-distortion
function are denoted by ${\mathcal {RD}}^B_t$ and
$R^B_{sum,t}(\mathbf{D})$ respectively. For any fixed $\mathbf D$, We
define $R_{sum,\infty}(\mathbf{D}):=\lim_{t\rightarrow
\infty}R^A_{sum,t}(\mathbf D)=\lim_{t\rightarrow
\infty}R^B_{sum,t}(\mathbf D)$.
 {Results of this section for the sum-rate-distortion function can
be generalized to corresponding results for the weighted
sum-rate-distortion function and the directed sum-rate-distortion
region in the same manner that results for the minimum sum-rate
function were generalized to corresponding results for the weighted
minimum sum-rate function and the directed sum-rate region in
Sections~\ref{subsection:weightedsumrate} and
\ref{subsection:directedregion}.}

The admissibility of a rate-distortion tuple can also be defined in
terms of the probability of excess distortion by replacing the
expected distortion conditions in Definition~\ref{def:ratedistortion}
by the conditions $\pP(d_A^{(n)}(\mathbf X, \mathbf Y,
\widehat{\mathbf Z}_A) > D_A) \leq \epsilon$ and
$\pP(d_B^{(n)}(\mathbf X, \mathbf Y, \widehat{\mathbf Z}_B) > D_B)
\leq \epsilon$. Although these conditions appear to be more
stringent\footnote{Any tuple which is admissible according to the
probability of excess distortion criteria is also admissible according
to the expected distortion criteria.}, it can be shown\footnote{Using
strong-typicality arguments in the proof of the achievability part of
the single-letter characterization of the rate-distortion region.}
that they lead to the same operational rate-distortion region.  For
simplicity, we focus on the expected distortion conditions as in
Definition~\ref{def:ratedistortion}.

\subsection{Characterization of $R_{sum,t}^A(p_{XY},\mathbf{D})$ and $\rho_t^A(p_{XY},\mathbf{D})$ for finite
  $t$ \cite{Journal_interactive}}

The single-letter characterization of $R_{sum,t}^A(p_{XY},\mathbf{D})$
is given by
\begin{equation}
R^{A}_{sum,t}(p_{XY},\mathbf D) = \min_{(p_{U^t|XY},\hat g_A,\hat
g_B)\in \mathcal{P}_t^A(p_{XY},\mathbf{D})}
[I(X;U^t|Y)+I(Y;U^t|X)],
\label{eqn:ratedistortion}
\end{equation}
where $\mathcal{P}_t^A(p_{XY},\mathbf{D})$ is the set of all tuples $(p_{U^t|XY},\hat
g_A,\hat g_B)$ such that (i) $\hat g_A$
and $\hat g_B$ are deterministic functions and $E\left[d_A(X,Y,\hat g_A(U^t,X))\right]\leq D_A, \mbox{ }
E\left[d_B(X,Y,\hat g_B(U^t,Y))\right]\leq D_B\}$, (ii) for
$i=1,\ldots,t$, if $i$ is odd, $U_i-(X,U^{i-1})-Y$ forms a Markov
chain, otherwise $U_i-(Y,U^{i-1})-X$ forms a Markov chain, and (iii) $|\mathcal{U}_1|,\ldots,|\mathcal{U}_t|$ are finite alphabets whose cardinality satisfy (\ref{eqn:cardinality}). Compared to
(\ref{eqn:minsumrate}), the expected distortion constraints  {have
replaced} the conditional entropy constraints in
(\ref{eqn:ratedistortion}). The rate reduction functional is defined
as follows.
\begin{equation}
\rho^A_{t}(p_{XY},\mathbf D):=H(X|Y)+H(Y|X)-R^{A}_{sum,t}(\mathbf
D)\nonumber = \max_{(p_{U^t|XY},\hat g_A, \hat
g_B)\in\mathcal{P}_t^A(p_{XY},\mathbf{D}) }[H(X|Y,U^t)+H(Y|X,U^t)].
\label{eqn:rhodistortion}
\end{equation}

For $t=0$, let  {$\mathcal P_{f_A f_B D}:=\{(p_{XY}, {\mathbf D}):
\exists~ \hat g_A, \hat g_B, s.t.\ E\left[d_A(X,Y,\hat
g_A(X))\right]\leq D_A, E\left[d_B(X,Y,\hat g_B(Y))\right]\leq
D_B\}$}. Then we have
\begin{equation*}
R_{sum,0}(p_{XY},\mathbf D)= \left\{
\begin{array}{cc}
0, & \mbox{if } (p_{XY},\mathbf D)\in\; \mathcal P_{f_A f_B D}, \\
+ \infty, & \mbox{otherwise.}
\end{array}
\right.
\end{equation*}
\begin{equation}\label{eqn:rho0distortion}
\rho_{0}(p_{XY},\mathbf D)= \left\{
\begin{array}{cc}
H(X|Y)+H(Y|X), & \mbox{if } (p_{XY},\mathbf D) \in\; \mathcal P_{f_A
  f_B D},\\
- \infty, & \mbox{otherwise.}
\end{array}
\right.
\end{equation}

\subsection{Characterization of  $R_{sum,\infty}(p_{XY},\mathbf D)$} \label{subsection:rdmain}

We can use the same technique as in Section~\ref{section:general} to
characterize the functional $\rho_{\infty}(p_{XY}, \mathbf D)$.
%

%
\vspace{1ex}
\begin{definition}{\it (Marginal-\-perturbations-\-distortion-\-concave,
$\rho_0$-\-majorizing family of functionals
$\mathcal{F}_D(\mathcal{P}_{XY})$)}\label{def:FD}
Let $\mathcal P_{XY}$ be any marginal-perturbations-closed family of
joint pmfs on $\Delta(\mathcal X \times \mathcal Y)$. The set of
marginal-perturbations-distortion-concave, $\rho_0$-majorizing family
of functionals $\mathcal{F}_D(\mathcal{P}_{XY})$ is the set of all the
functionals $\rho: \mathcal P_{XY}\times \mathcal{D}^2 \rightarrow
\rR$ satisfying the following conditions:
    \begin{enumerate}
    \item $\rho_0$-majorization: $\forall p_{XY}\in\; \mathcal P_{XY}$
      and $\forall \mathbf{D} \in \mathcal{D}^2$, $\rho(p_{XY},
      \mathbf{D})\geq \rho_0(p_{XY}, \mathbf{D})$.
    \item Concavity with respect to $X$-marginal perturbations and
      distortion vector: $\forall p_{XY}\in\; \mathcal{P}_{XY}$,
      $\rho$ is concave on $\mathcal{P}_{Y|X}(p_{XY})\times
      \mathcal{D}^2$.
    \item Concavity with respect to $Y$-marginal perturbations and
      distortion vector: $\forall p_{XY}\in\; \mathcal{P}_{XY}$,
      $\rho$ is concave on $\mathcal{P}_{X|Y}(p_{XY})\times
      \mathcal{D}^2$.
    \end{enumerate}
\end{definition}
\vspace{1ex}

The following characterization of $\rho_{\infty}(p_{XY},\mathbf{D})$
is the generalization of Theorem~\ref{thm:functioncomp} to the
rate-distortion problem.

\vspace{1ex}
\begin{theorem}\label{thm:ratedistortion}
(i) $\rho_{\infty}(p_{XY},\mathbf D) \in\; \mathcal
F_D(\mathcal{P}_{XY})$.
(ii) For all $\rho \in\; \mathcal F_D(\mathcal{P}_{XY})$ and $\forall
(p_{XY},\mathbf D) \in\; \mathcal P_{XY}\times \mathcal{D}^2$, we have
$\rho_{\infty}(p_{XY},\mathbf D) \leq \rho(p_{XY},\mathbf D)$.
\end{theorem}
\vspace{1ex}

The proof of Theorem~\ref{thm:ratedistortion} is parallel to that of
Theorem~\ref{thm:functioncomp}.

\begin{proof}
First we establish the relation between $\rho^A_t$ and $\rho^B_{t-1}$
as follows.
 {
\begin{lemma}\label{lemma:connectionrd}
\begin{itemize}
\item[(i)] For all $t\in \zZ^+$ and all $(p_{XY},\mathbf D) \in
 \mathcal{P}_{XY} \times \mathcal D^2$,
 \begin{equation}
   \rho^A_t(p_{XY},\mathbf{D}) =
   \max_{p_{U_1|X}}\left\{\max_{\scriptstyle \forall u_1\in \mathcal
   U_1, \mathbf{D}_{u_1}\in \mathcal{D}^2:\atop \scriptstyle
   E [\mathbf{D}_{U_1}]=\mathbf{D}} \left\{ \sum_{u_1\in\;
   \supp(p_{U_1})} p_{U_1}(u_1)
   \rho^B_{t-1}(p_{XY|U_1}(\cdot,\cdot|u_1),
   \mathbf{D}_{u_1})\right\}\right\}.
   \label{eqn:convexifyrd}
 \end{equation}
\item[(ii)] For all $t\in \zZ^+$ and all $(q_{XY},\mathbf D) \in
 \mathcal{P}_{XY} \times \mathcal D^2$, $\rho^A_t$ is concave on
 $\mathcal{P}_{Y|X}(q_{XY})\times \mathcal D^2$.
\item[(iii)] For all $t\in \zZ^+$ and all $(q_{XY},\mathbf D) \in
 \mathcal{P}_{XY} \times \mathcal D^2$, if $\rho: \mathcal{P}_{XY}
 \times \mathcal D^2 \rightarrow \rR$ is concave on
 $\mathcal{P}_{Y|X}(q_{XY})\times \mathcal D^2$ and if for all
 $(p_{XY}, \mathbf{D}) \in \mathcal{P}_{Y|X}(q_{XY}) \times \mathcal
 D^2$, $\rho^B_{t-1}(p_{XY},\mathbf D) \leq \rho(p_{XY},\mathbf D)$,
 then for all $(p_{XY}, \mathbf{D}) \in \mathcal{P}_{Y|X}(q_{XY})
 \times \mathcal D^2$, $\rho^A_t(p_{XY},\mathbf D) \leq
 \rho(p_{XY},\mathbf D)$.
\textcolor{myblue}{\item[(iv)] The results of parts (i) -- (iii) above also hold if $A$ is swapped with $B$ and simultaneously, $\mathcal{P}_{Y|X}$ and $p_{U_1|X}$ are replaced by $\mathcal{P}_{X|Y}$ and $p_{U_1|Y}$ respectively.}
\end{itemize}
\end{lemma}
\begin{proof}
See Appendix~\ref{app:lemma3proof}.
\end{proof}
\vspace{1ex}
}


 {Equipped with Lemma~\ref{lemma:connectionrd}, the rest of the proof of
Theorem~\ref{thm:ratedistortion} is parallel to that of
Theorem~\ref{thm:functioncomp},} except that all the rate reduction
functionals depend on $(p_{XY},\mathbf D)$ instead of only $p_{XY}$.
\end{proof}

 {The intuition underlying the proof of
Theorem~\ref{thm:ratedistortion} is essentially conveyed by the
first paragraph of the proof of Theorem~\ref{thm:functioncomp}.} The
main difference is that for each realization $U_1 = u_1$, the
distortion vector $\mathbf{D}_{u_1}$ in the $(t-1)$-message
subproblem could be different from the original distortion vector
$\mathbf{D}$. The only constraint that $\{\mathbf{D}_{u_1}\}_{u_1\in
\mathcal U_1}$ needs to satisfy is that $\sum_{u_1}
\mathbf{D}_{u_1}p_{U_1}(u_1) = \mathbf{D}$ holds. Therefore, we need
to consider  {{\em joint}} convex combinations of the distortion
vector and the marginal source distributions.

The counterparts of Corollaries~\ref{cor:hypograph} to
\ref{cor:optimalitytest} in the rate-distortion problem also hold.
The proofs are parallel to those of Corollaries~\ref{cor:hypograph} to
\ref{cor:optimalitytest}.

\vspace{1ex}
\begin{corollary}\label{cor:hypographrd}
\emph{(Constructing $\rho^A_t$ and $\rho^B_t$ from $\rho^B_{t-1}$
and $\rho^A_{t-1}$  {respectively})} For all $t\in \zZ^+$ and
 {all} $p_{XY}\in \mathcal{P}_{XY}$, we have
(i) $\mbox{ch}\left(\mbox{hypo}_{(\mathcal P_{Y|X}(p_{XY})\times
\mathcal D^2)} \rho_{t-1}^B\right)$ $= \mbox{hypo}_{(\mathcal
P_{Y|X}(p_{XY})\times \mathcal D^2)} \rho_t^A$ and
(ii) $\mbox{ch}\left(\mbox{hypo}_{(\mathcal P_{X|Y}(p_{XY})\times
\mathcal D^2)} \rho_{t-1}^A\right) = \mbox{hypo}_{(\mathcal
P_{X|Y}(p_{XY})\times \mathcal D^2)} \rho_t^B$.
\end{corollary}
\vspace{1ex}
\begin{corollary}\label{cor:concavityoptimalityrd}
\emph{(Concavity and optimality of $\rho^A_t$  {and $\rho^B_t$})}
For all $t\in \zZ^+$, the following conditions are equivalent:
\begin{itemize}
\item[(i)] $\forall p_{XY}\in \mathcal{P}_{XY}, \forall \mathbf D\in
\mathcal D^2, \rho^A_t(p_{XY},\mathbf D)=\rho_{\infty}(p_{XY},\mathbf
D)$,
\item[(ii)] $\forall p_{XY}\in \mathcal{P}_{XY}, \forall \mathbf D\in
\mathcal D^2, \rho^A_t(p_{XY},\mathbf D)=\rho^B_{t+1}(p_{XY},\mathbf
D)$,
\item[(iii)] $\forall p_{XY}\in \mathcal{P}_{XY}$, $\rho^A_t$ is
concave on $\mathcal{P}_{X|Y}(p_{XY})\times \mathcal D^2$.
\end{itemize}
 {For all $t\in \zZ^+$, the following three conditions are also
equivalent:
\begin{itemize}
\item[(iv)] $\forall p_{XY}\in \mathcal{P}_{XY}, \forall \mathbf D\in
\mathcal D^2, \rho^B_t(p_{XY},\mathbf D)=\rho_{\infty}(p_{XY},\mathbf
D)$,
\item[(v)] $\forall p_{XY}\in \mathcal{P}_{XY}, \forall \mathbf D\in
\mathcal D^2, \rho^B_t(p_{XY},\mathbf D)=\rho^A_{t+1}(p_{XY},\mathbf
D)$,
\item[(vi)] $\forall p_{XY}\in \mathcal{P}_{XY}$, $\rho^B_t$ is
concave on $\mathcal{P}_{Y|X}(p_{XY})\times \mathcal D^2$.
\end{itemize}
}
\end{corollary}
\vspace{1ex}
\begin{corollary}\label{cor:optimalitytestrd}
\emph{(Optimality test for an achievable sum-rate distortion
function)} Let $R^*$ be a sum-rate distortion function which is
achievable using an arbitrary number of messages. If $\rho^* :=
(H(X|Y) + H(Y|X) - R^*)\in \mathcal F_D(\mathcal P_{XY})$, then
$R^*=\textcolor{myblue}{R_{sum,\infty}}$.
\end{corollary}
\vspace{1ex}

\subsection{Iterative algorithm for computing $R_{sum,t}^A(p_{XY},{\bf
  D})$ and $R_{sum,\infty}(p_{XY},{\bf D})$}
  \label{subsection:rditera}
Corollary~\ref{cor:hypographrd} suggests the following algorithm
which is similar to the one presented in
Section~\ref{section:itera}.

\textbf{Algorithm to evaluate $R_{sum,t}^A(p_{XY},\mathbf D)$ and
$R_{sum,t}^B(p_{XY},\mathbf D)$}
\begin{itemize}
  \item \textbf{Initialization:} Choose a
    marginal-perturbations-closed family $\mathcal P_{XY}$ containing
    all joint source pmfs of interest. Define $\rho_0^A(p_{XY},
    \mathbf D) = \rho_0^B(p_{XY}, \mathbf D) = \rho_0(p_{XY}, \mathbf
    D)$ by equation (\ref{eqn:rhodistortion}) in the domain $\mathcal
    P_{XY}\times \mathcal{D}^2$. Choose a cover for $\mathcal P_{XY}$
    made up of $X$-marginal perturbation sets, denoted by $\{\mathcal
    P_{Y|X}(p_{XY})\}_{p_{XY}\in\; \mathcal A}$, where $\mathcal
    A\subseteq \mathcal P_{XY}$. Also choose a cover for $\mathcal
    P_{XY}$ made up of $Y$-marginal perturbation sets, denoted by
    $\{\mathcal P_{X|Y}(p_{XY})\}_{p_{XY}\in\; \mathcal B}$, where
    $\mathcal B\subseteq \mathcal P_{XY}$.
  \item \textbf{Loop:} For $\tau=1$ through $t$ do the following.\\
  For every $p_{XY}\in\; \mathcal A$, do the following in the set
  $\mathcal P_{Y|X}(p_{XY})\times \mathcal{D}^2$.
  \begin{itemize}
    \item Construct $\mbox{hypo}_{\mathcal P_{Y|X}(p_{XY})\times
      \mathcal{D}^2}\rho_{\tau-1}^B$.
    \item Let $\rho_\tau^A$ be the upper boundary of
      $\mbox{ch}\left(\mbox{hypo}_{\mathcal P_{Y|X}(p_{XY})\times
      \mathcal{D}^2}\rho_{\tau-1}^B\right)$.
  \end{itemize}

  For every $p_{XY}\in\; \mathcal B$, do the following in the set
  $\mathcal P_{X|Y}(p_{XY})\times \mathcal{D}^2$.

  \begin{itemize}
    \item Construct $\mbox{hypo}_{\mathcal P_{X|Y}(p_{XY})\times
      \mathcal{D}^2} \rho_{\tau-1}^A$.
    \item Let $\rho_\tau^B$ be the upper boundary of
      $\mbox{ch}\left(\mbox{hypo}_{\mathcal P_{X|Y}(p_{XY})\times
      \mathcal{D}^2} \rho_{\tau-1}^A\right)$.
  \end{itemize}

   {Optimality test: \\
    if $\forall (p_{XY}, \mathbf{D}) \in \mathcal{P}_{XY} \times
    \mathcal D^2$, $\rho^A_{\tau}(p_{XY},\mathbf
    D)=\rho^B_{\tau-1}(p_{XY},\mathbf D)$, then set $\rho_t^A =
    \rho_t^B = \rho^A_{\tau}$ and exit the loop.\\
    if $\forall (p_{XY}, \mathbf{D}) \in \mathcal{P}_{XY} \times
    \mathcal D^2$, $\rho^B_{\tau}(p_{XY},\mathbf
    D)=\rho^A_{\tau-1}(p_{XY},\mathbf D)$, then set $\rho_t^A =
    \rho_t^B = \rho^A_{\tau}$ and exit the loop.
}
  \item \textbf{Output:} $R_{sum,t}^A(p_{XY},\mathbf
    D)=H(X|Y)+H(Y|X)-\rho_t^A(p_{XY},\mathbf D)$, and
    $R_{sum,t}^B(p_{XY},\mathbf
    D)=H(X|Y)+H(Y|X)-\rho_t^B(p_{XY},\mathbf D)$.
\end{itemize}

In  {a computer} implementation, we  {will} need to discretize the
set $\mathcal P_{XY}\times\mathcal{D}^2$.
$R_{sum,\infty}(p_{XY},\mathbf D)$ can, in principle, be evaluated
to any precision by running this algorithm to a large enough value
of $t$, until the change between $\rho_{t-1}^A(p_{XY},\mathbf D)$
and $\rho_t^A(p_{XY},\mathbf D)$ is below a certain threshold. In
the special case $t=1$, and $d_A\equiv 0$, the interactive problem
reduces to the Wyner-Ziv problem  {(with a general coupled
distortion metric)}. If we further assume that $|\mathcal Y| = 1$,
the Wyner-Ziv problem reduces to the single-terminal rate-distortion
problem. Therefore, the algorithm described above can be used to
evaluate the single-terminal and Wyner-Ziv rate-distortion functions
as special cases.

\section{The benefit of interaction for lossy source reproduction}\label{section:Kaspi}

\subsection{Introducing the unresolved question}
The interactive rate-distortion problem defined in
Section~\ref{subsection:rdproblem} reduces to a lossy source
reproduction problem when $d_A(x,y,z_A)$ depends only on $(y,z_A)$ and
$d_B(x,y,z_B)$ depends only on $(x,z_B)$. In this special case, the
single-letter characterization of the sum-rate-distortion function was
given in \cite{Kaspi1985} for all finite number of messages. Yet,
whether more messages can strictly improve the sum-rate-distortion
function was left unresolved. If the goal is to reproduce both sources
{\em losslessly} at each terminal (zero distortion) then there is no
advantage in using multiple messages; two messages are sufficient and
the minimum sum-rate cannot be reduced by using more than two
messages.\footnote{If only one of the sources is required to be
losslessly reproduced at the other terminal then one message is
sufficient and the minimum sum-rate cannot be improved by using more
than one message.
} If, however, the goal is changed to losslessly {\em compute
functions} of sources at each terminal, then multiple messages can
decrease the minimum sum-rate by an arbitrarily large factor
\cite{OrlitskyRoche,ISIT08}. Therefore, the key unresolved question
pertains to {\em lossy source reproduction}: can multiple messages
strictly decrease the minimum sum-rate for a given (nonzero)
distortion? This question was unresolved even when only one source
needs to be reproduced with nonzero distortion.

In this section, we construct the first example which shows that two
messages can strictly improve the one-message (Wyner-Ziv)
rate-distortion function. The example also shows that the ratio of
the one-message rate to the two-message sum-rate can be arbitrarily
large and simultaneously the ratio of the backward rate to the
forward rate in the two-message sum-rate can be arbitrarily small.
{The rate reduction functional and its properties, specifically
Corollary~\ref{cor:concavityoptimalityrd}, play an important role in
the construction of this example.}  In
Section~\ref{subsection:Zamirexample} we provide another example
where in addition to the above properties, the one-message
rate-distortion function can be arbitrarily large and the
two-message sum-rate can be arbitrarily small.

In Section~\ref{subsection:improvement} and
Section~\ref{subsection:largeratio} we consider a lossy source
reproduction problem where only terminal $B$ needs to reproduce
$\mathbf X$ within distortion level $D=D_B$, and terminal $A$ is not
required to reproduce anything ($d_A\equiv 0$). In
Theorem~\ref{thm:improvement}, we will use
Corollary~\ref{cor:concavityoptimalityrd} to show that there exist
$p_{XY}, d$, and $D$ for which $R_{sum,1}^A(p_{XY},D) >
R_{sum,2}^B(p_{XY},D)$. We will do this by (i) choosing $p_{X|Y}$ so
that $X$ and $Y$ are symmetrically correlated binary random variables
with $\pP(Y \neq X) = p$, (ii) taking $d(x,\hat{x})$ to be the binary
erasure distortion function, (iii) selecting a value for $D$, and (iv)
showing that $\rho_1^A(p_{X|Y}p_Y,D)$ is not concave with respect to
$p_{Y}$, which implies that condition (iii) of
Corollary~\ref{cor:concavityoptimalityrd} does not hold for $t=1$.
Therefore condition (ii) does not hold: $\rho_1^A(p_{XY},D) \neq
\rho_2^B(p_{XY},D)$, which would imply that
$R_{sum,1}^A(p_{XY},D)>R_{sum,2}^B(p_{XY},D)$. In
Theorem~\ref{thm:largeratio} we will show that for certain values of
parameters $p$ and $D$, the two-message sum-rate can be split in such
a way that the ratio $R_1/R_2$ is arbitrarily small and simultaneously
the ratio $R_{sum,1}^A/(R_1+R_2)$ is arbitrarily large. This will be
proved by explicitly constructing auxiliary variables $V_1,V_2$
\footnote{In Section~\ref{section:Kaspi} the auxiliary variable in the
one-message problem is denoted by $U$ and the auxiliary variables the
two-message problems are denoted by $V_1,V_2$. The purpose is to avoid
confusion of auxiliary variables for different problems.} and decoding
function $\hat g_B$. While the explicit construction of $V_1,V_2$ and
$\hat g_B$ in the proof of Theorem~\ref{thm:largeratio} may make the
implicit proof of Theorem~\ref{thm:improvement} seem redundant, it is
unclear how the explicit construction can be generalized to other
families of source distributions and distortion functions. The
approach followed in the proof of Theorem~\ref{thm:improvement}, on
the other hand, provides an efficient method to {\em test} whether the
best two-message scheme can strictly outperform the best one-message
scheme for {\em more general} distributed source coding and function
computation problems. The implicit proof naturally points to an
explicit construction and was, in fact, the path taken by the authors
to arrive at the explicit construction. In
Section~\ref{subsection:Zamirexample} we extend a noninteractive
rate-distortion problem in \cite{Zamir_EAP} to an interactive problem,
 {to construct an} example where the one-message rate-distortion
function can be arbitrarily large and the two-message sum-rate can be
arbitrarily small.

\subsection{Implicit proof of the benefit of
interaction}\label{subsection:improvement}
\begin{theorem}\label{thm:improvement}
For the interactive rate-distortion problem where only terminal $B$
reproduces source sequence $\mathbf X$, there exists a distortion
function $d$, a joint distribution $p_{XY}$, and a distortion level
$D$ for which $R_{sum,1}^A(p_{XY},D)>R_{sum,2}^B(p_{XY},D)$.
\end{theorem}

\begin{proof}
According to Corollary~\ref{cor:concavityoptimalityrd}, to prove
Theorem~\ref{thm:improvement}, it is sufficient to show there exist
$p_{X|Y}$, $d$, and $D$ for which $\rho_1^A(p_{X|Y}p_Y,D)$ is not
concave with respect to $p_Y$. In particular, it is sufficient to
show that there exist $p_{Y,1}$ and $p_{Y,2}$ such that
\begin{equation}\label{eqn:notconcave}
\rho_1^A\left(p_{X|Y}\frac{p_{Y,1}+p_{Y,2}}{2},D\right)<
\frac{\rho_1^A\left(p_{X|Y}p_{Y,1},D\right)+\rho_1^A\left(p_{X|Y}p_{Y,2},D\right)}{2}.
\end{equation}
Let $\mathcal{X} = \mathcal{Y} = \{0,1\}$, and
$\widehat{\mathcal{X}} = \{0,1,e\}$. Let $d$ be the binary erasure
distortion function, i.e., $d: \{0,1\}\times \{0,e,1\}\rightarrow
\{0,1,\infty\}$ and for $i=0,1$, $d(i,i)=0$, $d(i,1-i)=\infty$, and
$d(i,e)=1$ \footnote{Although the distortion function is assumed to
take finite values in Section~\ref{section:rd}, the results can be
extended to some non-finite distortion measures including the binary
erasure distortion. See \cite[Problem~2.2.13]{Csi}.}. Let
$p_{Y,1}(1) = 1 - p_{Y,1}(0) = p_{Y,2}(0) = 1 - p_{Y,2}(1) = q$,
i.e., $p_{Y,1} = $ Bernoulli$(q)$ and $p_{Y,2} = $ Bernoulli$(\bar
q)$. Let $p_{X|Y}$ be the conditional pmf of the binary symmetric
channel with crossover probability $p$, i.e., $p_{X|Y}(1|0) =
p_{X|Y}(0|1) = p$. Let $p_Y:=(p_{Y,1}+p_{Y,2})/2$ which is
Bernoulli$(1/2)$. The joint distribution $p_{XY}=p_Y p_{X|Y}$ is the
joint pmf of a pair of doubly symmetric binary sources (DSBS) with
parameter $p$, i.e., if $p_{xy}$ denotes $p_{XY}(x,y)$, then $p_{00}
= p_{11} = \bar p/2$ and $p_{01} = p_{10} = p/2$. For these choices
of $p_{X|Y}$, $p_{Y,1}$, $p_{Y,2}$, $p_Y$, and $d$, we will analyze
the left and right sides of (\ref{eqn:notconcave}) step by step
through a sequence of propositions and establish the strict
inequality for a suitable choice of $D$. The proofs of all the
propositions are given in Appendix~\ref{app:proofs_Kaspisquestion}.

\noindent $\bullet$ \emph{Left-side of (\ref{eqn:notconcave}):} We
have
\begin{equation}\label{eqn:generalrho1}
\rho_1^A (p_{XY},D)= \max_{p_{U|X},\hat g_B:~ E[d(X,\hat
g_B(U,Y))]\leq D} \{ H(X|Y,U)+H(Y|X)\}.
\end{equation}
For the binary erasure distortion and a full-support joint source
pmf taking values in binary alphabets, (\ref{eqn:generalrho1})
simplifies to the expression given in Proposition~\ref{lem:erasure}.
\begin{proposition}\label{lem:erasure} If $\mathcal X=\mathcal Y=\{0,1\}$,
$\supp(p_{XY})=\{0,1\}^2$, $d$ is the binary erasure distortion, and
$D\in \rR$, then $\rho_{1} = \max_{p_{U|X}} (H(X|Y,U) + H(Y|X))$,
where $\mathcal U=\{0,e,1\}$ and
\begin{equation}
p_{U|X}(u|x)= \left\{
\begin{array}{ll}
\alpha_{0e}, & \mbox{ if } x=0, u=e, \\
1-\alpha_{0e}, & \mbox{ if } x=0, u=0, \\
\alpha_{1e}, & \mbox{ if } x=1, u=e, \\
1-\alpha_{1e}, & \mbox{ if } x=1, u=1,\\
0, & \mbox{ otherwise,}
\end{array}\right.\label{eqn:pUX}
\end{equation}
where $\alpha_{0e}, \alpha_{1e}\in [0,1]$ satisfy
$E[d(X,U)]=p_X(0)\alpha_{0e}+p_X(1)\alpha_{1e}\leq D$.
\end{proposition}

The expression for $\rho_1^A$ further simplifies to the one in
Proposition~\ref{prop:erasure} by using $p_{U|X}$ given by
(\ref{eqn:pUX}) in (\ref{eqn:generalrho1}).
\begin{proposition} \label{prop:erasure}
If $\mathcal X=\mathcal Y=\{0,1\}$, $\supp(p_{XY})=\{0,1\}^2$, $d$
is the binary erasure distortion, and $D\in \rR$, then
\begin{equation}
\rho_1^A(p_{XY},D)=\max_{\scriptstyle
\alpha_{0e},\alpha_{1e}\in[0,1]: \atop \scriptstyle
\phi(p_{XY},\alpha_{0e},\alpha_{1e})\leq D}
\psi(p_{XY},\alpha_{0e},\alpha_{1e}), \label{eqn:rho1}
\end{equation}
where
\begin{eqnarray*}
\lefteqn{\psi(p_{XY},\alpha_{0e},\alpha_{1e})}\\&:=&(p_{00}\alpha_{0e}+p_{10}\alpha_{1e})h_2\left(\frac{p_{00}\alpha_{0e}}{p_{00}\alpha_{0e}+p_{10}\alpha_{1e}}\right)\\
&&+(p_{01}\alpha_{0e}+p_{11}\alpha_{1e})h_2\left(\frac{p_{01}\alpha_{0e}}{p_{01}\alpha_{0e}+p_{11}\alpha_{1e}}\right)\\
&&+(p_{00}+p_{01})h_2\left(\frac{p_{00}}{p_{00}+p_{01}}\right)+(p_{11}+p_{10})h_2\left(\frac{p_{11}}{p_{11}+p_{10}}\right),
\end{eqnarray*}
$\phi(p_{XY},\alpha_{0e},\alpha_{1e}):=p_X(0) \alpha_{0e}+p_X(1)
\alpha_{1e}$.
\end{proposition}

Finally, for a DSBS with parameter $p$ and the binary erasure
distortion, $\rho_1^A$ reduces to the compact expression in
Proposition~\ref{prop:DSBS}.
\begin{proposition}\label{prop:DSBS}
If $d$ is the binary erasure distortion, $D\in [0,1]$, and $p_{XY}$
is the joint pmf of a DSBS with parameter $p$, then
\begin{equation}
\rho_1^A(p_{XY},D) =(1+D)h_2(p).\label{eqn:rho1DSBS}
\end{equation}
\end{proposition}

\noindent $\bullet$ \emph{Right-side of (\ref{eqn:notconcave}):}
Solving the rate reduction functionals in the right-side of
(\ref{eqn:notconcave}) requires solving the maximization problem
(\ref{eqn:rho1}) for asymmetric distributions $p_{X|Y}p_{Y,1}$ and
$p_{X|Y}p_{Y,2}$. Exactly solving this problem is cumbersome but it
is easy to provide a lower bound for the maximum as follows.
\begin{proposition}\label{prop:asymsource}
If $d$ is the binary erasure distortion, $p_{Y,1}$ is
Bernoulli$(q)$, $p_{Y,2}$ is Bernoulli$(\bar q)$, and $p_{X|Y}$ is
the conditional pmf of the binary symmetric channel with crossover
probability $p$, then the inequality
\begin{equation}
\frac{\rho_1^A(p_{X|Y}p_{Y,1},D)+\rho_1^A(p_{X|Y}p_{Y,2}, D)}{2}\geq
  C(p,q,\alpha_{0e},1)\label{eqn:rho1andC}
\end{equation}
holds for $D=\eta(p,q,\alpha_{0e},1)$, where
\begin{eqnarray}
C(p,q,\alpha_{0e},\alpha_{1e})&:=&\psi(p_{X|Y}p_{Y,1},\alpha_{0e},\alpha_{1e}),\nonumber\\
\eta(p,q,\alpha_{0e},\alpha_{1e})&:=&\phi(p_{X|Y}p_{Y,1},\alpha_{0e},\alpha_{1e}).
\nonumber
\end{eqnarray}
\end{proposition}
\begin{remark}\label{remark:achieveC} The rate-distortion tuple
$(H(X|Y)+H(Y|X)-C(p,q,\alpha_{0e},1),\eta(p,q,\alpha_{0e},1))$ is
  admissible for one-message source coding for joint source pmf
  $p_{X|Y}p_{Y,1}$ and corresponds to choosing $p_{U|X}$ given by
  (\ref{eqn:pUX}) with $\alpha_{1e}=1$ and the decoding function
  $g(u,y)=u$. Since this choice of $p_{U|X}$ and $\hat g_B$ may be
  suboptimal, $C(p,q,\alpha_{0e},1)$ is only a lower bound for the
  rate reduction functional.
\end{remark}

\noindent $\bullet$ \emph{Comparing left and right sides of
(\ref{eqn:notconcave}):} The left-side of (\ref{eqn:notconcave}) and
the lower bound of the right-side of (\ref{eqn:notconcave}) can be
compared as follows.
\begin{proposition}\label{prop:takinglimit}
Let $d$ be the binary erasure distortion, $p_Y$ be Bernoulli$(1/2)$,
and $p_{X|Y}$ be the binary symmetric channel with parameter $p$.
For all $q\in(0,1/2)$ and all $\alpha_{0e}\in (0,1)$, there exists $
p\in(0,1)$ such that the strict inequality $\rho_1^A(p_{XY},D) <
C(p,q,\alpha_{0e},1)$ holds for $D=\eta(p,q,\alpha_{0e},1)$.
\end{proposition}

Since the left-side of (\ref{eqn:notconcave}) is strictly less than
a lower bound of the right-side of (\ref{eqn:notconcave}), the
strict inequality (\ref{eqn:notconcave}) holds, which completes the
proof of Theorem~\ref{thm:improvement}.
\end{proof}

\subsection{Explicit proof of the benefit of
interaction}\label{subsection:largeratio}
Theorem~\ref{thm:largeratio} quantifies the multiplicative reduction
in the sum-rate that is possible with two messages.
\begin{theorem}\label{thm:largeratio}
If $d$ is the binary erasure distortion and $p_{XY}$ the joint pmf
of a DSBS with parameter $p$, then for all $L >0$ there exists an
admissible two-message rate-distortion tuple $(R_1,R_2, D)$ such
that $R_{sum,1}^A(p_{XY},D)/(R_1+R_2)>L$ and $R_1/R_2<1/L$.
\end{theorem}

\begin{proof}
The following single-letter characterization of $\mathcal{RD}^B_2$
was established in \cite{Kaspi1985}:
\begin{eqnarray}
{\mathcal {RD}}^B_2 = &\{&\!\!\!~(R_1,R_2,D)~|~\exists \
p_{V_1|Y},p_{V_2|XV_1},\hat g_B,
s.t.\nonumber\\
&&R_1 \geq I(Y;V_1|X),\nonumber\\
&&R_2 \geq I(X;V_2|Y,V_1), \nonumber\\
&& \eE[d(X,\hat g_B(V_1,V_2,Y))]\leq D ~\},\label{eqn:rateregion}
\end{eqnarray}
where $V_1\in \mathcal V_1$ and $V_2\in \mathcal V_2$ are auxiliary
random variables with bounded alphabets, such that the Markov chains
$V_1 - Y - X$ and $V_2 - (X,V_1) - Y$ hold, and $\hat g_B: \mathcal
V_1 \times \mathcal V_2 \times \mathcal Y\rightarrow
\widehat{\mathcal X}$ is a deterministic single-letter decoding
function.

We will explicitly construct $p_{V_1|Y},p_{V_2|XV_1}$, and $\hat
g_B$ which lead to an admissible tuple $(R_1,R_2,D)$. Let
$p_{V_1|Y}$ be the conditional pmf of the binary symmetric channel
with crossover probability $q$.  Let the conditional distribution
$p_{V_2|XV_1}(v_2|x,v_1)$ have the form described in
Table~\ref{tab:pV2} and let $\hat g_B(v_1,v_2,y):=v_2$.
\begin{table}[!htb]
\vglue -0.3cm
  \centering
 \caption{Conditional distribution $p_{V_2|XV_1}$}
\vglue -0.3cm
  \begin{tabular}{|c|c|c|c|}
  \hline
   $p_{V_2|XV_1}$ & $v_2=0$ & $v_2=e$& $v_2=1$ \\
   \hline
  $x=0,v_1=0$ & $1-\alpha$ & $\alpha$ & $0$ \\
  \hline
  $x=1,v_1=0$ & $0$ & $1$ & $0$\\
  \hline
  $x=0,v_1=1$ & $0$ & $1$ & $0$ \\
  \hline
  $x=1,v_1=1$ & $0$ & $\alpha$ & $1-\alpha$\\
  \hline
\end{tabular}
\label{tab:pV2} \vglue -0.3cm
\end{table}

The corresponding rate-distortion tuple can be shown to satisfy the
following property  {which is proved in
Appendix~\ref{app:proofs_Kaspisquestion}.}
\begin{proposition}\label{prop:largeratio}
Let $d$ be the binary erasure distortion and let $p_{XY}$ be the
joint pmf of a DSBS with parameter $p$. For $p_{V_1|Y},
p_{V_2|XV_1}$, and $\hat g_B$ as described above, and all $L >0$,
there exist parameters $p,q,\alpha$ such that the two-message
rate-distortion tuple $(R_1,R_2, D)$ given by $R_1 = I(Y;V_1|X)$,
$R_2 = I(X;V_2|Y,V_1)$, $D = E[d(X,V_2)]$ satisfies
$R_{sum,1}^A(p_{XY},D)/(R_1+R_2)>L$ and $R_1/R_2<1/L$.
\end{proposition}


This completes the proof of Theorem~\ref{thm:largeratio}.
\end{proof}

The conditional pmfs $p_{V_1|Y}$ and $p_{V_2|XV_1}$ in the proof of
Theorem~\ref{thm:largeratio} are related to the conditional pmf
$p_{U|X}$ in the proof of Theorem~\ref{thm:improvement} as follows.
Given $V_1=0$, the conditional distribution
$p_{XYV_2|V_1}(x,y,v_2|0)=p_{Y,1}(y) p_{X|Y}(x|y)p_{U|X}(v_2|x)$,
where $p_{U|X}$ is given by (\ref{eqn:pUX}) with
$\alpha_{0e}=\alpha$ and $\alpha_{1e}=1$. Given $V_1=1$, the
conditional distribution $p_{XYV_2|V_1}(x,y,v_2|1)=p_{Y,2}(y)
p_{X|Y}(x|y)p_{U|X}(v_2|x)$, where $p_{U|X}$ is given by
(\ref{eqn:pUX}) with $\alpha_{1e}=\alpha$ and $\alpha_{0e}=1$.
Conditioning on $V_1$, in effect, decomposes the two-message problem
into two one-message problems that were analyzed in the proof of
Theorem~\ref{thm:improvement}.

\subsection{Example showing $R_{sum,1}^A$ can be arbitrarily large and $R_{sum,2}^B$ can be arbitrarily small}
\label{subsection:Zamirexample}
In the example described in Section~\ref{subsection:largeratio}, the
multiplicative gain $R_{sum,1}^A/(R_1+R_2)$ is shown to be arbitrarily
large. The additive gain $R_{sum,1}^A-(R_1+R_2)$, however, is not
shown to be large. In this subsection we provide an example where
$R_{sum,1}^A$ can be arbitrarily large and the two-message sum-rate
$(R_1+R_2)$ can be arbitrarily small, so that both
$R_{sum,1}^A/(R_1+R_2)$ and $R_{sum,1}^A-(R_1+R_2)$ are arbitrarily
large. Interestingly, $R_1/R_2$ can be arbitrarily small at the same
time. The noninteractive one-message version of this example was
provided in \cite{Zamir_EAP}. In this paper we extend it to an
interactive problem to demonstrate the benefit of interaction.

To construct the example, we need to introduce the notion of a
\emph{planar difference set} used in \cite{Zamir_EAP}. Let $\mathcal
G$ be an Abelian group. A subset $\mathcal Z \subseteq \mathcal G$
is a planar difference set of $\mathcal G$ if \textcolor{myblue}{it is the largest possible subset such that} for all non-zero $d\in
\mathcal G$ the equation $z_1 - z_2 = d$ has exactly one solution
for $z_1, z_2 \in \mathcal Z$. Although not every Abelian group has
a subset that is a planar different set, it is true that for any
prime number $p$ and any positive integer $m$, a planar difference
set of size $\alpha = p^m+1$ exists in the group of integers modulo
$(\alpha^2-\alpha+1)$ (see \cite{Zamir_EAP}). Thus, there exist
planar difference sets of arbitrarily large size.

For our example, we take $\mathcal G$ to be the Abelian group of
integers with addition modulo $(\alpha^2-\alpha+1)$, where $\alpha =
p^m+1$, $p$ prime and $m$ a positive integer. Let $\mathcal Z$ be a
planar difference set of $\mathcal G$ of size $\alpha$. Let
$(W,S,K)\sim p_{WSK}:=p_W p_{S|W} p_K$, where
$p_W(2)=1-p_W(1):=\epsilon\in (0,1)$, $p_K:=\mbox{Unif}_{\mathcal
G}$, $p_{S|W}(\cdot|1):=\mbox{Unif}_{\mathcal Z}$, and
$p_{S|W}(\cdot|2):=\mbox{Unif}_{\mathcal G}$. Let $X:=(S+K)$ and
$Y:=(K,W)$.

Let $\{(X(i),Y(i))\}_{i=1}^n$ be $n$ iid samples of a two-component
discrete memoryless stationary source with joint pmf $p_{XY}$.
$\mathbf X = (X(1),\ldots,X(n))$ is available at terminal $A$ and
$\mathbf Y = (Y(1),\ldots,Y(n))$ is available at terminal $B$. Let
$\mathbf W = (W(1),\ldots,W(n))$ denote the second component of
source $\mathbf Y$. $B$ is required to reproduce $\mathbf X$ as
$\widehat {\mathbf X}$. The distortion function is defined as
follows.

\[d(x,\hat x)=\left\{
    \begin{array}{r@{\ , \quad}l}
      0 &\mbox{if }(x-\hat x)\in \mathcal Z\\
      \infty &\mbox{otherwise}.
    \end{array}
    \right.\]

The one-message Wyner-Ziv rate-distortion function for this problem
was evaluated in \cite{Zamir_EAP} (see Equation~(23) and the last
but one displayed equation in \cite{Zamir_EAP}) and is given by
\[\forall D>0, \hspace{1cm} R_{sum,1}^A(D)=(1-\epsilon)\left(1-\frac{1}{|\mathcal Z|}\right)\log_2 |\mathcal
Z|+\epsilon \left( \log_2|\mathcal G|-\log_2|\mathcal Z|\right).\]
Since $|\mathcal Z|=\alpha$ and $|\mathcal G|=\alpha^2-\alpha+1$, we
have
\[R_{sum,1}^A(D)=(1-\epsilon)\left(1-\frac{1}{\alpha}\right)\log_2\alpha+\epsilon \log_2 \left(\alpha-1+\frac{1}{\alpha}\right).\]


Let us now consider a two-message interactive code with initial
terminal $B$. In the first message, terminal $B$ sends $\mathbf W$
to $A$ at the rate $R_1=H(W)=h_2(\epsilon)$. For all the sample
indices $i\in \{1,\ldots,n\}$ such that $W(i)=1$ (which implies
$S(i) \in \mathcal Z$), terminal $A$ sends nothing and terminal $B$
generates $\widehat X(i) = K(i)$. Since for such sample indices
$d(X(i),K(i))=d(S(i),0)=0$ holds, the distortion at these samples is
zero. For the remaining sample indices, $W(i)=2$, and terminal $A$
sends $X(i)$ completely at rate $H(X)=\log_2|\mathcal G|$, and
terminal $B$ sets $\widehat X(i)=X(i)$. The distortion at these
samples is also zero. The rate of the second message is $R_2
=p_W(2)H(X)= \epsilon \log_2|\mathcal G|=\epsilon
\log_2(\alpha^2-\alpha+1)$. Hence $(R_1,R_2,D)=(h_2(\epsilon),
\epsilon \log_2(\alpha^2-\alpha+1), 0)$ is an admissible two-message
rate-distortion tuple.

Now take $\epsilon = 1/ (\log|\mathcal
G|)^2=1/(\log_2(\alpha^2-\alpha+1))^2$ and consider the asymptotic
behavior of $R_{sum,1}^A, (R_1/R_2), (R_1+R_2)$ as $\alpha$
increases. As $\alpha\rightarrow \infty$ we have
\begin{align*}
&R_{sum,1}^A(D)\sim \log_2 \alpha,\\
&\frac{R_1}{R_2}= \frac{h_2(\epsilon)}{\epsilon \log_2
(\alpha^2-\alpha+1)} \sim \frac{\log_2(1/\epsilon)}{\log_2
(\alpha^2-\alpha+1)} \sim \frac{\log_2\log_2 \alpha}{\log_2 \alpha},\\
&(R_1+R_2)\sim R_2=\epsilon \log_2 (\alpha^2-\alpha+1)=
\frac{1}{\log_2 (\alpha^2-\alpha+1)}\sim \frac{1}{2 \log_2 \alpha},
\end{align*}
where $f_1(\alpha)\sim f_2(\alpha)$ denotes $\lim_{\alpha\rightarrow
\infty} f_1(\alpha)/f_2(\alpha)=1$. Thus we have proved the
following theorem.
\begin{theorem}
There exists an interactive source reproduction problem, for which
the following properties hold simultaneously: (i) the Wyner-Ziv
rate-distortion function $R_{sum,1}^A$ is arbitrarily large, (ii)
the two-message sum-rate $(R_1+R_2)$ is arbitrarily small, and (iii)
the two-message sum-rate  {can be} split in a way that $(R_1/R_2)$
is arbitrarily small.
\end{theorem}

\section{Concluding remarks}\label{section:conclusion}

\textcolor{myblue}{In this work, we studied a two-terminal
  interactive function computation problem with alternating messages
  within the framework of distributed block source coding theory. We
  developed a new limit-free single-letter characterization of the
  infinite-message sum-rate-distortion function by viewing it as a
  functional of the joint source distribution and distortion levels
  and studying its convex-geometric properties. This led to an
  optimality test for any achievable rate-distortion functional and an
  iterative algorithm for computing the infinite-message limit. This
  allowed us to fully characterize the limit for computing the Boolean
  AND function of independent at one or both terminals and to
  construct the first examples that demonstrate that the Wyner-Ziv
  rate-distortion function can be strictly improved using two
  messages.}

\textcolor{myblue}{Several questions remain unresolved. First, it is
  unclear if a limit-free characterization can be developed for a {\em
    single} joint source pmf and a specified pair of distortion
  levels. The current characterization is for a whole family of pmfs
  that is closed under marginal-\-perturbations. We have seen that for
  some pmfs, the infinite-message limit is reached using a finite
  number of messages. Is there a simple way to test this for a given
  pmf? Is there a simple way to estimate the number of messages needed
  to reach the limit?
For computing the Boolean AND function of independent sources at one
or both terminals, we used a particular infinite-sequence of auxiliary
random variables which, for some pmfs, suggests the need for an
infinite number of infinitesimal-rate messages. Is this fundamentally
necessary or is it an artifact of our particular construction?
Algorithmic issues were not the focus of this work. A formal
complexity-analysis of the iterative algorithm would be an interesting
direction of research. Here we established pointwise convergence of
the algorithm but did not characterize the rate of convergence. There
is some empirical evidence to suggest that the convergence is uniform
and quadratic but it is unclear if this is true in general. Another
important question that is only empirically addressed here is the
impact of the discretization step-size in a computer implementation of
the iterative algorithm.}

\textcolor{myblue}{The success of the approach described in this paper
  hinges on the availability of a finite-message single-letter
  characterization. A fascinating open question is whether the
  functional characterization can be arrived at from
  first-\-principles without access to the finite-message single
  letter characterization. If so, then it may pave the way for
  analyzing other multi-terminal interactive source coding problems
  where single-letter characterizations are rarely available.
Our present approach can be extended (with suitable modifications) to
handle the so-called collocated networks \cite{Journal_col} where
source nodes observe {\em independent} sources and make a sequence of
noiseless broadcasts that are heard by all other nodes and a sink node
that wishes to compute a multivariable function of all the
sources. More general network problems seem daunting at this time. One
approach to make progress would be to develop ``interactive cutset
bounds'' and ``collocated network bounds'' as in \cite{Journal_col}.
Connections to entropic regions in network coding (\cite{YeungISIT},
\cite[Chapter 21]{networkcoding}) are also worth exploring.}


\section*{Acknowledgment}
The authors would like to thank Prof.~David Casta\~{n}\'{o}n for
discussions related to marginal-\-perturbations-\-closed families of
pmfs, Dr.~Vinod Prabhakaran for introducing them to the unresolved
question in \cite{Kaspi1985}, and Prof.~Ram Zamir for introducing them
to the example in Section~\ref{subsection:Zamirexample} and for
suggesting that it might provide another example where the Wyner-Ziv
rate-distortion function can be strictly improved.


\appendices
\renewcommand{\theequation}{\thesection.\arabic{equation}}
\setcounter{equation}{0}

\section{\label{app:lemma1proof}Proof of Lemma~\ref{lemma:connection}}

\textcolor{myblue}{First we define two collections of conditional pmfs for convenience:
\begin{equation*}
  \mathcal{P}_{ent,t}(p_{XY},f_A,f_B)  :=  \{p_{U^t|XY}:
  \mbox{condition (i) (conditional entropy constraint) of Fact~\ref{fact:finite} holds}~\}, 
\end{equation*}
\begin{equation*}
  \mathcal{P}_{mc,t}^A := \{p_{U^t|XY}:
  \mbox{conditions (ii) and (iii) (Markov chains and cardinality bounds) of Fact~\ref{fact:finite} hold}~ \}. 
\end{equation*}
Then we have $\mathcal{P}_t^A
(p_{XY},f_A,f_B):=\mathcal{P}_{mc,t}^A \cap
\mathcal{P}_{ent,t}(p_{XY},f_A,f_B)$.}

(i) For all $t\in \zZ^+$ and all $p_{XY}\in \mathcal{P}_{XY}$, we have
\begin{eqnarray*}
  \rho^A_t(p_{XY}) &=&
      \max_{ p_{U^t|XY} \in\; \mathcal{P}_t^A(p_{XY}) }
      \left\{ H(X|Y,U^t) + H(Y|X,U^t) \right\} \nonumber \\
  &=&\max_{ p_{U_1|X} }
       \left\{ \hspace{-10ex}
         \max_{ \scriptstyle p_{U_2^t|XYU_1}: \atop \scriptstyle
       \hspace{10ex} p_{U_1|X} p_{U_2^t|XYU_1} \in\;
       \mathcal{P}_t^A(p_{XY}) }
           \hspace{-10ex} \left\{ H(X|Y,U^t) + H(Y|X,U^t) \right\}
       \right\} \nonumber \\
  &\stackrel{(a)}{=}& \max_{p_{U_1|X}}
       \left\{ \hspace{-5ex} \sum_{ \hspace{5ex}u_1\in\;
       \supp(p_{U_1}) } \hspace{-5ex}  p_{U_1}(u_1)
         \left\{ \hspace{-14.5ex}
       \max_{ \scriptstyle \hspace{4ex}
         p_{U_2^t|XYU_1}(\cdot|\cdot,\cdot,u_1): \atop
         \scriptstyle \hspace{14.5ex} p_{U_1|X} p_{U_2^t|XYU_1}
         \in\; \mathcal{P}_t^A(p_{XY|U_1}(\cdot,\cdot|u_1)) }
         \hspace{-11.5ex} \left\{ H(X|Y,U_2^t,U_1=u_1)
%
%
%
  +H(Y|X,U_2^t,U_1=u_1) \right\} \right\} \right\} \nonumber \\
  &\stackrel{(b)}{=}& \max_{p_{U_1|X}}\left\{ \sum_{u_1\in\;
  \supp(p_{U_1})} p_{U_1}(u_1)\:
  \rho^B_{t-1}(p_{XY|U_1}(\cdot,\cdot|u_1))\right\}.
\end{eqnarray*}

Step (a) follows from the ``law of total conditional entropy'' with
the additional observation that conditioned on $U_1= u_1$,
$(H(X|Y,U_2^t,U_1=u_1) + H(Y|X,U_2^t,U_1=u_1))$ only depends on
$p_{U_2^t|XYU_1}(\cdot|\cdot,\cdot,u_1)$. Step (b) is due to the
observation that for a fixed $p_{U_1|X}$, conditioned on $U_1=u_1$,
(i) $p_{U_1|X}p_{U_2^t|XYU_1}\in\; \mathcal P_{mc,t}^A$ iff
$p_{U_2^t|XYU_1}\in \mathcal P_{mc,t-1}^B$ and (ii)
$p_{U_1|X}p_{U_2^t|XYU_1}\in\; \mathcal
P_{ent,t}(p_{XY,u_1},f_A,f_B)$,  {where $p_{XY,u_1} :=
p_{XY|U_1}(\cdot,\cdot|u_1)$,} iff $p_{U_2^t|XYU_1}$ $\in\mathcal
P_{ent,t-1}(p_{XY,u_1},f_A,f_B)$. Therefore,
$p_{U_1|X}p_{U_2^t|XYU_1}\in\; \mathcal P_t^A(p_{XY,u_1})$ iff
$p_{U_2^t|XYU_1}\in\; \mathcal P_{t-1}^B(p_{XY,u_1})$.

%
%

(ii) For an arbitrary $q_{XY}\in\; \mathcal{P}_{XY}$, consider two
arbitrary joint pmfs $p_{XY,1}, p_{XY,0}\in\;
\mathcal{P}_{Y|X}(q_{XY})$.  For every $\lambda \in (0,1)$, let
$p_{XY,\lambda}:= \lambda p_{XY,1}+ \bar \lambda p_{XY,0}$. Due to
Remark~\ref{rem:margpertbsets}(i), $p_{XY,\lambda}\in\;
\mathcal{P}_{Y|X}(q_{XY})$.  We need to show that
$\rho^A_t(p_{XY,\lambda}) \geq
\lambda\:\rho^A_t(p_{XY,1})+\bar{\lambda}\:\rho^A_t(p_{XY,0})$. For
$i=0,1$, let $p_{U_1|X,i}$ be the conditional pmf that maximizes the
objective function in (\ref{eqn:convexify}) for source pmf
$p_{XY,i}$. Let $p_{U_1,i}$ and $p_{XY|U_1,i}$ denote, {respectively,}
the $U_1$-marginal and conditional {pmfs} of
$p_{XY,i}p_{U_1|X,i}$. Thus we have $\rho^A_t(p_{XY,i})=\sum_{u_1}
p_{U_1,i}(u_1) \rho^B_{t-1}(p_{XY|U_1,i}(\cdot,\cdot|u_1))$. Define a
new auxiliary variable $V\in \mathcal{U}_1\times\{0,1\}$ with pmf
$p_{V}(u_1,1):=\lambda p_{U_1,1}(u_1)$ and
$p_{V}(u_1,0):=\bar{\lambda} p_{U_1,0}(u_1)$. Let
$p_{XY|V}(\cdot,\cdot|u_1,i):=p_{XY|U_1,i}(\cdot,\cdot|u_1)$. Then the
$XY$-marginal pmf of $p_V p_{XY|V}$ is $p_{XY,\lambda}$. We have,
\begin{eqnarray*}
\lambda\:\rho^A_t(p_{XY,1})+\bar{\lambda}\:\rho^A_t(p_{XY,0})&=&\lambda
\sum_{u_1}p_{U_1,1}(u_1)
\rho^B_{t-1}(p_{XY|U_1,1}(\cdot,\cdot|u_1))+\bar{\lambda}\sum_{u_1}p_{U_1,0}(u_1)
\rho^B_{t-1}(p_{XY|U_1,0}(\cdot,\cdot|u_1))\\
  &=&\sum_{u_1,i}p_V(u_1,i)\rho^B_{t-1}(p_{XY|V}(\cdot,\cdot|u_1,i))\\
  &\leq& \rho^A_t(p_{XY,\lambda}),
\end{eqnarray*}
where the last step is because (\ref{eqn:convexify}) holds for
source pmf $p_{XY,\lambda}$. \vspace{1ex}

(iii)
\textcolor{myblue}{
First we make the following observation.
}

\vspace{1ex}
\textcolor{myblue}{\begin{remark}\label{rem:lemma1i}
For all $u_1\in\; \supp(p_{U_1})$, $p_{XY|U_1}(\cdot,\cdot|u_1)\in\;
\mathcal{P}_{Y|X}(p_{XY})$. This is confirmed by noting that since
$Y-X-U_1$ is a Markov chain,  {$\forall (x,u_1) \in\;
\supp(p_{X}p_{U_1|X})$}
we have $p_{Y|XU_1}(y|x,u_1)=p_{Y|X}(y|x)$ (see the paragraph after
Definition~\ref{def:xperturb}).
\end{remark}}
\vspace{1ex}

For all $p_{XY}\in \mathcal{P}_{Y|X}(q_{XY})$, we have
\begin{eqnarray*}
\rho^A_t(p_{XY})&\stackrel{(c)}{=}&\max_{p_{U_1|X}}\left\{
\sum_{u_1} p_{U_1}(u_1)\:
\rho^B_{t-1}(p_{XY|U_1}(\cdot,\cdot|u_1))\right\}\\
  &\stackrel{(d)}{\leq}& \max_{p_{U_1|X}}\left\{ \sum_{u_1}
p_{U_1}(u_1)\: \rho(p_{XY|U_1}(\cdot,\cdot|u_1))\right\}\\
    &\stackrel{(e)}{\leq}&\rho(p_{XY}),
\end{eqnarray*}
 {where the equality (c) follows from (\ref{eqn:convexify}), the
inequality (d) is true because $\rho^B_{t-1}(p_{XY}) \leq
\rho(p_{XY})$, and the final inequality (e) is true due to the
following reasons (i)
$p_{XY|U_1}(\cdot,\cdot|u_1)\in\mathcal{P}_{Y|X}(p_{XY})\subseteq
\mathcal{P}_{Y|X}(q_{XY})$ due to Remarks~\ref{rem:margpertbsets} and
\ref{rem:lemma1i}, (ii) $\rho$ is concave on
$\mathcal{P}_{Y|X}(q_{XY})$, and (3) Jensen's inequality.}

\textcolor{myblue}{(iv) This part is proved by reversing the roles of terminal $A$ and $B$ in parts (i) -- (iii).}
\hspace*{\fill}~\QED\\

\textcolor{myblue}{\section{Proof of
Proposition~\ref{prop:concavity}} \label{app:proofconcavity}}

\textcolor{myblue}{We first prove part (i) of the proposition. Let
  $(X_{[0]},Y_{[0]})$ and $(X_{[1]},Y_{[1]})$ be independent random
  tuples with $(X_{[\theta]},Y_{[\theta]}) \sim p_{XY_{[\theta]}} \in
  \Delta(\mathcal{X}\times\mathcal{Y})$, for $\theta=0,1$. Let $\Theta
  \sim$ Ber$(\lambda), \lambda \in (0,1)$, be independent of
  $(X_{[0]},Y_{[0]})$ and $(X_{[1]},Y_{[1]})$. Then
  $(X_{[\Theta]},Y_{[\Theta]})$ has the mixture pmf: $\lambda
  p_{XY_{[1]}}+(1-\lambda)p_{XY_{[0]}} =: p_{XY_{[\lambda]}}$.}

\textcolor{myblue}{Due to (\ref{eqn:minsumrate}), for any $t$, we have}
\textcolor{myblue}{\begin{eqnarray*}
R_{sum,t}^A(p_{XY_{[\lambda]}})&=& \min_{p_{U^t|XY_{[\Theta]}}\in
\mathcal{P}_t^A \left(p_{XY_{[\lambda]}}\right)}I(X_{[\Theta]};U^t|Y_{[\Theta]})+I(Y_{[\Theta]};U^t|X_{[\Theta]})\nonumber\\
&\stackrel{(a)}{=}& \min_{p_{U^t|XY_{[\Theta]}}\in
\mathcal{P}_t^A \left(p_{XY_{[\lambda]}}\right)}I(X_{[\Theta]},\Theta;U^t|Y_{[\Theta]})+I(Y_{[\Theta]},\Theta;U^t|X_{[\Theta]})\nonumber\\
&\geq& \min_{p_{U^t|XY_{[\Theta]}}\in
\mathcal{P}_t^A \left(p_{XY_{[\lambda]}}\right)}I(X_{[\Theta]};U^t|Y_{[\Theta]},\Theta)+I(Y_{[\Theta]};U^t|X_{[\Theta]},\Theta)\nonumber\\
&\geq& \lambda \left(\min_{p_{U^t|XY_{[\Theta]}}\in
\mathcal{P}_t^A \left(p_{XY_{[\lambda]}}\right)}I(X_{[\Theta]};U^t|Y_{[\Theta]},\Theta=1)+I(Y_{[\Theta]};U^t|X_{[\Theta]},\Theta=1)\right)\\
&&+(1-\lambda)
\left(\min_{p_{U^t|XY_{[\Theta]}}\in
\mathcal{P}_t^A \left(p_{XY_{[\lambda]}}\right)}I(X_{[\Theta]};U^t|Y_{[\Theta]},\Theta=0)+I(Y_{[\Theta]};U^t|X_{[\Theta]},\Theta=0)\right)\\
&\stackrel{(b)}{\geq} & \lambda \left(\min_{p_{U^t|XY_{[\Theta]}}\in
\mathcal{P}_t^A \left(p_{XY_{[1]}}\right)}I(X_{[\Theta]};U^t|Y_{[\Theta]},\Theta=1)+I(Y_{[\Theta]};U^t|X_{[\Theta]},\Theta=1)\right)\\
&&+(1-\lambda)
\left(\min_{p_{U^t|XY_{[\Theta]}}\in
\mathcal{P}_t^A \left(p_{XY_{[0]}}\right)}I(X_{[\Theta]};U^t|Y_{[\Theta]},\Theta=0)+I(Y_{[\Theta]};U^t|X_{[\Theta]},\Theta=0)\right)\\
&=& \lambda R_{sum,t}^A(p_{XY_{[1]}}) + (1-\lambda)
R_{sum,t}^A(p_{XY_{[0]}}).
\end{eqnarray*}}
\textcolor{myblue}{Step (a) is because $\Theta-(X_{[\Theta]},Y_{[\Theta]})-U^t$
forms a Markov chain. Step (b) is because
$\mathcal{P}_t^A\left(p_{XY_{[\lambda]}}\right)\subseteq
\mathcal{P}_t^A\left(p_{XY_{[1]}}\right)\bigcap
\mathcal{P}_t^A\left(p_{XY_{[0]}}\right)$, which is due to the
following three reasons. (i) $H(f_A(X_{[\Theta]},Y_{[\Theta]})|U^t)=0$ implies
$H(f_A(X_{[\Theta]},Y_{[\Theta]})|U^t,\Theta)=0$, which in turn implies
$H(f_A(X_{[\Theta]},Y_{[\Theta]})|U^t,\Theta=\theta)=0$ for $\theta=0,1$.
The same property holds if $f_A$ is replaced by $f_B$. (ii) The Markov chain constraints induce the following factorization structure
on the conditional pmf of $U^t$ given $X^m$: $p_{U^t|XY}(u^t|x,y) = p_{U_1|X}(u_1|x) \cdot
  p_{U_2|YU_1}(u_2|y,u_1)
  \cdot\:p_{U_3|XU^2}(u_3|x,u^2) \ldots$, which is common for
all source pmfs. (iii) The cardinality bounds are common for all
source pmfs.}

\textcolor{myblue}{Part (ii) of the proposition immediately follows
from part (i) by taking the limit $t\rightarrow \infty$.}

\section{\label{app:achievabililty_example2}Achievability of $R^*$ in
  Section~\ref{subsection:oneside}}

The achievability proof of $R^*(p,q)$ for $(p,q)\in (0,1/2]^2$ uses
the same technique {that was used} in \cite[Sec.~IV.F]{ISIT08}.  {If
  $R^*(p,q)$ can be shown to be an achievable sum-rate for $(p,q)\in
  (0,1/2]^2$, then $R^*(1-p,q)$ will be an achievable sum-rate for
$(p,q)\in [1/2,1)\times(0,1/2]$. This is because when $p \geq 1/2$,
$X^c \sim$ Ber$(1-p)$ (with $(1-p)\leq 1/2$) so that an achievable
coding scheme for $(p,q) \in (0,1/2]^2$ can be used to enable $B$ to
compute $\mathbf X^c \wedge \mathbf Y$ and therefore also $\mathbf X
\wedge \mathbf Y = (\mathbf X^c \wedge \mathbf Y)^c \wedge \mathbf
Y$. For $1/2\leq q \leq 1$, the rate $H(X) = h_2(p)$ is an achievable
sum-rate because it corresponds to a coding scheme in which $\mathbf
X$ is completely reproduced at $B$ by sending a single message of rate
$h_2(p)$ from $A$ to $B$. We will now show that $R^*(p,q)$ is an
achievable sum-rate for $(p,q)\in (0,1/2]^2$.}  Since we are
    interested in the limit $t \rightarrow \infty$, it is sufficient
    to consider only even-valued $t$.

\textcolor{myblue}{The proof of achievability of $R^*(p,q)$ is based
  on random coding and random binning arguments
  \cite{CoverThomas}. Choosing the auxiliary random variables $U^t$ in
  (\ref{eqn:minsumrate}) plays a central role in the generation of
  random codebooks in random coding, and in determining the sizes of
  bins in random binning. As $t\rightarrow \infty$, an unbounded
  number of auxiliary random variables need to be chosen. In the next
  paragraph we describe a systematic procedure to construct $U^t$
  which ensures that (i) the constructed $U^t$ satisfy the
  requirements in Fact~\ref{fact:finite} even when $t$ grows without
  bound, and (ii) the achievable sum-rate converges as $t\rightarrow
  \infty$.}

\textcolor{myblue}{Our construction of $U^t$ is based upon two extra
  auxiliary random variables $V_x$ and $V_y$.} {Let $(V_x,V_y) \sim
  \mbox{Unif}_{[0,1]^2}$ be real auxiliary random variables and define
  $X := 1_{[1-p,1]}(V_x)$ and $Y :=1_{[1-q,1]}(V_y)$. Then $p_{XY}$ is
  indistinguishable from the joint source pmf, i.e., $p_X(1) = 1 -
  p_X(0) = p$, $p_Y(1) = 1 - p_Y(0) = q$ and $X\Perp Y$. We will
  interpret the symbols $0$ and $1$ as real zero and real one
  respectively as needed. This interpretation will allow us to express
  Boolean arithmetic in terms of real arithmetic, e.g., $X \wedge Y$
  (Boolean AND) $= XY$ (real multiplication).} Define a {\em
  rate-allocation curve} $\Gamma$ parametrically by $\Gamma :=
\{(\alpha(s), \beta(s)), 0 \leq s \leq 1\}$ where $\alpha$ and $\beta$
are real, nondecreasing, absolutely continuous functions with
$\alpha(0)=\beta(0)=0$, $\alpha(1)=(1-p)$, and
$\beta(1)\in[0,1-q]$. We note that in \cite[Sec.~IV.F]{ISIT08} where
the AND function {was required to be} computed at both terminals
rather than {at} only terminal $B$, $\Gamma$ {was required} to satisfy
a different condition $\beta(1)=(1-q)$ {and this lead} to a different
admissible sum-rate {expression}. The significance of $\Gamma$ will
become clear later. Now choose a partition of $[0,1]$, $0 = s_0 < s_1
<\ldots <s_{t/2-1}<s_{t/2}=1$, such that
$\max_{i=1,\ldots,t/2}(s_i-s_{i-1})<\Delta_t$.  For $i=1,\ldots,t/2$,
define $t$ auxiliary random variables as follows,
\[
U_{2i-1}:= 1 _{[\alpha(s_i),1]\times [\beta(s_{i-1}),1]}(V_x,V_y),~
U_{2i}:= 1_{[\alpha(s_i),1]\times [\beta(s_i),1]}(V_x,V_y).
\]

\begin{figure*}[!htb]
\begin{center}
\scalebox{0.5}{\input{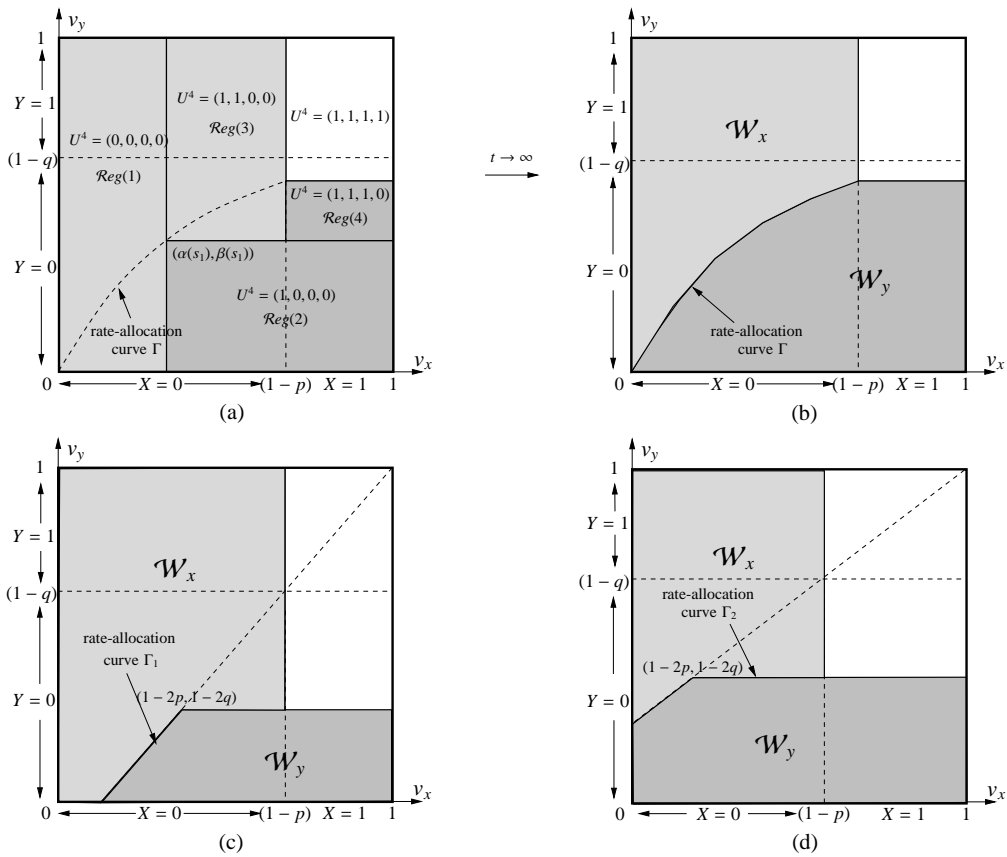}}
\end{center}
\caption{\sl (a) $4$-message interactive code (b) $\infty$-message
interactive code (c) $\infty$-message interactive code for $0<p \leq
q\leq 1/2$ with rate-allocation curve $\Gamma_1$ (d)
$\infty$-message interactive code for $0<q \leq p\leq 1/2$ with
rate-allocation curve $\Gamma_2$. \label{fig:integral}}
\end{figure*}

In Figure~\ref{fig:integral}(a), $(V_x,V_y)$ is uniformly distributed
on the unit square and $U^t$ are defined to be $1$ in rectangular
regions which are nested.  The following properties can be verified:

\begin{itemize}
\item[$P1$:] $U_1 \geq U_2 \geq \ldots \geq U_t$.
\item[$P2$:] \textcolor{myblue}{$p_{U^t|XY}$ satisfies the conditional entropy constraint in Fact~\ref{fact:finite}, that is,}
$H(X \wedge Y| Y, U^t)=0$: since
$U_{t}=1_{[1-p,1] \times [\beta(1),1]}(V_x,V_y)$ and
$Y=1_{[1-q,1]}(V_y)$. Therefore $U_{t}\wedge Y=1_{[1-p,1] \times
[1-q,1]}(V_x,V_y)=X \wedge Y$.
\item[$P3$:] \textcolor{myblue}{$p_{U^t|XY}$ satisfies the Markov chain constraints in Fact~\ref{fact:finite}}: for example,
consider $U_{2i} - (Y,U^{2i-1})- X$. $U_{2i-1}=0 \Rightarrow
U_{2i}=0$ and the Markov chain holds. $U_{2i-1}= Y =1 \Rightarrow
(V_x, V_y) \in [\alpha(s_i),1] \times [1-q,1] \Rightarrow U_{2i}=1$
and the Markov chain holds. Given $U_{2i-1}=1, Y=0$, $(V_x,V_y) \sim
\mbox{Unif}_{[\alpha(s_i),1] \times [\beta(s_{i-1}),1-q]}
\Rightarrow$ $V_x $ and $V_y$ are conditionally independent. Thus $X
\Perp U_{2i}| (U_{2i-1}=1, Y=0)$ because $X$ is a function of only
$V_x$ and $U_{2i}$ is a function of only $V_y$ upon conditioning. So
the Markov chain $U_{2i} - (Y,U^{2i-1})- X$ holds in all situations.
 \item[$P4$:] $(Y,U_{2i}) \Perp X |U_{2i-1}=1$: this can be proved by
 the same method as in $P3$.
\end{itemize}
$P2$ and $P3$ show that $p_{U^t|XY}\in \mathcal{P}_t^A(p_{XY})$.

For $i = 1,\ldots,t/2$, the $(2i)$-th rate is given by
\begin{eqnarray*}
\lefteqn{I(Y;U_{2i}|X,U^{2i-1}) =} &&\\
&\stackrel{P1}{=}&I(Y;U_{2i}|X,U_{2i-1}=1)p_{U_{2i-1}}(1)\\
&\stackrel{P4}{=}&I(Y;U_{2i}|U_{2i-1}=1)p_{U_{2i-1}}(1)\\
&=&H(Y|U_{2i-1}=1)p_{U_{2i-1}}(1)-H(Y|U_{2i},
U_{2i-1}=1)p_{U_{2i-1}}(1)
\\
&\stackrel{(a)}{=}&H(Y|U_{2i-1}=1)p_{U_{2i-1}}(1)-H(Y|U_{2i}=1)p_{U_{2i}}(1)\\
&=&
(1-\alpha(s_i))\left((1-\beta(s_{i-1}))h_2\left(\frac{q}{1-\beta(s_{i-1})}\right)
\right.\\ &&\left.
-(1-\beta(s_i))h_2\left(\frac{q}{1-\beta(s_i)}\right)\right)
\\
&\stackrel{(b)}{=}& (1-\alpha(s_i)) \int_{\beta(s_{i-1})}^{\beta(s_i)} \log_2 \left(\frac{1-v_y}{1-q-v_y}\right) d v_y \\
&=&
\int\!\!\!\!\int_{[\alpha(s_i),1]\times[\beta(s_{i-1}),\beta(s_i)]}
w_y(v_y,q) d v_x d v_y,
\end{eqnarray*}
where step (a) is due to property $P4$ and because
$(U_{2i-1},U_{2i})=(1,0) \Rightarrow Y=0$, hence $H(Y|U_{2i},
U_{2i-1}=1)p_{U_{2i-1}}(1) =
H(Y|U_{2i}=1,U_{2i-1}=1)p_{U_{2i},U_{2i-1}}(1,1)\stackrel{P1}{=}H(Y|U_{2i}=1)p_{U_{2i}}(1)$,
and step (b) is because
\[
\frac{\partial}{\partial v_y} \left( -(1-v_y)h_2\left(
\frac{q}{1-v_y}\right) \right) = \log_2
\left(\frac{1-v_y}{1-q-v_y}\right)=: w_y(v_y,q).
\]
The $2i$-th rate can thus be expressed as a 2-D integral of a weight
function $w_y$ over the rectangular region ${\mathcal
  Reg}(2i):=[\alpha(s_i),1]\times[\beta(s_{i-1}),\beta(s_i)]$ (a
horizontal bar in Figure~\ref{fig:integral}(a)). Therefore, the sum of
rates of all messages sent from terminal $B$ to terminal $A$ is the
integral of $w_y$ over the union of all the corresponding horizontal
bars in Figure~\ref{fig:integral}(a). Similarly, the sum of rates of
all messages sent from terminal $A$ to terminal $B$ can be expressed
as the integral of another weight function
$w_x(v_x,p):=\log_2((1-v_x)/(1-p-v_x))$ over the union of all the
vertical bars in Figure~\ref{fig:integral}(a).

Now let $t\rightarrow \infty$ such that $\Delta_t \rightarrow 0$.
Since $\alpha$ and $\beta$ are absolutely continuous,
$(\alpha(s_i)-\alpha(s_{i-1}))\rightarrow 0$ and
$(\beta(s_i)-\beta(s_{i-1}))\rightarrow 0$.  The union of the
horizontal (resp.~vertical bars) in Figure~\ref{fig:integral}(a) tends
to the region ${\mathcal W}_y$ (resp.~${\mathcal W}_x$) in
Figure~\ref{fig:integral}(b).  Hence an achievable infinite-message
sum-rate given by
\begin{equation}
\int\!\!\!\!\int_{{\mathcal W}_x} w_x(v_x,p) dv_x
dv_y+\int\!\!\!\!\int_{{\mathcal W}_y} w_y(v_y,q) dv_x
dv_y\label{eqn:integralrate}
\end{equation}
depends on only the rate-allocation curve $\Gamma$ which coordinates
the progress of source descriptions at $A$ and $B$. When $0<p \leq
q\leq1/2$, choose $\Gamma=\Gamma_1$ to be the piecewise linear curve
connecting $(0,0)$, $(1-p/q,0)$, $(1-2p,1-2q)$, and $(1-p,1-2q)$ in
that order (see Figure~\ref{fig:integral}(c)). When $0<q \leq p\leq
1/2$, choose $\Gamma=\Gamma_2$ to be the piecewise linear curve
connecting $(0,0)$, $(0,1-q/p)$, $(1-2p,1-2q)$, and $(1-p,1-2q)$ in
that order (see Figure~\ref{fig:integral}(d)). For these two choices
of the rate-allocation curve, (\ref{eqn:integralrate}) can be
evaluated in closed form and is given by the expressions in the first
two cases of (\ref{eqn:Rstaroneside}), which completes the proof.

\begin{remark}
The two curves $\Gamma_1$ and $\Gamma_2$ were specifically chosen to
minimize the value of (\ref{eqn:integralrate}).  {The (nontrivial)
proof of this fact is omitted because it is not needed to show that
$R^*$ in (\ref{eqn:Rstaroneside}) is an achievable sum-rate.}
\end{remark}

\section{\label{app:lemma3proof}Proof of Lemma~\ref{lemma:connectionrd}}

(i) For all $t\in \zZ^+$ and all $p_{XY}\in \mathcal{P}_{XY}$, we have
\begin{eqnarray*}
\lefteqn{\rho^A_t(p_{XY},
\mathbf{D})}\nonumber\\&=&\max_{(p_{U^t|XY},\hat g_A, \hat
g_B)\in\mathcal{P}_t^A(p_{XY},\mathbf{D})} \left\{ H(X|Y,U^t)+H(Y|X,U^t)\right\}\nonumber\\
&=& \max_{p_{U_1|X}} \left\{\hspace{-11ex} \max_{\scriptstyle
(p_{U_2^t|XYU_1},\hat g_A, \hat g_B): \atop \hspace{11ex}
\scriptstyle (p_{U_1|X} p_{U_2^t|XYU_1},\hat g_A, \hat g_B)
\in\mathcal{P}_t^A(p_{XY},\mathbf{D})} \hspace{-11ex}
\left\{ H(X|Y,U^t)+H(Y|X,U^t)\right\}\right\}\nonumber\\
&\stackrel{(a)}{=}& \max_{p_{U_1|X}} \left\{\max_{\scriptstyle
\mathbf{D}_{u_1}\in \mathcal{D}^2,\forall u_1\in \mathcal U_1:\atop
\scriptstyle
E [\mathbf{D}_{U_1}]=\mathbf{D} } \left\{\hspace{-26ex}
\max_{\scriptstyle (p_{U_2^t|XYU_1},\hat g_A, \hat g_B): \atop
\hspace{26ex} \scriptstyle (p_{U_1|X}(u_1|\cdot)
p_{U_2^t|XYU_1}(\cdot|\cdot, \cdot,u_1), \hat g_A, \hat g_B)
\in\mathcal{P}_t^A(p_{XY|U_1}(\cdot,\cdot|u_1),\mathbf{D}_{u_1})}
\hspace{-20ex}
\left\{ H(X|Y,U^t)+H(Y|X,U^t)\right\}\right\}\right\}\nonumber\\
&\stackrel{(b)}{=}& \max_{p_{U_1|X}} \left\{\max_{\scriptstyle
\mathbf{D}_{u_1}\in \mathcal{D}^2,\forall u_1\in \mathcal U_1:\atop
\scriptstyle
E [\mathbf{D}_{U_1}]=\mathbf{D} } \left\{ \hspace{-6ex}
\sum_{\hspace{6ex} u_1\in \supp(p_{U_1})}\hspace{-6ex} p_{U_1}(u_1)
\left\{\hspace{-21ex} \max_{\scriptstyle
(p_{U_2^t|XYU_1^*}(\cdot|\cdot,\cdot,u_1),\hat g_A(u_1\ldots), \hat
g_B(u_1\ldots)): \atop \hspace{21ex} \scriptstyle
(p_{U_1^*|X}(u_1|\cdot) p_{U_2^t|XYU_1^*}(\cdot|\cdot, \cdot,u_1),
\hat g_A(u_1\ldots), \hat g_B(u_1\ldots))
\in\mathcal{P}_t^A(p_{XY|U_1}(\cdot,\cdot|u_1),\mathbf{D}_{u_1})}
\hspace{-20ex}
\left\{ H(X|Y,U_2^t,U_1=u_1)+H(Y|X,U_2^t,U_1=u_1)\right\}\right\}\right\}\right\}\nonumber\\
&\stackrel{(c)}{=}& \max_{p_{U_1|X}}\left\{\max_{\scriptstyle
\mathbf{D}_{u_1}\in \mathcal{D}^2,\forall u_1\in \mathcal U_1:\atop
\scriptstyle
E [\mathbf{D}_{U_1}]=\mathbf{D}} \left\{ \sum_{u_1\in\;
\supp(p_{U_1})} p_{U_1}(u_1)
\rho^B_{t-1}(p_{XY|U_1}(\cdot,\cdot|u_1),
\mathbf{D}_{u_1})\right\}\right\}.
\end{eqnarray*}

In step (a) we replaced the overall distortion constraints
$E[d_A(X,Y,\hat g_A(U^t,X))]\leq D_{A}$ and $E[d_B(X,Y,\hat
g_B(U^t,Y))]\leq D_{B}$ by the individual distortion constraints
$E[d_A(X,Y,$ $\hat g_A(U_1,U_2^t,X))|U_1=u_1]\leq D_{A,u_1}$ and
$E[d_B(X,Y,\hat g_B(U_1,U_2^t,Y))|U_1=u_1]\leq D_{B,u_1}$ for all
$u_1\in \mathcal U_1$, and maximized the objective function over all
the possibilities of the individual distortion levels $\mathbf
D_{u_1}$ satisfying $E[\mathbf D_{U_1}]=\sum_{u_1}\mathbf
D_{u_1}p_{U_1}(u_1)=\mathbf D$. Step (b) follows from the ``law of
total conditional entropy'' with the additional observation that
conditioned on $U_1=u_1$, $(H(X|Y,U_2^t, U_1=u_1)+H(Y|X,U_2^t,
U_1=u_1))$ only depends on $p_{U_2^t|XYU_1}(\cdot|\cdot,\cdot,u_1)$,
$\hat g_A(u_1,\ldots)$, and $\hat g_B(u_1,\ldots)$. Step (c) is due
to the observation that for a fixed $p_{U_1|X}$, conditioned on
$U_1=u_1$, $(p_{U_1|X} p_{U_2^t|XYU_1}, \hat g_A, \hat g_B)
\in\mathcal{P}_t^A(p_{XY,u_1},\mathbf{D}_{u_1})$ iff
$(p_{U_2^t|XYU_1}, \hat g_A, \hat g_B)
\in\mathcal{P}_{t-1}^B(p_{XY,u_1},\mathbf{D}_{u_1})$,  {where
$p_{XY,u_1} := p_{XY|U_1}(\cdot,\cdot|u_1)$}. As discussed in
Remark~\ref{rem:lemma1i}  \textcolor{myblue}{in Appendix~\ref{app:lemma1proof}},  {for all $u_1\in\; \supp(p_{U_1})$,
$p_{XY|U_1}(\cdot,\cdot|u_1)\in\; \mathcal{P}_{Y|X}(p_{XY})$.}
\vspace{1ex}

(ii) For an arbitrary $q_{XY}\in\; \mathcal{P}_{XY}$, consider two
arbitrary joint pmfs $p_{XY,1}, p_{XY,0}\in\;
\mathcal{P}_{Y|X}(q_{XY})$, and two arbitrary distortion vectors
$\mathbf D_1, \mathbf D_0\in \mathcal D^2$.  For every $\lambda \in
(0,1)$, let $(p_{XY,\lambda},\mathbf D_\lambda):= \lambda
(p_{XY,1},\mathbf D_1)+ \bar \lambda (p_{XY,0},\mathbf D_0)$. We
need to show that $\rho^A_t(p_{XY,\lambda},\mathbf D_\lambda) \geq
\lambda\:\rho^A_t(p_{XY,1},\mathbf
D_1)+\bar{\lambda}\:\rho^A_t(p_{XY,0},\mathbf D_0)$. For $i=0,1$,
let $p_{U_1|X,i}$ and $\{\mathbf D_{u_1,i}\}_{u_1\in \mathcal U_1}$
be the conditional pmf and individual distortion vectors that
maximize the objective function in (\ref{eqn:convexifyrd}) for
source pmf $p_{XY,i}$ and distortion level $\mathbf D_i$. Let
$p_{U_1,i}$ and $p_{XY|U_1,i}$ denote,  {respectively,} the
$U_1$-marginal and conditional  {pmfs} of $p_{XY,i}p_{U_1|X,i}$.
Therefore we have $\rho^A_t(p_{XY,i},\mathbf D_i)=\sum_{u_1}
p_{U_1,i}(u_1) \rho^B_{t-1}(p_{XY|U_1,i}(\cdot,\cdot|u_1),\mathbf
D_{u_1,i})$. Define a new auxiliary variable $V\in
\mathcal{U}_1\times\{0,1\}$ such that $p_{V}(u_1,1):=\lambda
p_{U_1,1}(u_1)$ and $p_{V}(u_1,0):=\bar{\lambda} p_{U_1,0}(u_1)$.
Let $p_{XY|V}(\cdot,\cdot|u_1,i):=p_{XY|U_1,i}(\cdot,\cdot|u_1)$.
Then the $XY$-marginal pmf of $p_V p_{XY|V}$ is $p_{XY,\lambda}$.
Let the individual distortion vectors $\mathbf D_v$ for $v=(u_1,i)$
be $\mathbf D_{u_1,i}$. The overall distortion vector is $E[\mathbf
D_V]=\mathbf D_\lambda$.  We have,
\begin{eqnarray*}
  \lefteqn{\lambda\:\rho^A_t(p_{XY,1},\mathbf
    D_1)+\bar{\lambda}\:\rho^A_t(p_{XY,0},\mathbf D_0)}\\
  &=&\lambda \sum_{u_1}p_{U_1,1}(u_1)
  \rho^B_{t-1}(p_{XY|U_1,1}(\cdot,\cdot|u_1),\mathbf
  D_{u_1,1})+\bar{\lambda}\sum_{u_1}p_{U_1,0}(u_1)
  \rho^B_{t-1}(p_{XY|U_1,0}(\cdot,\cdot|u_1),\mathbf D_{u_1,0})\\
  &=&\sum_{u_1,i}p_V(u_1,i)\rho^B_{t-1}(p_{XY|V}(\cdot,\cdot|u_1,i),\mathbf
  D_{u_1,i})\\
  &\leq& \rho^A_t(p_{XY,\lambda},\mathbf D_\lambda),
\end{eqnarray*}
where the last step is because (\ref{eqn:convexifyrd}) holds for
source pmf $p_{XY,\lambda}$ and distortion level $\mathbf
D_\lambda$. \vspace{1ex}

(iii) For all $p_{XY}\in \mathcal{P}_{Y|X}(q_{XY})$, we have
\begin{eqnarray*}
\rho^A_t(p_{XY},\mathbf{D})&\stackrel{(d)}{=}&
\max_{p_{U_1|X}}\left\{\max_{\scriptstyle \forall u_1\in \mathcal U_1,
\mathbf{D}_{u_1}\in \mathcal{D}^2:\atop \scriptstyle
E [\mathbf{D}_{U_1}]=\mathbf{D}} \left\{ \sum_{u_1} p_{U_1}(u_1)
\rho^B_{t-1}(p_{XY|U_1}(\cdot,\cdot|u_1),
\mathbf{D}_{u_1})\right\}\right\}\\
&\stackrel{(e)}{\leq}& \max_{p_{U_1|X}}\left\{\max_{\scriptstyle
\forall u_1\in \mathcal U_1, \mathbf{D}_{u_1}\in \mathcal{D}^2:\atop
\scriptstyle
E [\mathbf{D}_{U_1}]=\mathbf{D}} \left\{ \sum_{u_1} p_{U_1}(u_1)
\rho(p_{XY|U_1}(\cdot,\cdot|u_1),
\mathbf{D}_{u_1})\right\}\right\}\\
&\stackrel{(f)}{\leq}&\rho(p_{XY}),
\end{eqnarray*}
 {where the equality (d) follows from (\ref{eqn:convexifyrd}), the
inequality (e) is true because $\rho^B_{t-1} \leq \rho$, and the final
inequality (f) is true due to the following reasons: (i)
$p_{XY|U_1}(\cdot,\cdot|u_1)\in\mathcal{P}_{Y|X}(p_{XY})\subseteq
\mathcal{P}_{Y|X}(q_{XY})$ due to Remarks~\ref{rem:margpertbsets} and
\ref{rem:lemma1i}, (ii) $\rho$ is concave on
$\mathcal{P}_{Y|X}(q_{XY})\times \mathcal D^2$, and (iii) Jensen's
inequality.}

\textcolor{myblue}{(iv) This part is proved by reversing the roles of terminal $A$ and $B$ in parts (i) -- (iii).}
\hspace*{\fill}~\QED\\

\section{\label{app:proofs_Kaspisquestion}Proofs of propositions in Section~\ref{section:Kaspi}}

\noindent{\em Proof of Proposition~\ref{lem:erasure}:}
Given a general $p_{U|X}$ and $\hat g_B$ satisfying the original
constraint in (\ref{eqn:generalrho1}), we will construct $U^*$
satisfying the stronger constraints in Proposition~\ref{lem:erasure}
with an objective function that is not less than the original one as
follows.

Without loss of generality, we assume $\supp(p_U)=\mathcal U$. For
$i=0,1$, let $\mathcal U_i:=\{u \in \mathcal U: p_{X|U}(i|u)=1\}$.
Let $\mathcal U_e:=\{u\in \mathcal U: p_{X|U}(1|u)\in(0,1)\}$. Then
$\{\mathcal U_1, \mathcal U_0, \mathcal U_e\}$ forms a partition of
$\mathcal U$. For each $u\in \mathcal U_e$, since
$p_{XY|U}(x,y|u)>0$ for all $(x,y)\in \{0,1\}^2$, it follows that
$\hat g_B(u,y=0)=\hat g_B(u,y=1)=e$ must hold, because otherwise
$E(d(X,\hat g_B(U,Y)))=\infty$. But for every $u\in \mathcal U_i$,
$i=0,1$, $\hat g_B(u,y)$ may equal $i$ or $e$ but not $(1-i)$ to get
a finite distortion. When we replace $\hat g_B$ by
\begin{equation*}
\hat g_B^*(u,y)= \left\{
\begin{array}{ll}
i, & \mbox{ if } u\in \mathcal U_i, i=0,1, \\
e, & \mbox{ if } u\in \mathcal U_e,
\end{array}\right.
\end{equation*}
the distortion for $u\in \mathcal U_i, i=0,1,$ is reduced to zero,
and the distortion for $u\in \mathcal U_e$ remains unchanged.
Therefore we have $E(d(X,\hat g_B^*(U,Y)))\leq E(d(X,\hat
g_B(U,Y)))\leq D$. We note that $\hat g_B^*(U,Y)$ is completely
determined by $U$. Let $U^* :=\hat g_B^*(U,Y)$. Then $U^*=i$ iff
$U\in \mathcal U_i, i=\{0,1,e\}$. The objective function
$H(X|Y,U)+H(Y|X) = H(X|Y,U,U^*)+H(Y|X) \leq H(X|Y,U^*)+H(Y|X)$,
which
completes the proof. \hspace*{\fill}~\QED\\

\noindent{\em Proof of Proposition~\ref{prop:DSBS}:}

For a fixed $p_{XY}$, $H(X|Y,U)+H(Y|X)$ is concave with respect to
$p_{XYU}$ and therefore also $p_{U|X}$. Since $p_{U|X}$ is linear
with respect to $(\alpha_{0e},\alpha_{1e})$,
$\psi(p_{XY},\alpha_{0e},\alpha_{1e})=H(X|Y,U)+H(Y|X)$ is concave
with respect to $(\alpha_{0e},\alpha_{1e})$.

The maximum in (\ref{eqn:rho1}) can be achieved along the axis of
symmetry given by $\alpha_{1e}=\alpha_{0e}$ because (i) $\psi$ and
$\phi$ are both symmetric with respect to $\alpha_{0e}$ and
$\alpha_{1e}$, i.e.,
$\psi(p_{XY},\alpha_{0e},\alpha_{1e})=\psi(p_{XY},\alpha_{1e},\alpha_{0e})$
and
$\phi(p_{XY},\alpha_{0e},\alpha_{1e})=\phi(p_{XY},\alpha_{1e},\alpha_{0e})$,
and (ii) $\psi(p_{XY},\alpha_{0e},\alpha_{1e})$ is a concave
function of $(\alpha_{0e},\alpha_{1e})$. When $D\in [0,1]$,
$\rho_1^A$ can be further simplified as follows.
\begin{equation*}
\rho_1^A(p_{XY},D) =\max_{\scriptstyle \alpha_{0e}=\alpha_{1e}\in
[0,D]} \psi(p_{XY},\alpha_{0e},\alpha_{1e})=(1+D)h_2(p),
\end{equation*}
which completes the proof. \hspace*{\fill}~\QED\\

\noindent{\em Proof of Proposition~\ref{prop:asymsource}:}
\begin{table}[!htb]
\vglue -0.3cm
  \centering
 \caption{Joint distribution $p_{X|Y}p_{Y,1}$}
\vglue -0.3cm
  \begin{tabular}{|c|c|c|}
  \hline
   $p_{X|Y}p_{Y,1}$ & $y=0$ & $y=1$ \\
   \hline
  $x=0$ & $\bar p\bar q$ & $pq$ \\
  \hline
  $x=1$ & $p\bar q$ & $\bar p q$ \\
  \hline
\end{tabular}
\label{tab:pXY1} \vglue -0.3cm
\end{table}
For the joint pmf $p_{X|Y}p_{Y,1}$ summarized in
Table~\ref{tab:pXY1}, functions $\psi$ and $\eta$ simplify even
further to special functions of $(p,q,\alpha_{0e},\alpha_{1e})$ as
follows:
\begin{eqnarray}
C(p,q,\alpha_{0e},\alpha_{1e})&=&\psi(p_{X|Y}p_{Y,1},\alpha_{0e},\alpha_{1e})
\nonumber\\
&=&\bar q(\bar p  \alpha_{0e}+p \alpha_{1e})h_2\left(\frac{\bar p
\alpha_{0e}}{\bar p \alpha_{0e}+ p
\alpha_{1e}}\right)\nonumber\\
&&+q(p \alpha_{0e}+\bar p
\alpha_{1e})h_2\left(\frac{p\alpha_{0e}}{p\alpha_{0e}+\bar p
  \alpha_{1e}}\right)\nonumber\\
&&+(\bar p \bar q + p q) h_2\left(\frac{\bar p \bar q}{\bar p \bar q
+
p q}\right)\nonumber\\
&&+(\bar p q+p \bar q)h_2\left(\frac{\bar p q}{\bar p q+p \bar
q}\right),\label{eqn:Cpq}\\
\eta(p,q,\alpha_{0e},\alpha_{1e})&=&\phi(p_{X|Y}p_{Y,1},\alpha_{0e},\alpha_{1e})
\nonumber\\
&=&(\bar p \bar q + p q)\alpha_{0e}+(\bar p q+p \bar
q)\alpha_{1e}.\nonumber
\end{eqnarray}
Observe that $C(p,q,\alpha_{0e},\alpha_{1e})=C(p,\bar
q,\alpha_{1e},\alpha_{0e})$, and $\eta(p,q,\alpha_{0e},\alpha_{1e})
= \eta(p,\bar q,\alpha_{1e},\alpha_{0e})$ hold. Therefore we have
\begin{eqnarray*}
\rho_1^A(p_{X|Y}p_{Y,2},D)&=&\max_{\scriptstyle
\alpha_{0e},\alpha_{1e}\in[0,1]: \atop \scriptstyle \eta(p,\bar q,
\alpha_{0e},\alpha_{1e})\leq D} C(p,\bar
q,\alpha_{0e},\alpha_{1e})\\
&=&\max_{\scriptstyle \alpha_{0e},\alpha_{1e}\in[0,1]: \atop
\scriptstyle
\eta(p,q, \alpha_{1e},\alpha_{0e})\leq D} C(p,q,\alpha_{1e},\alpha_{0e})\\
&=&\rho_1^A(p_{X|Y}p_{Y,1},D).
\end{eqnarray*}
It follows that
\begin{eqnarray}
\frac{\rho_1^A(p_{X|Y}p_{Y,1},D)+\rho_1^A(p_{X|Y}p_{Y,2},
  D)}{2}&=&\rho_1^A(p_{X|Y}p_{Y,1},D)\nonumber\\
  &\geq& C(p,q,\alpha_{0e},1)\nonumber
\end{eqnarray}
holds for $D=\eta(p,q,\alpha_{0e},1)$. \hspace*{\fill}~\QED

\noindent{\em Proof of Proposition~\ref{prop:takinglimit}:}

Since $D=\eta(p,q,\alpha_{0e},1)\in[0,1]$ always holds, we have
$\rho_1^A(p_{XY},D)=(1+D)h_2(p)$ due to (\ref{eqn:rho1DSBS}). We
will show that for any fixed $q\in(0,1/2)$ and $\alpha_{0e}\in
(0,1)$, $\lim_{p\rightarrow 0} C(p,q,\alpha_{0e},1)/h_2(p)>
\lim_{p\rightarrow 0} (1+D)$ holds, which implies that $\exists
p\in(0,1)$ such that $C(p,q,\alpha_{0e},1)/h_2(p)>(1+D)$, which, in
turn, implies Proposition~\ref{prop:takinglimit}. It is convenient
to use the following lemma to analyze the limits.
\begin{lemma}\label{lem:entropyratio}
Let $f(p)$ be a function differentiable around $p=0$ such that
$f(0)=0$ and $f'(0)>0$. Then
\[\lim_{p\rightarrow 0} \frac{h_2(f(p))}{h_2(p)}=f'(0)\]
\end{lemma}
\begin{proof}
Applying the l'H\^{o}pital rule several times, we have
\begin{eqnarray*}
\lim_{p\rightarrow 0} \frac{h_2(f(p))}{h_2(p)}&=&\lim_{p\rightarrow
0}
\frac{\ln(1-f(p))-\ln f(p)}{\ln(1-p)-\ln p } f'(0)\\
&=&\lim_{p\rightarrow 0} \frac{\ln f(p)}{\ln p}
f'(0)\\&=&\lim_{p\rightarrow 0} \frac{p}{f(p)} (f'(0))^2\\&=&f'(0),
\end{eqnarray*}
which completes the proof of Lemma~\ref{lem:entropyratio}.
\end{proof}

Applying Lemma~\ref{lem:entropyratio}, we have
\begin{eqnarray}
 &&\lim_{p\rightarrow 0}
\frac{C(p,q,\alpha_{0e},1)}{h_2(p)} =2 - q(1-\alpha_{0e}),\label{eqn:limitC}\\
&&\lim_{p\rightarrow 0} (1+D) =
2-\bar q (1- \alpha_{0e}),\label{eqn:limitrho1}\\
&&\lim_{p\rightarrow 0}
\left(\frac{C(p,q,\alpha_{0e},1)}{h_2(p)}-(1+D)\right) =
(1-2q)(1-\alpha_{0e}).\nonumber
\end{eqnarray}
Therefore for any $\alpha_{0e}\in(0,1)$ and $q\in(0,1/2)$, there
exists a small enough $p$ such that $C(p,q,\alpha_{0e},1)>(1+D)$
holds, which completes the proof.\hspace*{\fill}~\QED\\

%

\noindent{\em Proof of Proposition~\ref{prop:largeratio}:}

For the rate-distortion tuple $(R_1,R_2,D)$ corresponding to the
choice of $p_{V_1|Y}$, $p_{V_2|XV_1}$ and $\hat g_B$ described in
the proof of Theorem~\ref{thm:largeratio}, we have (i) $R_1 =
I(Y;V_1|X)=H(Y|X)- C_2(p,q)$, where $C_2(p,q)$ is the sum of the
last two terms in (\ref{eqn:Cpq}); (ii)
$R_2=I(X;V_2|Y,V_1)=2h_2(p)-C(p,q,\alpha,1)-R_1$; and (iii)
$D=\eta(p,q,\alpha,1)$. It follows that
\begin{eqnarray*}
\lim_{p\rightarrow 0} \frac{R_{1}}{h_2(p)} \!\!\!&=&\!\!\! 0,\\
\lim_{p\rightarrow 0} \frac{R_{2}}{h_2(p)} \!\!\!&=&\!\!\!
2-\lim_{p\rightarrow 0}
\frac{C(p,q,\alpha,1)}{h_2(p)}-\lim_{p\rightarrow 0}
\frac{R_1}{h_2(p)} =q(1-\alpha).
\end{eqnarray*}
Therefore for all $q>0$ and $\alpha\in (0,1)$, we have
\begin{equation}\lim_{p\rightarrow
0}\frac{R_1}{R_2}=0.\label{eqn:R1R2}
\end{equation}
For the one-message rate-distortion function, we have
$R_{sum,1}(p_{XY},D)=2h_2(p)-\rho_1^A(p_{XY},D)$, where
$\rho_1^A(p_{XY},D)$ is given by (\ref{eqn:rho1DSBS}). Therefore we
have
\[
\lim_{p\rightarrow 0} \frac{R_{sum,1}(p_{XY},D)}{h_2(p)} {=}
2-\lim_{p\rightarrow 0} \frac{\rho_1^A(p_{XY},D)}{h_2(p)} {=} \bar q
(1-\alpha),
\]
which implies that
\begin{equation}
\lim_{p\rightarrow 0} \frac{R_{sum,1}(p_{XY},D)}{R_1+R_2} =
\frac{\bar q}{q}. \label{eqn:Rsum1R1plusR2}
\end{equation}
For any $L >0$, we can always find a small enough $q>0$ such that
$\bar q/q > L+1$. Due to (\ref{eqn:R1R2}) and
(\ref{eqn:Rsum1R1plusR2}), there exists $p>0$ such that
$R_1/R_2<1/L$ and $R_{sum,1}/(R_1+R_2)>L$.\hspace*{\fill}~\QED

\begin{remark}
The convergence of the limit analyzed in
Lemma~\ref{lem:entropyratio} is actually slow, because the logarithm
function increases to infinity slowly. The consequence is that if
one chooses a small $q$ to get $R_{sum,1}/(R_1+R_2)$ close to the
limit $\bar q/q$, then $p$ needs to be very small. For example, when
$q=1/10, \alpha_{0e}=1/2$, $\bar q/q=9$, with $p=10^{-200}$, we get
$R_{sum,1}/R_{sum,2}^*\approx 8.16$. This, however, does not mean
that the benefit of multiple messages only occurs in extreme cases.
In  {computer simulations} we have observed that for the erasure
distortion, the gain for certain asymmetric sources can be much more
than that for the DSBS example analyzed in this paper. The DSBS
example was chosen in this paper only because it is easy to analyze.
\end{remark}

\footnotesize
\bibliographystyle{IEEEtran}
\bibliography{mybibfile}

\begin{thebibliography}{10}
\providecommand{\url}[1]{#1}
\csname url@samestyle\endcsname
\providecommand{\newblock}{\relax}
\providecommand{\bibinfo}[2]{#2}
\providecommand{\BIBentrySTDinterwordspacing}{\spaceskip=0pt\relax}
\providecommand{\BIBentryALTinterwordstretchfactor}{4}
\providecommand{\BIBentryALTinterwordspacing}{\spaceskip=\fontdimen2\font plus
\BIBentryALTinterwordstretchfactor\fontdimen3\font minus
  \fontdimen4\font\relax}
\providecommand{\BIBforeignlanguage}[2]{{%
\expandafter\ifx\csname l@#1\endcsname\relax
\typeout{** WARNING: IEEEtran.bst: No hyphenation pattern has been}%
\typeout{** loaded for the language `#1'. Using the pattern for}%
\typeout{** the default language instead.}%
\else
\language=\csname l@#1\endcsname
\fi
#2}}
\providecommand{\BIBdecl}{\relax}
\BIBdecl

\bibitem{ISIT08}
{N.~Ma and P.~Ishwar}, ``{Two-terminal distributed source coding with
  alternating messages for function computation},'' in \emph{Proc.~IEEE
  Int.~Symp.~Information~Theory}, {Toronto, Canada}, {{Jul.~6--11,}} 2008, pp.
  51--55, arXiv:0801.0756v4.

\bibitem{Journal_interactive}
N.~Ma and P.~Ishwar, ``Some results on distributed source coding for
  interactive function computation,'' \emph{IEEE Trans.~Inf.~Theory}, vol.~57,
  no.~9, pp. 6180 -- 6195, Sep. 2011.

\bibitem{Kaspi1985}
A.~H. Kaspi, ``Two-way source coding with a fidelity criterion,'' \emph{IEEE
  Trans.~Inf.~Theory}, vol.~31, no.~6, pp. 735--740, Nov. 1985.

\bibitem{Yao1979}
A.~C. Yao, ``Some complexity questions related to distributed computing,'' in
  \emph{Proc. 11th Annu. ACM Symp. Theory of Computing}, {Atlanta, GA},
  {{Apr.~30--May~2,}} 1979, pp. 209--213.

\bibitem{CommComplexity}
E.~Kushilevitz and N.~Nisan, \emph{Communication Complexity}.\hskip 1em plus
  0.5em minus 0.4em\relax Cambridge: Cambridge University Press, 1997.

\bibitem{Orlitskyworst1}
A.~Orlitsky, ``Worst-case interactive communication i: two messages are almost
  optimal,'' \emph{IEEE Trans.~Inf.~Theory}, vol.~36, no.~5, pp. 1111--1126,
  Sep. 1990.

\bibitem{Orlitskyworst2}
------, ``Worst-case interactive communication ii: two messages are not
  optimal,'' \emph{IEEE Trans.~Inf.~Theory}, vol.~37, no.~4, pp. 995--1005,
  Jul. 1991.

\bibitem{OrlitskyRoche}
A.~Orlitsky and J.~R. Roche, ``Coding for computing,'' \emph{IEEE
  Trans.~Inf.~Theory}, vol.~47, no.~3, pp. 903--917, Mar. 2001.

\bibitem{Journal_col}
N.~Ma, P.~Ishwar, and P.~Gupta, ``Interactive source coding for function
  computation in collocated networks,'' accepted for publication by \emph{IEEE
  Trans.~Inf.~Theory}.

\bibitem{YeungISIT}
X.~Yan, R.~W. Yeung, and Z.~Zhang, ``The capacity region for multi-source
  multi-sink network coding,'' in \emph{Proc.~IEEE
  Int.~Symp.~Information~Theory}, {Nice, France}, {{Jun.~24--29,}} 2007, pp.
  116--120.

\bibitem{networkcoding}
R.~W. Yeung, \emph{Information theory and network coding}.\hskip 1em plus 0.5em
  minus 0.4em\relax New York, New York: Springer Verlag, 2008.

\bibitem{ConvexAnalysis}
R.~T. Rockafellar, \emph{Convex Analysis}.\hskip 1em plus 0.5em minus
  0.4em\relax Princeton: Princeton University Press, 1970.

\bibitem{convexhull}
F.~P. Preparata and M.~I. Shamos, \emph{Computational Geometry: An
  Introduction}.\hskip 1em plus 0.5em minus 0.4em\relax New York, New York:
  Springer Verlag, 1985.

\bibitem{Bertsekas}
D.~P. Bertsekas, \emph{Nonlinear Programming}.\hskip 1em plus 0.5em minus
  0.4em\relax Belmont, Massachusetts: Athena Scientific, 1995.

\bibitem{Zamir_EAP}
A.~Cohen and R.~Zamir, ``Entropy amplification property and the loss for
  writing on dirty paper,'' \emph{IEEE Tran.~Info.~Theory}, vol.~54, no.~4, pp.
  1477--1487, Apr. 2008.

\bibitem{Csi}
I.~Csisz\'ar and J.~K\"orner, \emph{Information Theory: Coding Theorems for
  Discrete Memoryless Systems}.\hskip 1em plus 0.5em minus 0.4em\relax Hungary:
  Akad\'emiai Kiad\'o, 1986.

\bibitem{CoverThomas}
T.~M. Cover and J.~A. Thomas, \emph{Elements of Information Theory}.\hskip 1em
  plus 0.5em minus 0.4em\relax New York: Wiley, 1991.

\end{thebibliography}

\end{document}